%% file: main.tex
\documentclass[dvips,12pt,a4paper]{article}
\usepackage{graphicx}
\usepackage{styles/here}
\usepackage{amsmath}
\usepackage{amssymb}
\usepackage{fancyhdr}
\usepackage{hhline}
\usepackage{rotating}
\usepackage{color}
\include{definitions}

\begin{document}
\input{title}
\input{introduction}
\input{simulation}
\input{generator}
\input{analysis}
\input{conclusion}
\input{acknowledgments}
\appendix
\input{mass}

\input{books}
\end{document}

%% file: definitions.tex
\definecolor{mycolor}{gray}{0.8}

\def\Ps{M_\mathrm{Pl}}
\def\TPs{M_P}
\def\Mp{M_P}
\def\mbh{M_\mathrm{BH}}
\def\Rs{R_\mathrm{S}}
\def\Th{T_H}
\def\Mbhth{M_\mathrm{BH}^\mathrm{th}}
\def\mbhmin{M_\mathrm{BH}^\mathrm{min}}
\def\fbhth{f_\mathrm{BH}^\mathrm{th}}

\def\MBHTru{M_\mathrm{BH}^\mathrm{true}}

\def\Fb{fb$^{-1}$}
\def\Pb{pb$^{-1}$}

\def\ILu{$\int\mathcal{L}dt$}
\def\ILuD{$\int\mathcal{L}_{\mathrm{discovery}}dt$}

\def\epsmin{\varepsilon^{\mbhmin=1}}

\def\BmL{B-L}

\def\bgwj{$W^{\pm}q$}
\def\bggj{$\gamma q$}
\def\bgzj{$Zq$,$\gamma^{*}q$}
\def\bgqq{$qq$}

\def\bggg{$\gamma\gamma$}
\def\bggv{$\gamma V$}
\def\bgtt{$\ttpair$}
\def\bgwz{$W^{\pm}Z$}
\def\bgzz{$ZZ$}
\def\bgww{$W^{\pm}W^{\mp}$}

\newcommand{\mET}       {\not \!\!\! {E_T} }

\newcommand{\pt}        {p_{T}}
\newcommand{\pz}        {p_{z}}

\newcommand{\ttpair}        {{\mathrm t}\bar{\mathrm t}}

\def\etal{{\it et~al.}}
\def\ie{{i.e.}}

\newcommand{\enm}[2]{$#1\cdot 10^{#2}$}

\newcommand{\prl}[3]{Phys. Rev. Lett. {\bf #1}, #2 (#3)}
\newcommand{\prdd}[3]{Phys. Rev. D {\bf #1}, #2 (#3)}

\newcommand{\plb}[3]{Phys. Lett. B {\bf #1}, #2 (#3)}

%% file: title.tex
\begin{titlepage}

\begin{flushright}
    \large
    ATL-PHYS-2003-037
\end{flushright}

\bigskip

\bigskip

\bigskip

\begin{center}
{\bf \Huge
Study of Black Holes
with the ATLAS detector
at the LHC}
\end{center}

\bigskip

\bigskip

\bigskip

\begin{center}
{\large J.~Tanaka$^{1,\dagger}$, T.~Yamamura$^{2}$, S.~Asai$^{1}$, J.~Kanzaki$^{3}$}
\end{center}

\noindent $^{1}$ International Center for Elementary Particle Physics~(ICEPP), University of Tokyo
\newline
$^{2}$ Department of Physics, Faculty of Science, University of Tokyo
\newline
$^{3}$ High Energy Accelerator Research Organization~(KEK)
\newline
$^{\dagger}$ Mail address: Junichi.Tanaka@cern.ch
\newline

\begin{abstract}

We evaluate the potential of the ATLAS detector for discovering black holes produced at the LHC,
as predicted in models with large extra dimensions where
quantum gravity is at the TeV scale.
We assume that black holes decay by Hawking evaporation to all Standard Model particles democratically.
We comment on the possibility to estimate the Planck scale.

\end{abstract}


\end{titlepage}

%% file: introduction.tex
\section{Introduction}\label{sec:intro}
Black holes will be produced at the Large Hadron Collider~(LHC)
if the fundamental Planck scale is of order a TeV~\cite{BH,BH2}.
This scenario occurs in the model of large extra dimensions~\cite{ADD} proposed by
Arkani-Hamed, Dimopoulos and Dvali, where
the Standard Model~(SM) gauge and matter fields are confined to a 3-dimensional brane
while gravity is free to propagate in extra dimensions of large size.
This model was motivated by the necessity to solve the hierarchy problem.

Using Gauss' law, the fundamental Planck scale of the (4+$n$)-dimension $\TPs$ is related to
the 4-dimensional Planck scale $\Ps$~($\sim10^{19}$~GeV) by
\[
\Ps^2 \sim \TPs^{n+2}R^n,
\]
where $n$ is the number of extra dimensions and $R$ is the size of the compactified dimensions.
Assuming that the fundamental Planck scale $\TPs$ is the same as the electroweak scale~($\sim$1~TeV),
the case $n=1$ yields a very large $R$~($\sim10^{13}$~cm), ruled out by experiments.
For $n\ge2$, the size of $R$ is less than $\sim10^{-2}$~cm, which does not contradict results of gravitational experiments~\cite{GRAV}.
From astrophysical constraints~\cite{PDG}, the size of $\TPs$ is larger than $O$(TeV) and $O$(10~TeV)
for $n=2$ and $3$, respectively. In particular, the case of $n=2$ would be ruled out
from the point of view of the solution of the hierarchy problem~\cite{YU}.

We consider here black holes of mass $\mbh$ much larger than the fundamental Planck scale $\TPs$
since the phenomenology of black holes at $\mbh\sim\TPs$ is very complex and beyond the scope of the present study.

For the collision of two partons at a center-of-mass energy $\sqrt{\hat{s}}=\mbh$,
by classical arguments~\cite{BH} the total cross section is given by
\[
\sigma(\mbh) \sim \pi \Rs^2 = \frac{1}{\Mp^2} [ \frac{\mbh}{\Mp} ( \frac{8\Gamma (\frac{n+3}{2})}{n+2} ) ]^{\frac{2}{1+n}},
\]
where $\Rs$ is the Schwarzschild radius:
\[
\Rs = \frac{1}{\sqrt{\pi}\Mp} [ \frac{\mbh}{\Mp} ( \frac{8\Gamma (\frac{n+3}{2})}{n+2} ) ]^{\frac{1}{1+n}}.
\]
For proton-proton collisions at the LHC,
the differential cross section is given by
\begin{align}
\frac{ d\sigma (pp\to\mathrm{BH}+X) }{ d\mbh } = 
\frac{ dL }{ d\mbh } \hat{\sigma}(ab\to\mathrm{BH})|_{\hat{s}=\mbh^2}, \label{eq:bh-diffcross}
\end{align}
\[
\frac{ dL }{ d\mbh } = \frac{2\mbh}{s}\sum_{a,b}\int^1_{\mbh^2/s}
\frac{ dx_a }{ x_a } f_a(x_a)f_b(\frac{\mbh^2}{sx_a})
\]
\[
\hat{\sigma}(ab\to\mathrm{BH})|_{\hat{s}=\mbh^2} = \pi \Rs^2,
\]
where $a$ and $b$ are partons in protons and $f_i(x)$ are the parton distribution functions~(PDFs).
An exponential suppression of the geometrical cross section has been proposed by Voloshin~\cite{Voloshin}.
However since subsequent studies did not support this result~\cite{NoVoloshin},
we do not consider the effect here.

Black holes decay through several phases~\cite{BH2}:
{\it balding},
{\it Hawking evaporation}~({\it spin-down} and {\it Schwarzschild}) and
{\it Planck} phases.
Since the {\it Hawking evaporation} phase is expected to be the main phase of the decay,
we naively consider only this process and assume that a black hole evaporates until its mass becomes zero.

The radiation is characterized by the Hawking temperature $\Th$,
\[
\Th(\Mp,n,\mbh) = \Mp (\frac{ \Mp }{ \mbh }\frac{ n+2 }{ 8\Gamma (\frac{n+3}{2}) })^{\frac{1}{n+1}}
\frac{ n+1 }{ 4\sqrt{\pi} } = \frac{ n+1 }{ 4\pi\Rs }.
\]
The heavier the black hole, the colder are its decay products.
As it evaporates, its temperature increases but
we will ignore the time evolution of black holes in this phase.

We assume that the evaporation is described by black body radiation.
The energy spectrum of the decay products obeys the Boltzmann distribution~\cite{BH}:
\[
\frac{dN}{dE} \sim \frac{x^2}{e^x+c},
\]
where $x\equiv E/\Th$ and $c$ is a constant, which depends on the quantum statistics of the decay products, \ie, $c=-1, +1, 0$ for bosons, fermions and Boltzmann statistics, respectively.
In this study, we use $c=0$ for all particles.

Certain conservation laws are obeyed in the decay of back holes.
We assume that black holes decay ``democratically'', \ie, with
roughly equal probability to all of the SM particles.

In this paper, we present the discovery potential of the black holes at the LHC.
We describe a method of estimation of the fundamental Planck scale~$\TPs$ from the discovery potential.

The outline of this paper is as follows:
the simulation conditions and description of signal and background samples are given in Section~\ref{sec:simulation}.
Our original generator of the black holes is explained in detail in Section~\ref{sec:generator}.
In Section~\ref{sec:analysis}, we describe the event selection
and the reconstruction and finally,
a conclusion is given in Section~\ref{sec:conclusion}.

%% file: simulation.tex
\section{Simulation}\label{sec:simulation}
The generator, developed for the purpose of this study,
is described in Section~\ref{sec:generator}.
Initial state parton showers, hadronisation and decay are performed using PYTHIA~6.2~\cite{PYTHIA}.
All background samples are generated by PYTHIA~6.2.
CTEQ5L is used for the parton distribution function.

All samples are processed through the parameterized simulator, ATLFAST~\cite{ATLFAST}, of the ATLAS detector.
The energy resolutions and efficiencies for jet and particle reconstruction are
corrected using the results of the full detector simulation.

\subsection{Signal}
We have generated signal samples for the various values of $(\TPs,n)$ listed in Table~\ref{table:signals}.
We have generated black holes whose mass~$\mbh$ is larger than $\TPs$.
Factorization and renormalization scales are set to a mass~$\mbh$ of the generated black hole.
We call the lower limit of a black hole mass a threshold mass~$\Mbhth=\fbhth\times\TPs$.
We assume that the energy spectrum of all the products from the black hole decay obeys the Boltzmann distribution.

\begin{table}[H]
\caption{Cross section of signal samples for each $(\TPs,n)$ parameter.
$\fbhth$, which is described in the text, is 1 for all cases. The unit of $\TPs$ is TeV.}
\begin{center}
\begin{tabular}{cccc} \hline \hline
$\TPs,n$ & $\sigma$~(pb) & $\TPs,n$ & $\sigma$~(pb) \\ \hline
1,2 & $9.45\cdot 10^{3}$ & 5,2 & 0.662 \\
1,3 & $8.26\cdot 10^{3}$ & 5,3 & 0.603 \\
1,4 & $8.06\cdot 10^{3}$ & 5,5 & 0.625 \\
1,5 & $8.24\cdot 10^{3}$ & 5,7 & 0.699 \\
1,7 & $9.05\cdot 10^{3}$ & & \\
3,2 & 25.5 & 6,2 & 0.125 \\
3,3 & 22.9 & 6,3 & 0.114 \\
3,5 & 23.4 & 6,5 & 0.119 \\
3,7 & 26.0 & 6,7 & 0.133 \\
4,2 & 3.74 & 7,2 & 0.0229 \\
4,3 & 3.38 & 7,3 & 0.0210 \\
4,5 & 3.49 & 7,5 & 0.0220 \\
4,7 & 3.89 & 7,7 & 0.0247 \\ \hline \hline
\end{tabular}
\end{center}
\label{table:signals}
\end{table}


\subsection{Background}
Our background samples are listed in Table~\ref{table:backgrounds}.
We use kinematical cuts at generation level to save CPU time and
storage space of the data.
The lower limit for the energy in the center-of-mass
of initial partons, is set to 50~GeV for all the samples.
The lower limit for the transverse momentum in the rest frame
of initial partons~$\hat{p}_{T}^{min}$, is also used as shown in Table~\ref{table:backgrounds}.

\begin{table}[H]
\caption{Cross section of background samples for each mode.
``Proc'' is a process identification defined in the PYTHIA.}
\begin{center}
\begin{tabular}{cccc} \hline \hline
Process & $\sigma$~(pb) & $\hat{p}_{T}^{min}$ & Proc \\ \hline
\bgqq~($q=$quark,lepton,gluon) & $1.29\cdot 10^{4}$ & 280.0 & 11,12,13,28,53,68 \\
\bgtt & 493 & 10.0 & 81,82 \\
\bgww & 0.468 & 240.0 & 25 \\
\bgwz & 25.9 & 10.0 & 23 \\
\bgzz & 10.6 & 10.0 & 22 \\
\bggg & 229 & 10.0 & 18,114 \\
\bggv~($V=W^{\pm},\gamma^{*},Z$) & 280 & 10.0 & 19,20 \\
\bgwj & 73.4 & 240.0 & 16,31 \\
\bgzj & 31.5 & 240.0 & 15,30 \\
\bggj & 23.5 & 240.0 & 14,29,115 \\ \hline \hline
\end{tabular}
\end{center}
\label{table:backgrounds}
\end{table}


%% file: generator.tex
\section{Generator}\label{sec:generator}
A black hole generator was developed for this study, based on the
assumptions and approximations discussed in Section~\ref{sec:intro}\footnote{
Assumptions and approximations of our generator are similar with
a generator, TRUENOIR, described in \cite{LD}.
Another generator has recently been published~\cite{Cambridge}.}.

We require the following conditions for the decay of black holes:
\begin{itemize}
\item The number of fermions constrained by spin conservation
\item Four-Momentum conservation
\item Assumption of Boltzmann distribution for the energy spectrum of decay products
\item Color conservation
\item Assumption of democratic decay
\item Charge conservation
\item Option of conservation of the difference between the lepton number and the baryon number~($\BmL$ conservation)
\item Option of implementing Voloshin suppression~\cite{Voloshin}
\end{itemize}
We can select whether to apply the $\BmL$ conservation and the Voloshin suppression, but
they are switched off~(no $\BmL$ conservation and no Voloshin suppression) in this study.

For the production of the black holes, we use the Eq.~(\ref{eq:bh-diffcross}) of Section~\ref{sec:intro}.
Using the PDFs, the types of partons in the initial hard scattering are selected randomly.

\subsection{Spin Conservation}
We require that the number of fermions in the decay products of the black holes
be even or odd according to the parity of the number of initial state fermions.
Note that we do not consider gravitons in this study.

\subsection{Four-Momentum Conservation and Boltzmann Distribution}

We assume that the energy spectrum of the decay products follows a
Boltzmann distribution. However, since energy and momentum must be
conserved, all particles cannot be sampled randomly from that distribution.
We do not allow the case $N=2$ since it is fully constrained. For $N\ge 3$,
particle energies are successively sampled from the Boltzmann distribution
until the summed energy exceeds the mass of the black hole. The last
particle is then given the energy required by the energy constraint.
The first $N-2$ particles are assigned random directions and
the last two are emitted in a direction such as to conserve overall momentum.



\subsection{Color Conservation}

\begin{figure}[H]
\begin{center}
\resizebox{0.20\textwidth}{!}{\includegraphics{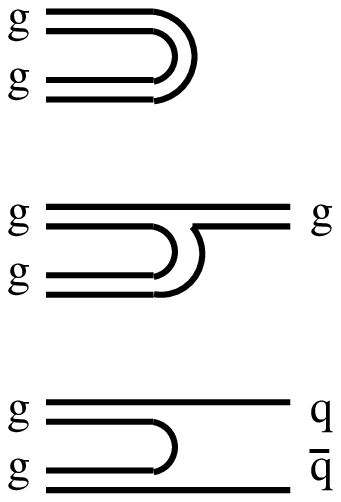}}
\hspace{1.5cm}
\resizebox{0.20\textwidth}{!}{\includegraphics{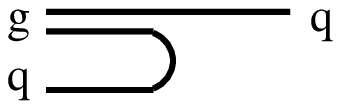}}
\end{center}
\caption{Color connection for $gg$~(left) and $gq$~(right).}
\label{fig:color-gg}
\end{figure}

Color is assumed conserved in the decay of black holes. 
Color is assigned to the initial partons and color connection is
implemented. As an example, if the initial partons are two gluons, we select
randomly either a $q\bar{q}$ pair or a gluon from the final state partons, if
present. 
We then connect the colors as shown in Figure~\ref{fig:color-gg}.
If no $q\bar{q}$ pair or gluon is present,
the color connection must be applied between the two initial state gluons,
if possible. Another example of color connection when
the initial partons are a quark and a gluon is shown in Figure~\ref{fig:color-gg}.

For the remaining partons, color connection is applied between pairs with
closest opening angle. If there is an odd gluon remaining, it is connected
to the $q\bar{q}$ pair or a $gg$ pairs or to the initial partons.

\subsection{Democratic Decay}
We assume a democratic decay, constrained by the conservation laws.
The number of degrees of freedom for each particle takes into account charge, spin and color.
We do not consider gravitons in this study.
The mass of the SM Higgs is set at 120~GeV.
Events are produced where the types of particles are chosen
randomly with the probabilities listed in Table~\ref{table:democratic-decay}
and are accepted only if conservation laws can be applied.

\begin{table}[H]
\caption{Degrees of freedom and assigned probability in the generator for each particle.}
\begin{center}
\begin{tabular}{ccc} \hline \hline
Particle & Degrees of freedom & Assigned probability \\ \hline
$g$~(gluon) & 8 & 0.0690 \\
$W$         & 6 & 0.0517 \\
$Z$         & 3 & 0.0259 \\
$\gamma$    & 2 & 0.0172 \\
lepton~($e,\mu,\tau$) & 4 & 0.0345 \\
neutrino~($\nu_e,\nu_\mu,\nu_\tau$) & 4 & 0.0345 \\
quark~($u,d,c,s,t,b$) & 12& 0.1034 \\
Higgs       & 1 & 0.0086 \\
Graviton    & 5 & 0.0000 \\ \hline \hline
\end{tabular}
\end{center}
\label{table:democratic-decay}
\end{table}

\subsection{Performance}
\begin{figure}[H]
\begin{center}
\resizebox{0.45\textwidth}{!}{\includegraphics{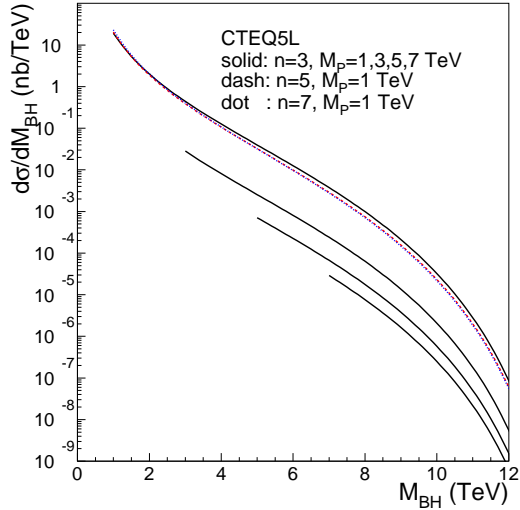}}
\end{center}
\caption{Differential cross section as a function of a black hole mass for each ($\TPs$,$n$) parameter.}
\label{fig:bh-diffcross}
\end{figure}

\begin{figure}[H]
\hspace{1.0cm} (a) ($\Mp$, $n$)=(1,3) \hspace{3.6cm} (b) ($\Mp$, $n$)=(1,7)
\vspace{-0.6cm}
\begin{center}
\resizebox{0.45\textwidth}{!}{\includegraphics{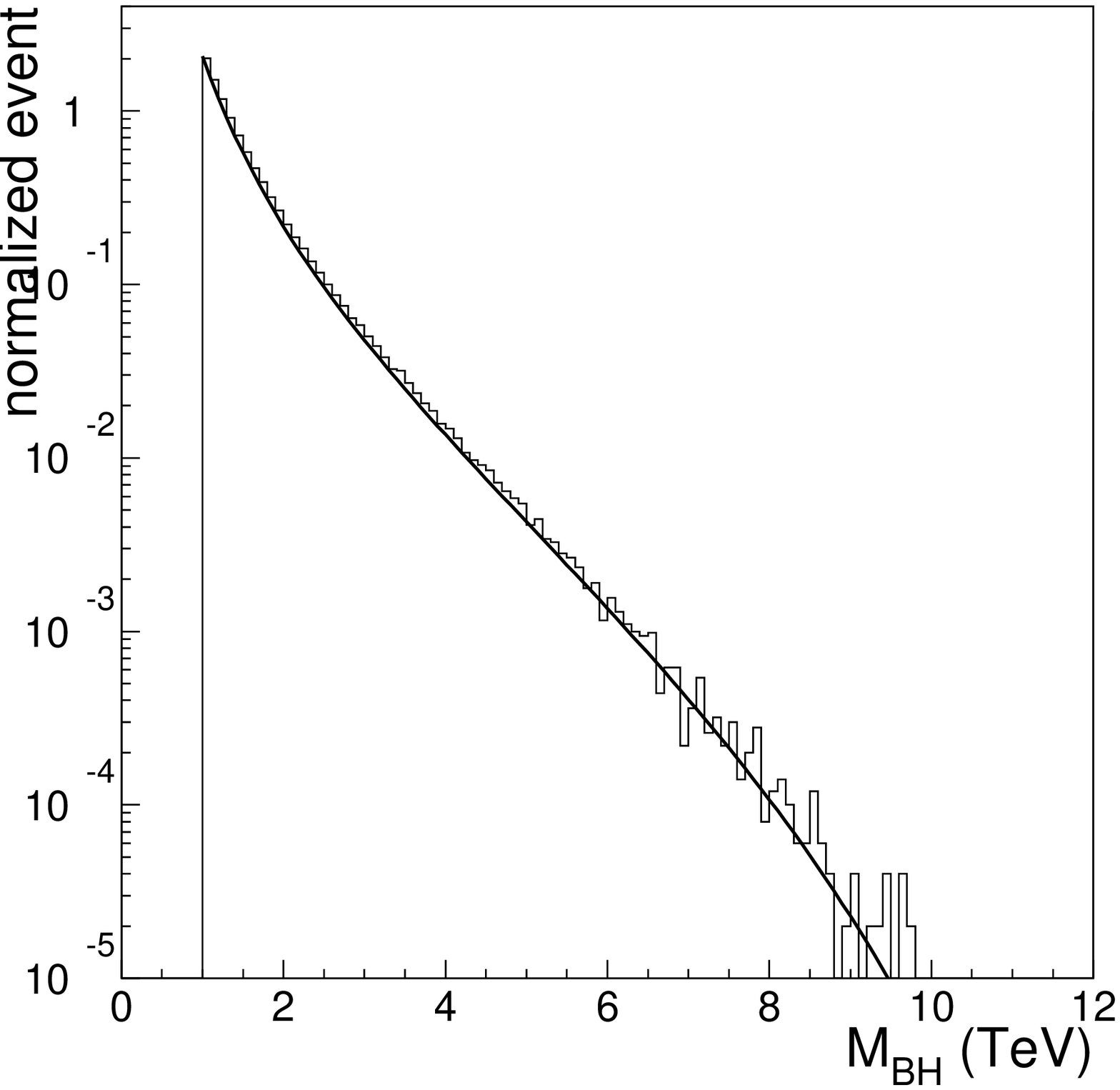}}
\resizebox{0.45\textwidth}{!}{\includegraphics{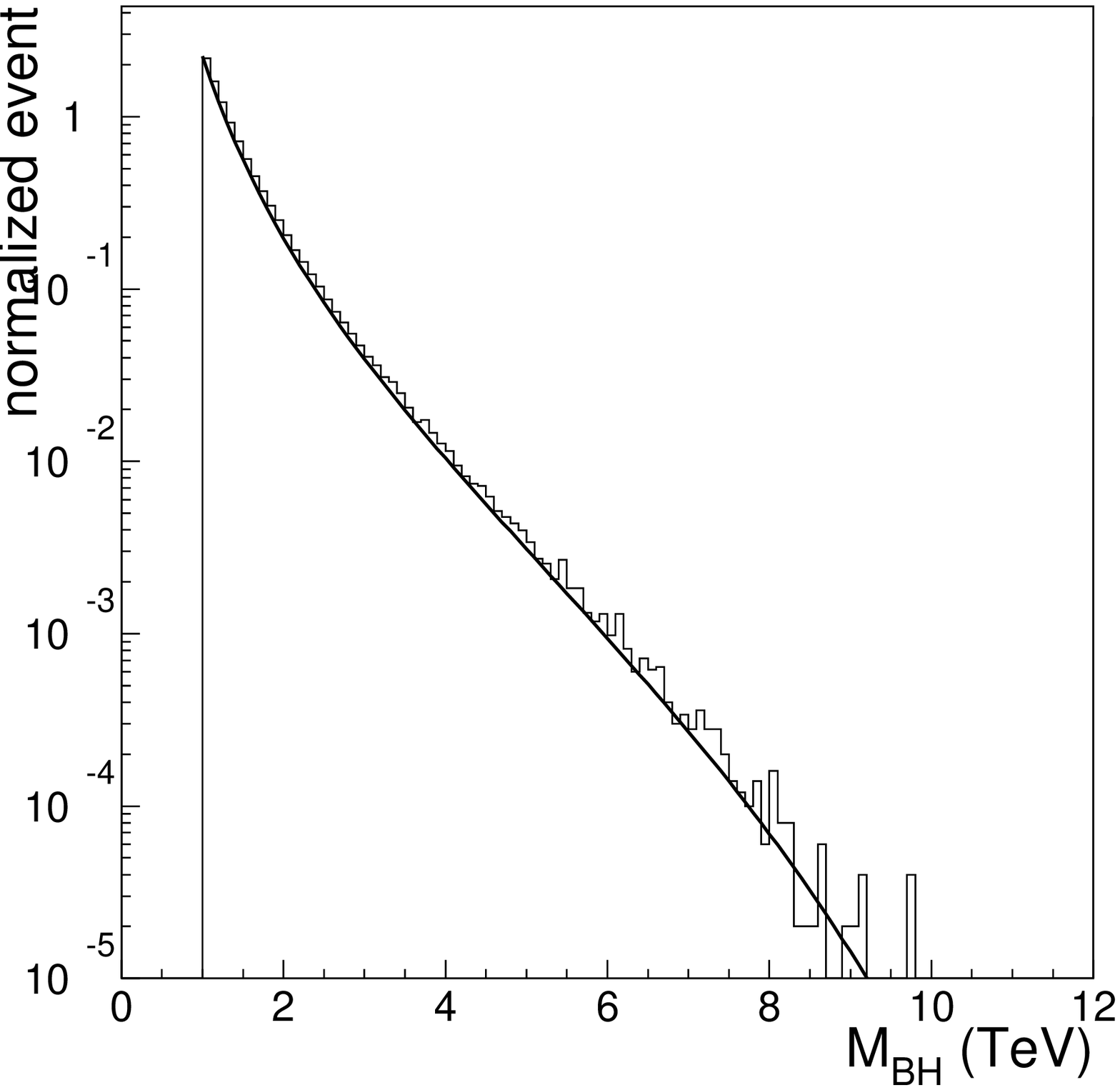}}
\end{center}
\hspace{1.0cm} (c) ($\Mp$, $n$)=(3,3) \hspace{3.6cm} (d) ($\Mp$, $n$)=(5,3)
\vspace{-0.6cm}
\begin{center}
\resizebox{0.45\textwidth}{!}{\includegraphics{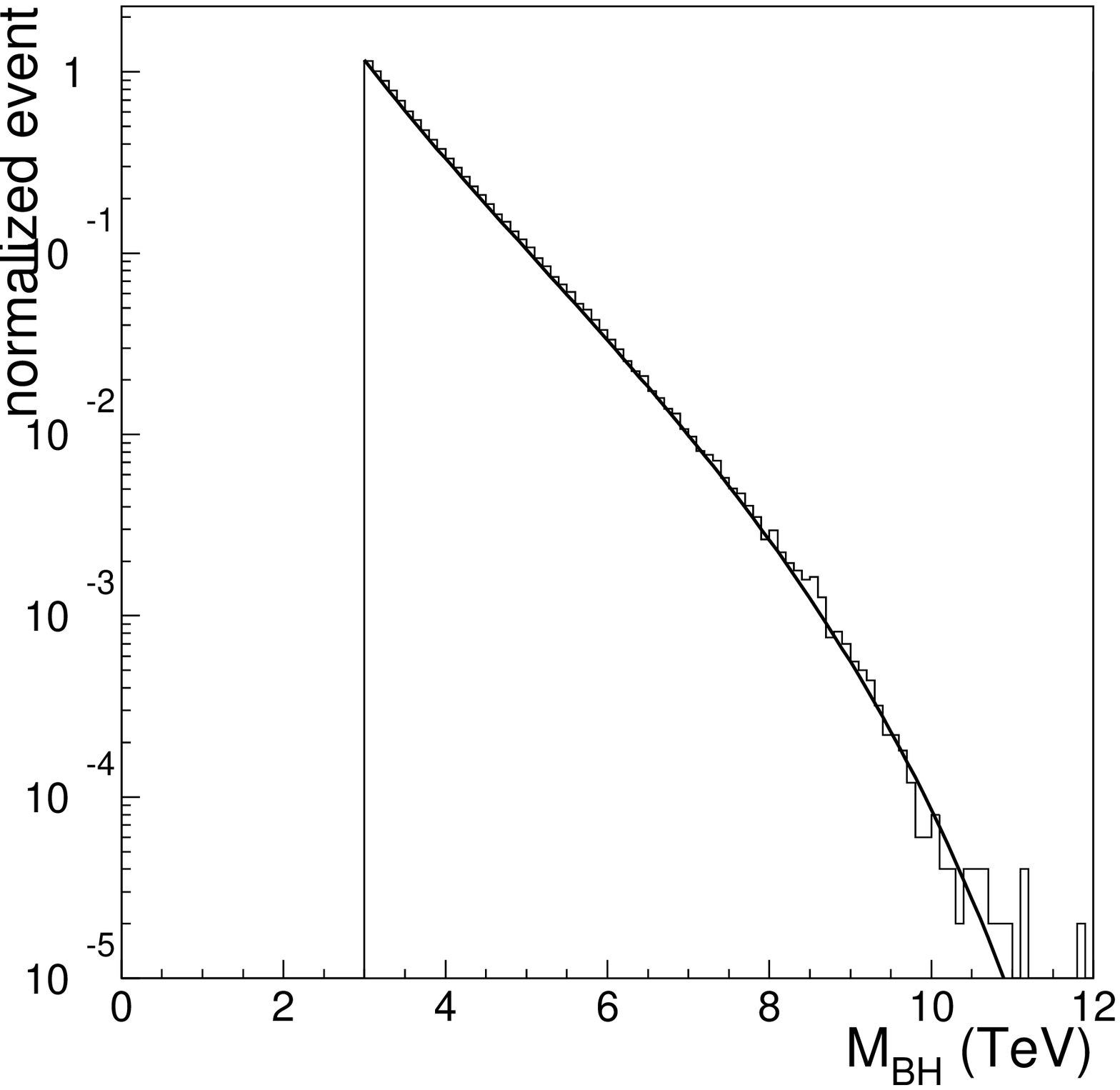}}
\resizebox{0.45\textwidth}{!}{\includegraphics{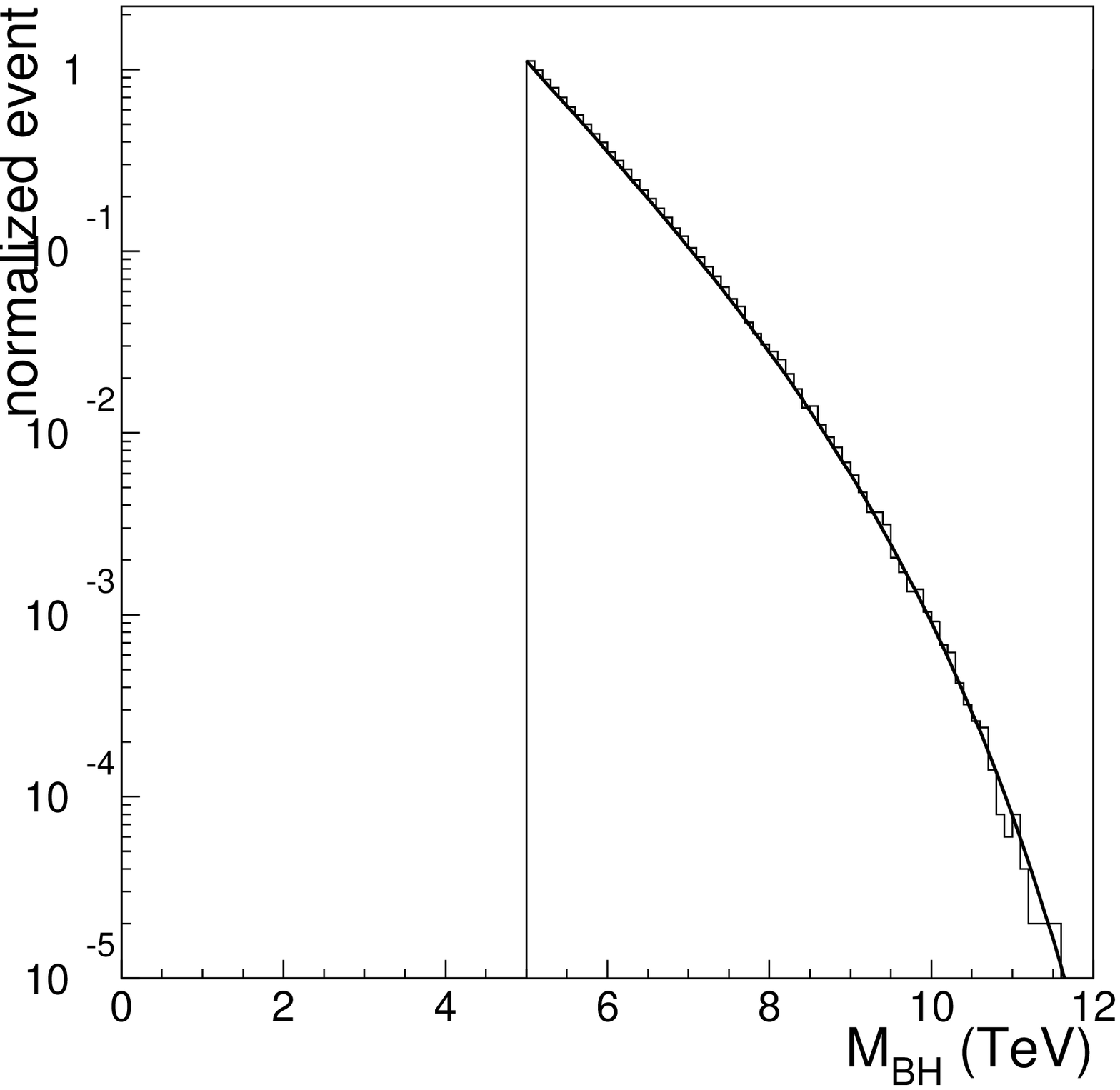}}
\end{center}
\caption{Mass distributions of the generated black holes.
The histogram shows generated black holes and the solid line shows an expected shape from Eq.~(\ref{eq:bh-diffcross}).}
\label{fig:bh-genmass-dist}
\end{figure}

Figure~\ref{fig:bh-diffcross} shows the differential cross sections,
calculated from Eq.~(\ref{eq:bh-diffcross}).
Figures~\ref{fig:bh-genmass-dist} show the shape of the mass distributions of generated black holes.
The distributions shown start at $\mbh > \TPs$ although,
as mentioned above, the validity of the model applies in the
region $\mbh \gg \TPs$.

\begin{figure}[H]
\hspace{0.3cm} (a) $\pz$ for (1,3,1) \hspace{1.8cm} (b) Charge \hspace{2.8cm} (c) Lepton number~($L$)
\vspace{-0.6cm}
\begin{center}
\resizebox{0.30\textwidth}{!}{\includegraphics{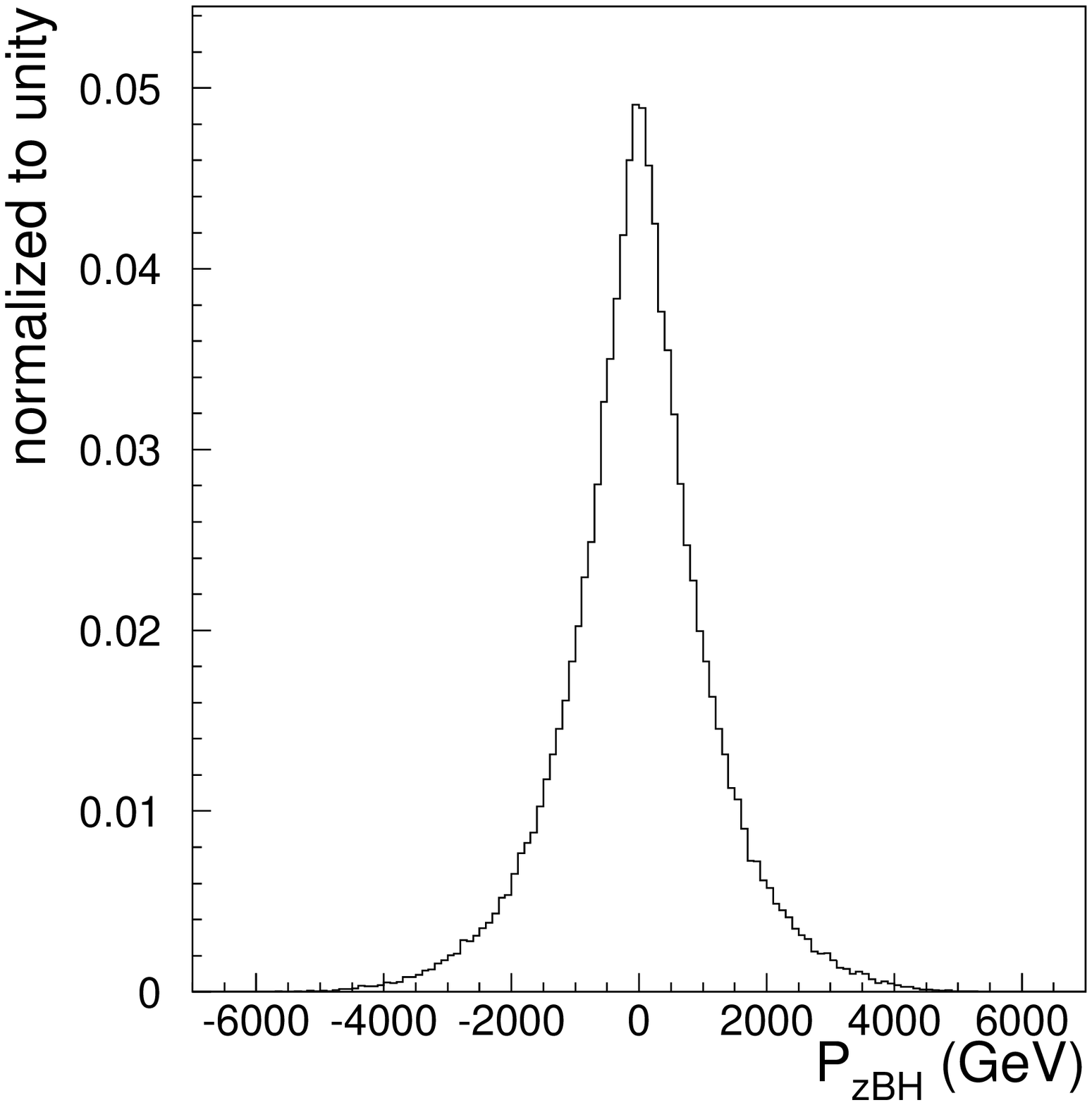}}
\resizebox{0.30\textwidth}{!}{\includegraphics{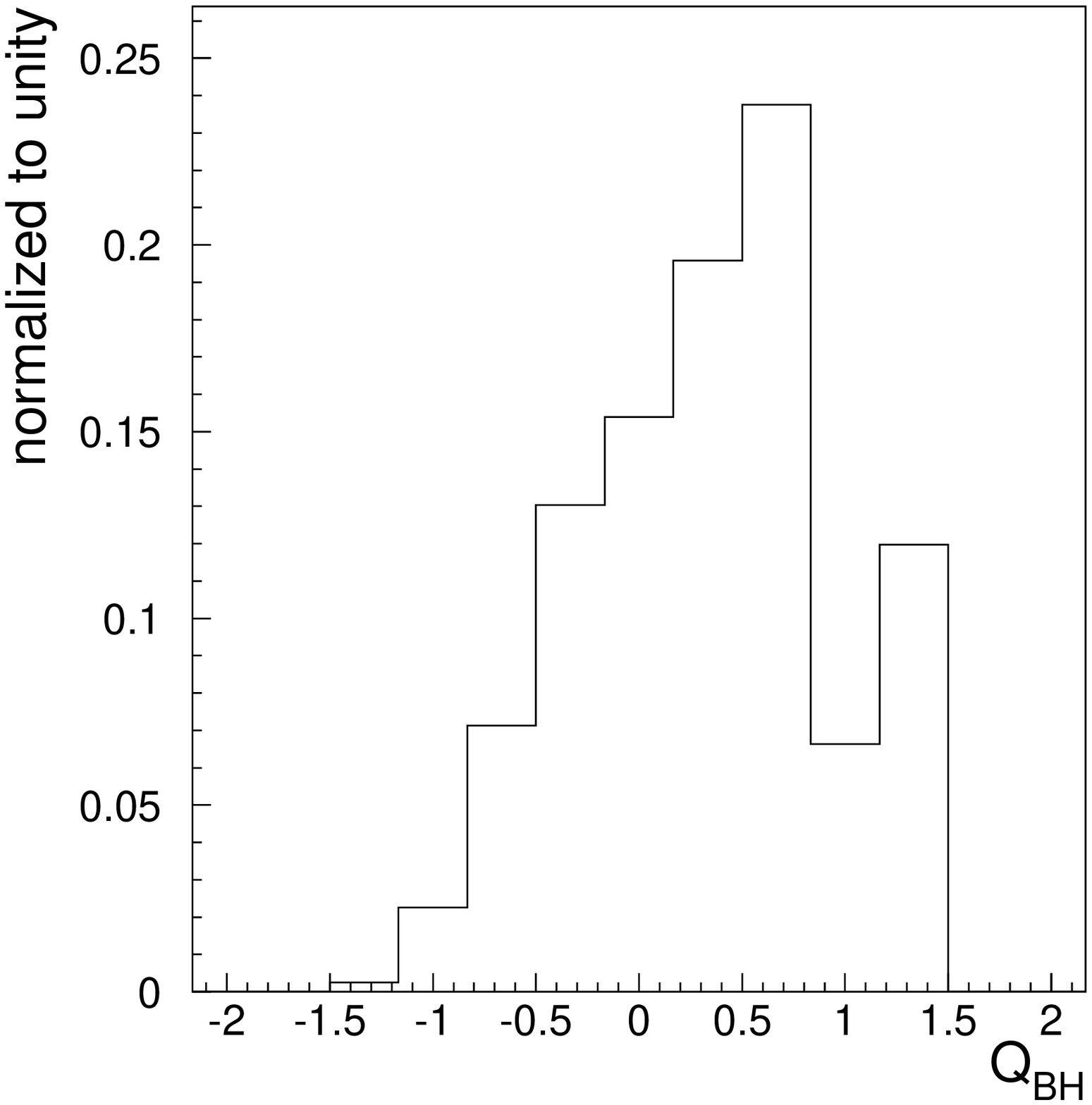}}
\resizebox{0.30\textwidth}{!}{\includegraphics{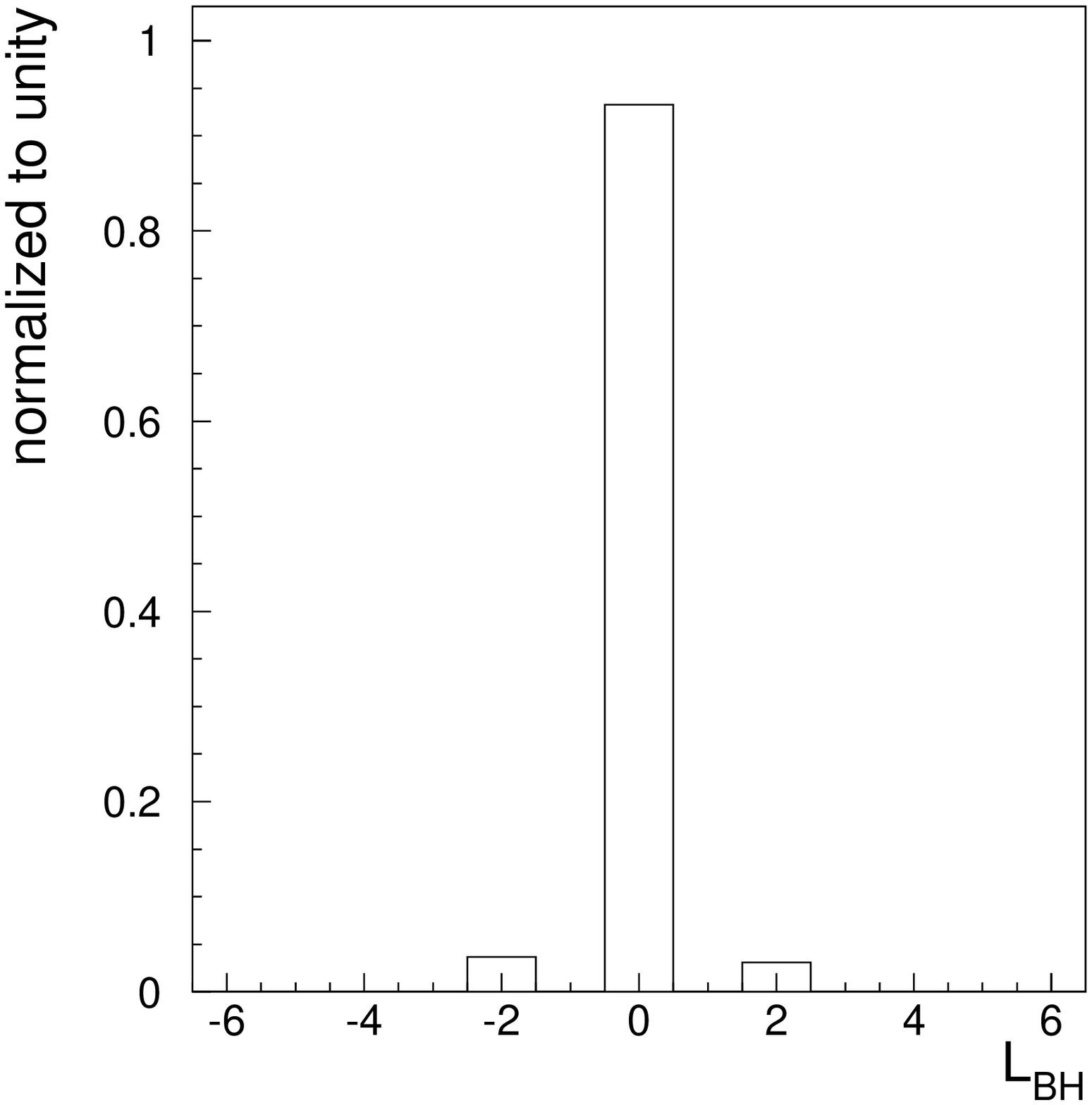}}
\end{center}
\hspace{0.3cm} (d) Baryon number~($B$) \hspace{1.0cm} (e) $\BmL$ \hspace{2.8cm} (f) Multiplicity
\vspace{-0.6cm}
\begin{center}
\resizebox{0.30\textwidth}{!}{\includegraphics{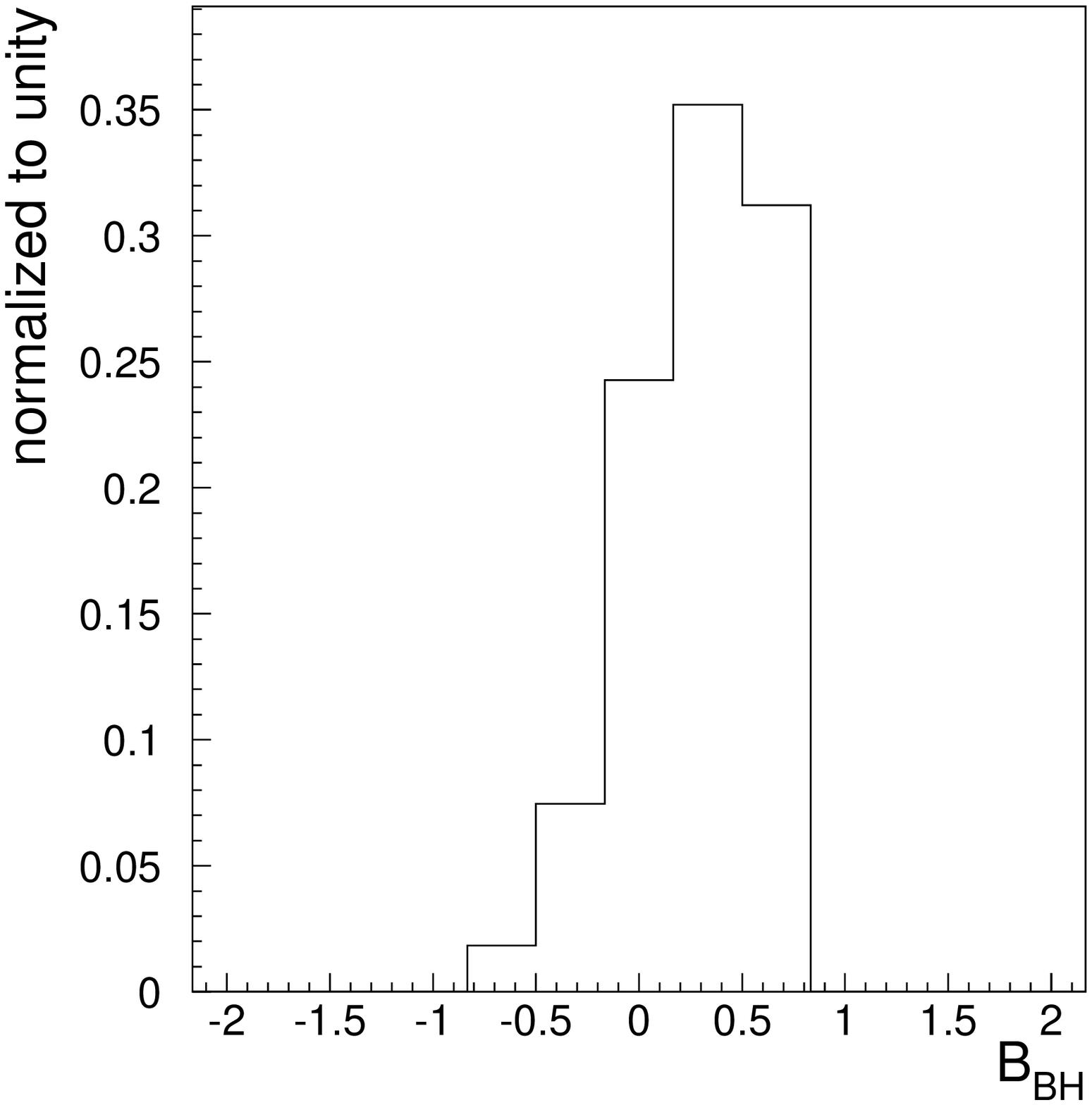}}
\resizebox{0.30\textwidth}{!}{\includegraphics{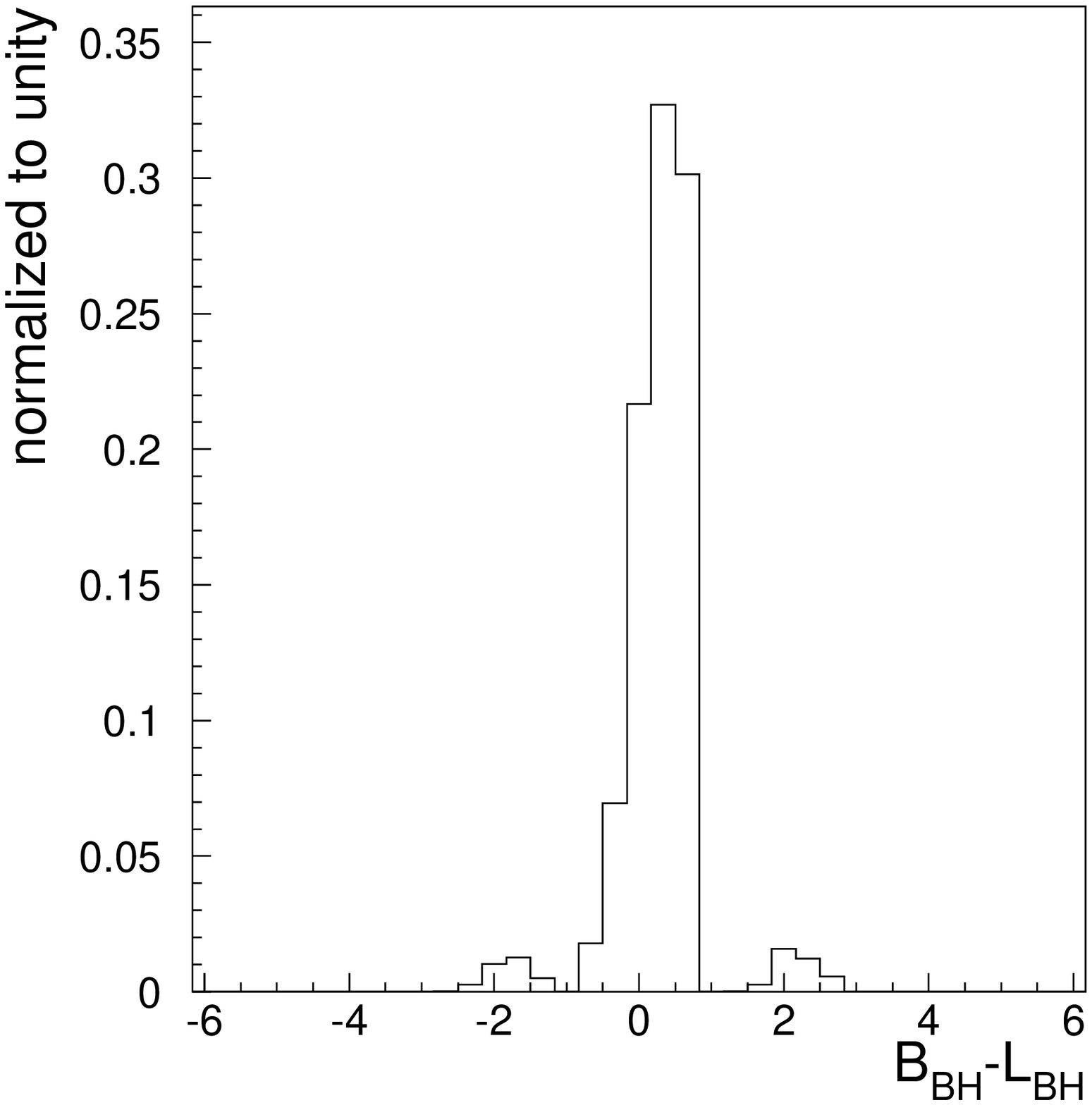}}
\resizebox{0.30\textwidth}{!}{\includegraphics{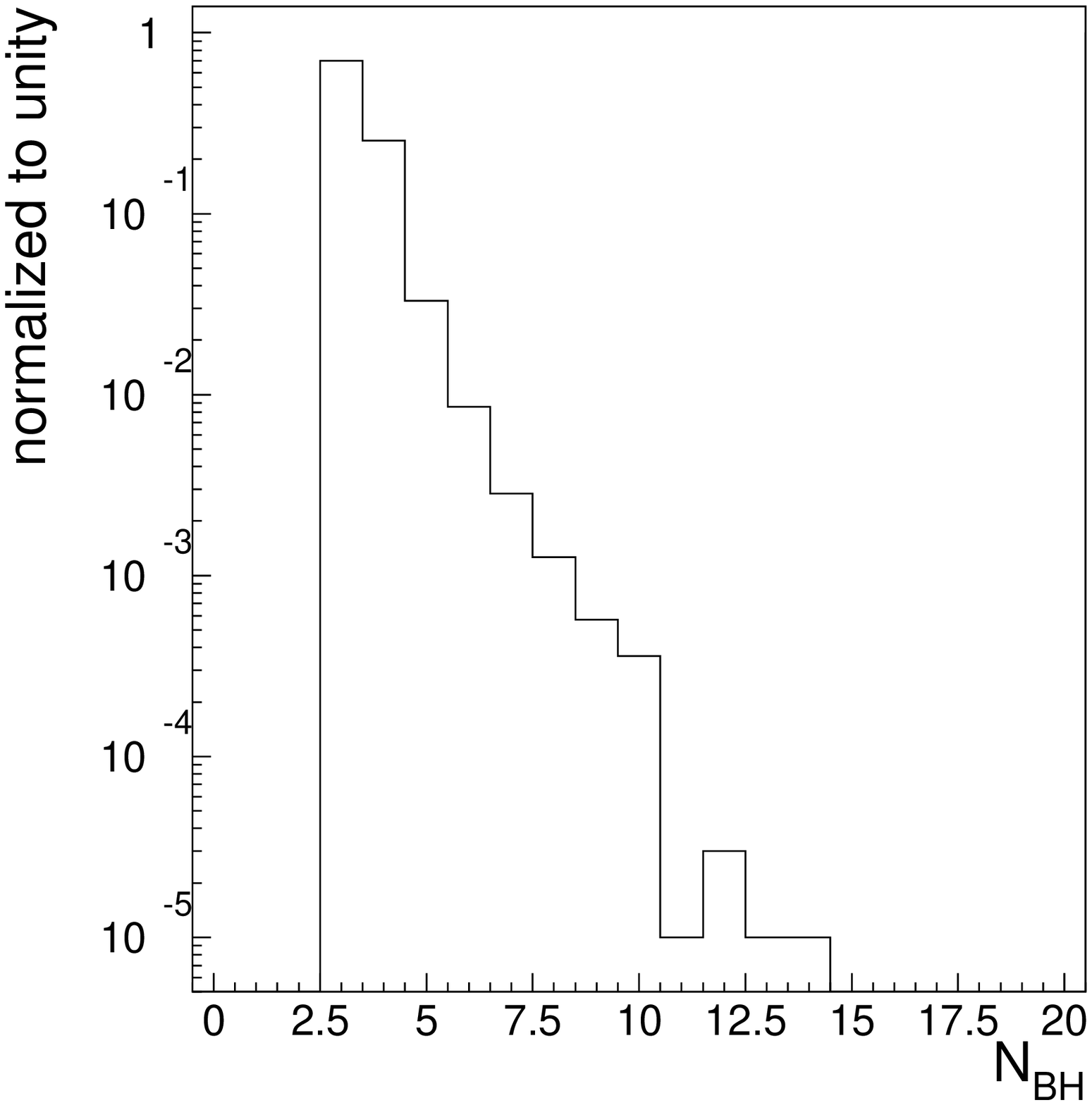}}
\end{center}
\hspace{2.7cm} (g) $L$ for (1,3,9) \hspace{1.5cm} (h) $L$ in no charge and spin conservation.
\vspace{-0.6cm}
\begin{center}
\resizebox{0.30\textwidth}{!}{\includegraphics{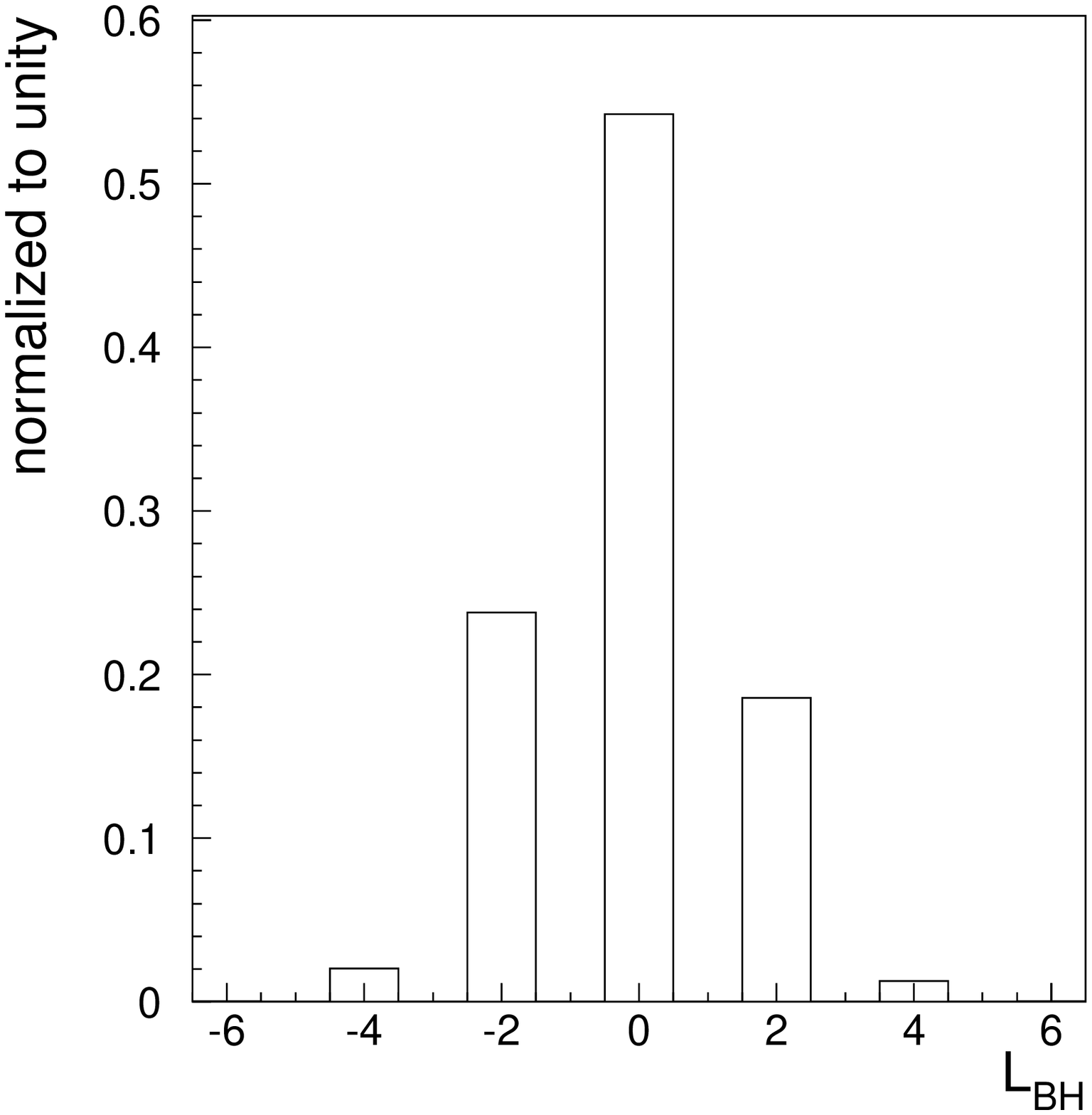}}
\resizebox{0.30\textwidth}{!}{\includegraphics{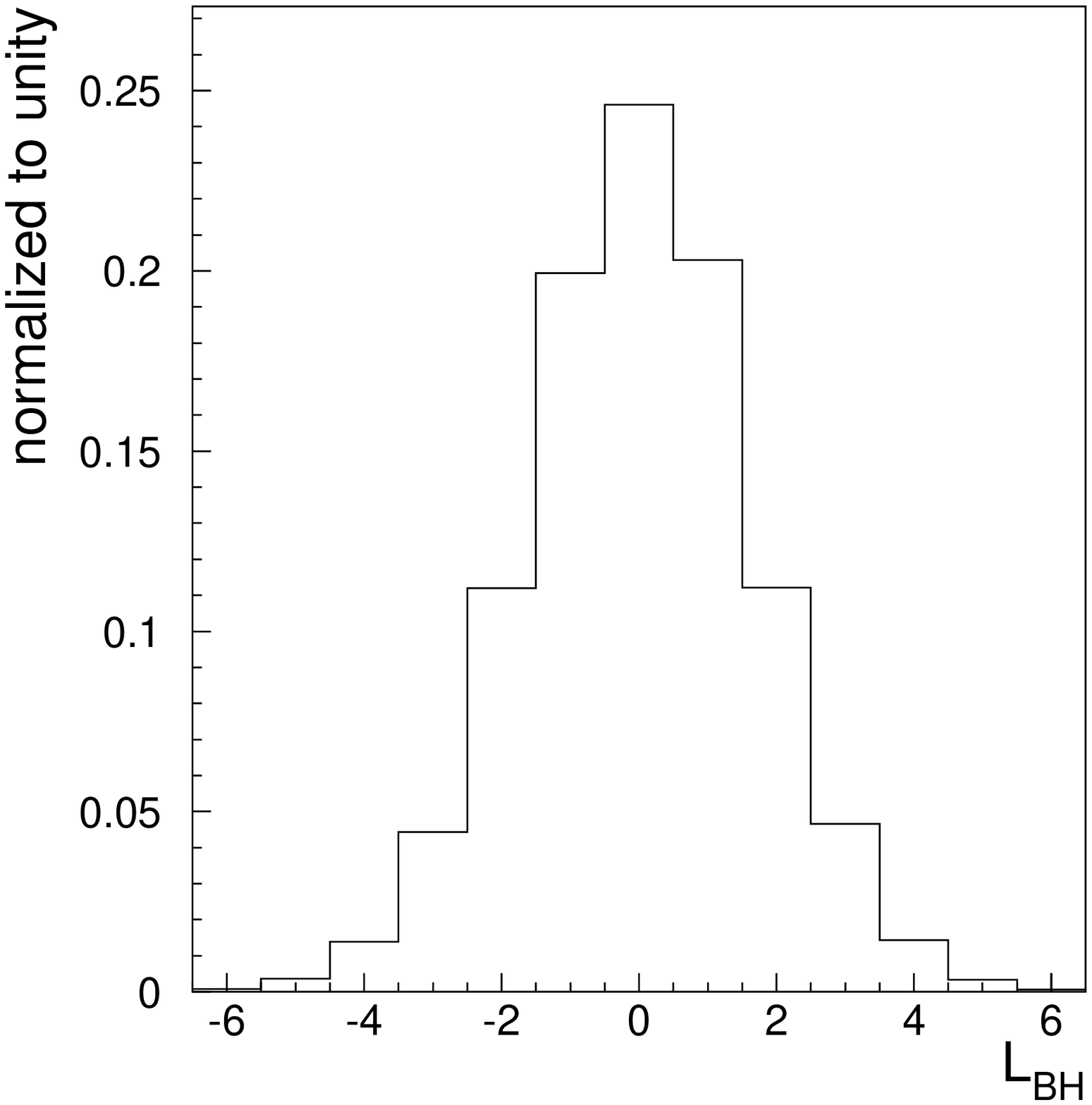}}
\end{center}
\hspace{2.7cm} (i) Multiplicity \hspace{2.2cm} (j) Multiplicity for (7,3,1)
\vspace{-0.6cm}
\begin{center}
\resizebox{0.30\textwidth}{!}{\includegraphics{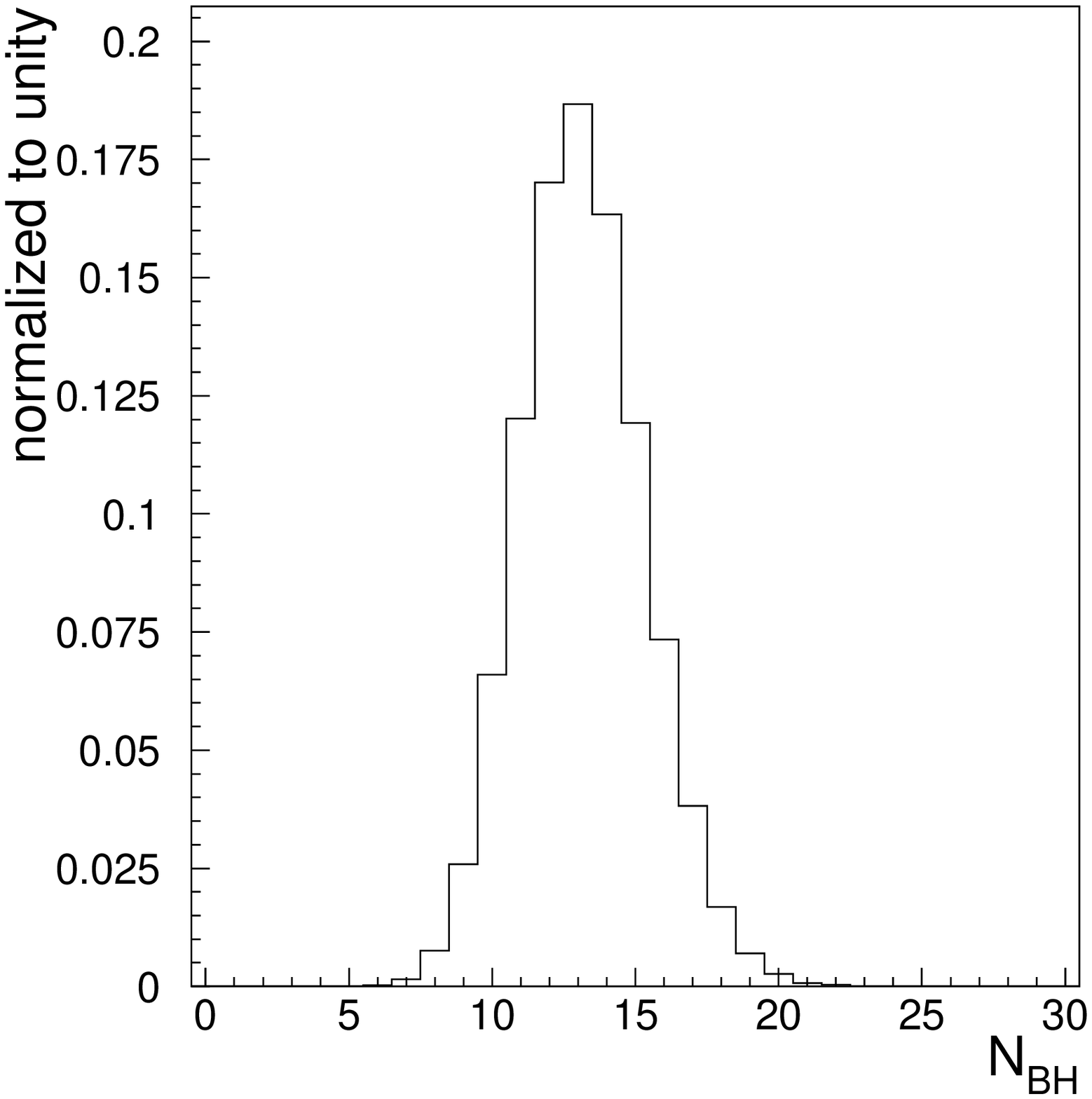}}
\resizebox{0.30\textwidth}{!}{\includegraphics{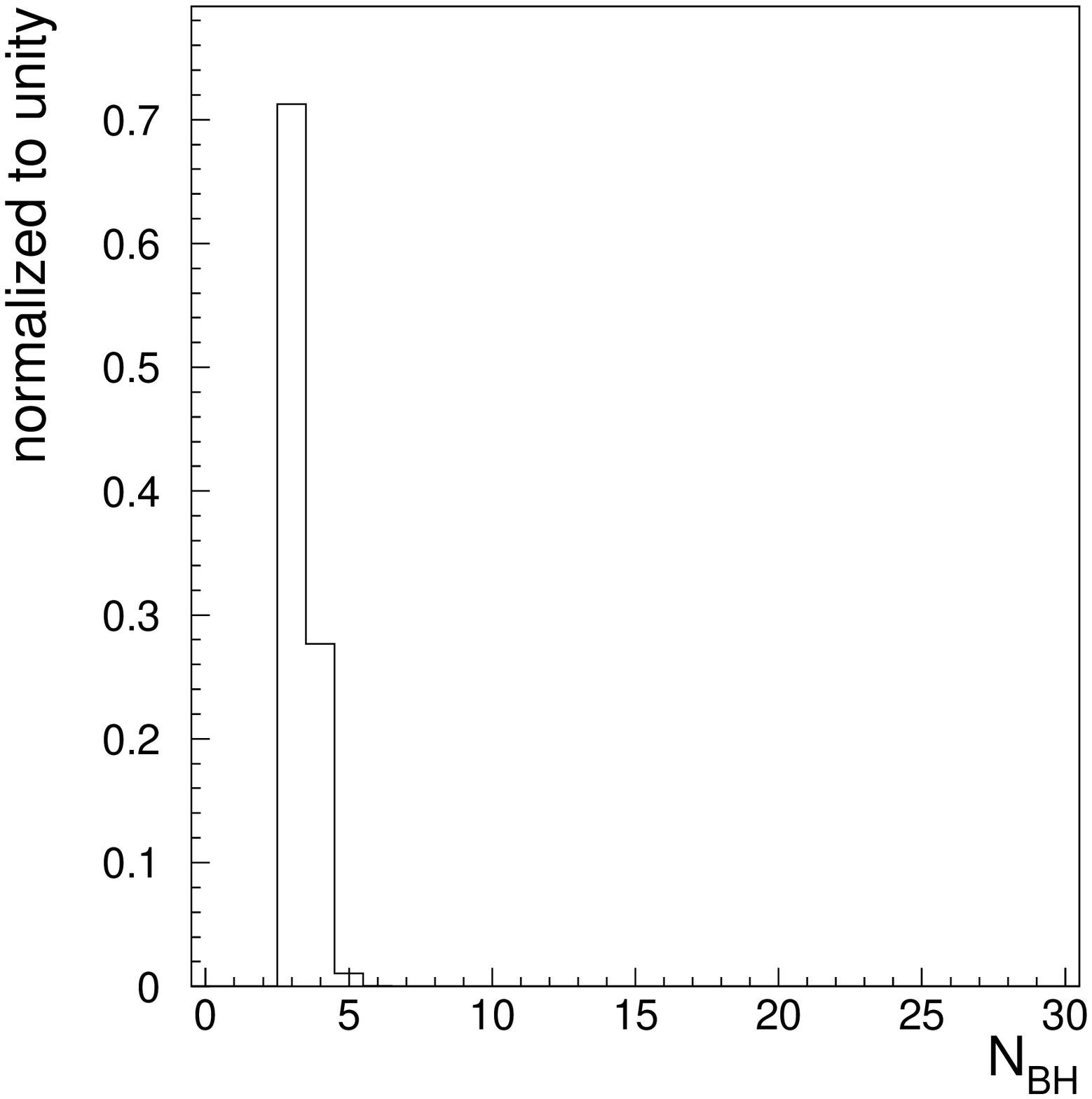}}
\end{center}
\caption{Distributions of generated black hole properties: $\pz$, charge, lepton number~($L$), baryon number~($B$), $\BmL$, and multiplicity of decay products.
(a)--(f) for ($\Mp$, $n$, $\fbhth$)=(1,3,1), (g)--(i) for ($\Mp$, $n$, $\fbhth$)=(1,3,9) and (j) for ($\Mp$, $n$, $\fbhth$)=(7,3,1).}
\label{fig:bh-properties}
\end{figure}

Figures~\ref{fig:bh-properties} show various distributions of the generated black holes:
$\pz$, charge, lepton number~($L$), baryon number~($B$), $\BmL$, and the multiplicity of decay products.
The absolute values of $L$ are always even because of the spin conservation as shown at
Figures~\ref{fig:bh-properties}~(c) and (g).
When spin conservation is turned off, they take on both even and odd values, as shown at Figures~\ref{fig:bh-properties}~(h).
Figures~\ref{fig:bh-properties}~(f),(i) and (j) indicate that the multiplicity of decay products
depends on $\mbh$ not $\TPs$.

\begin{figure}[H]
\hspace{0.7cm} (a) PDG code for (1,3,1) \hspace{2.6cm} (b) $|$PDG code$|$ for (1,3,1)
\vspace{-0.8cm}
\begin{center}
\resizebox{0.43\textwidth}{!}{\includegraphics{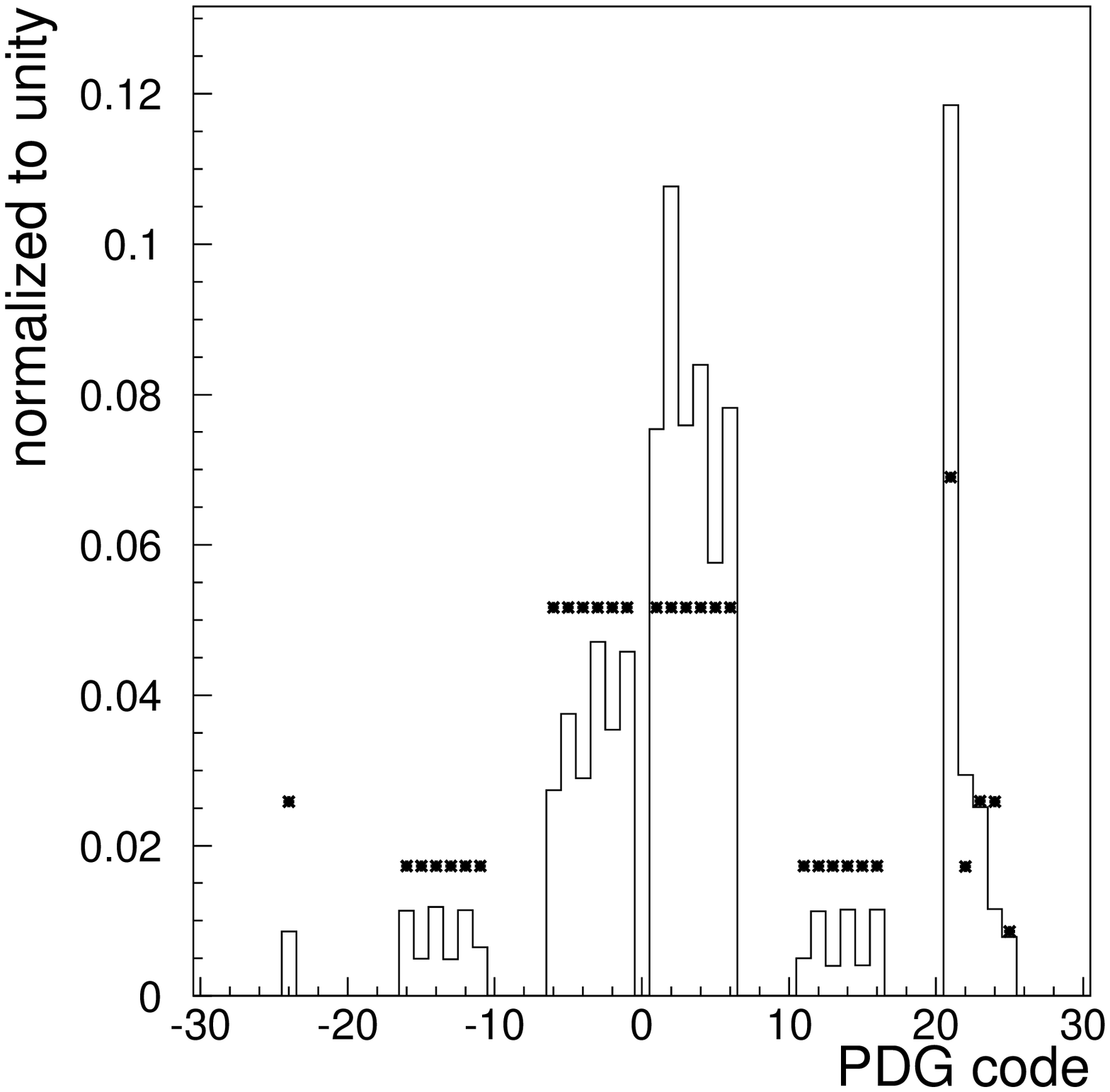}}
\resizebox{0.43\textwidth}{!}{\includegraphics{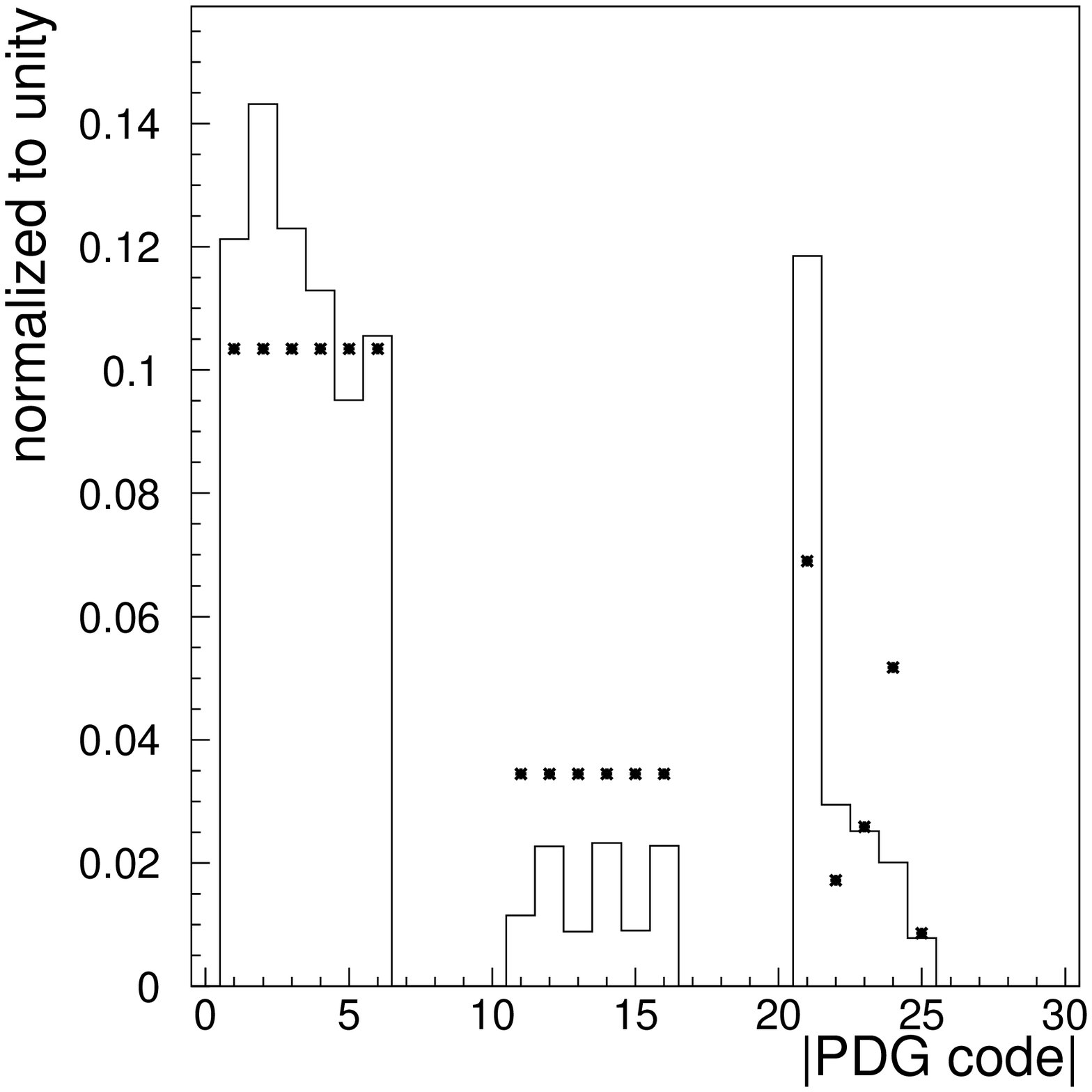}}
\end{center}
\hspace{0.7cm} (c) PDG code for (1,3,9) \hspace{2.6cm} (d) $|$PDG code$|$ for (1,3,9)
\vspace{-0.8cm}
\begin{center}
\resizebox{0.43\textwidth}{!}{\includegraphics{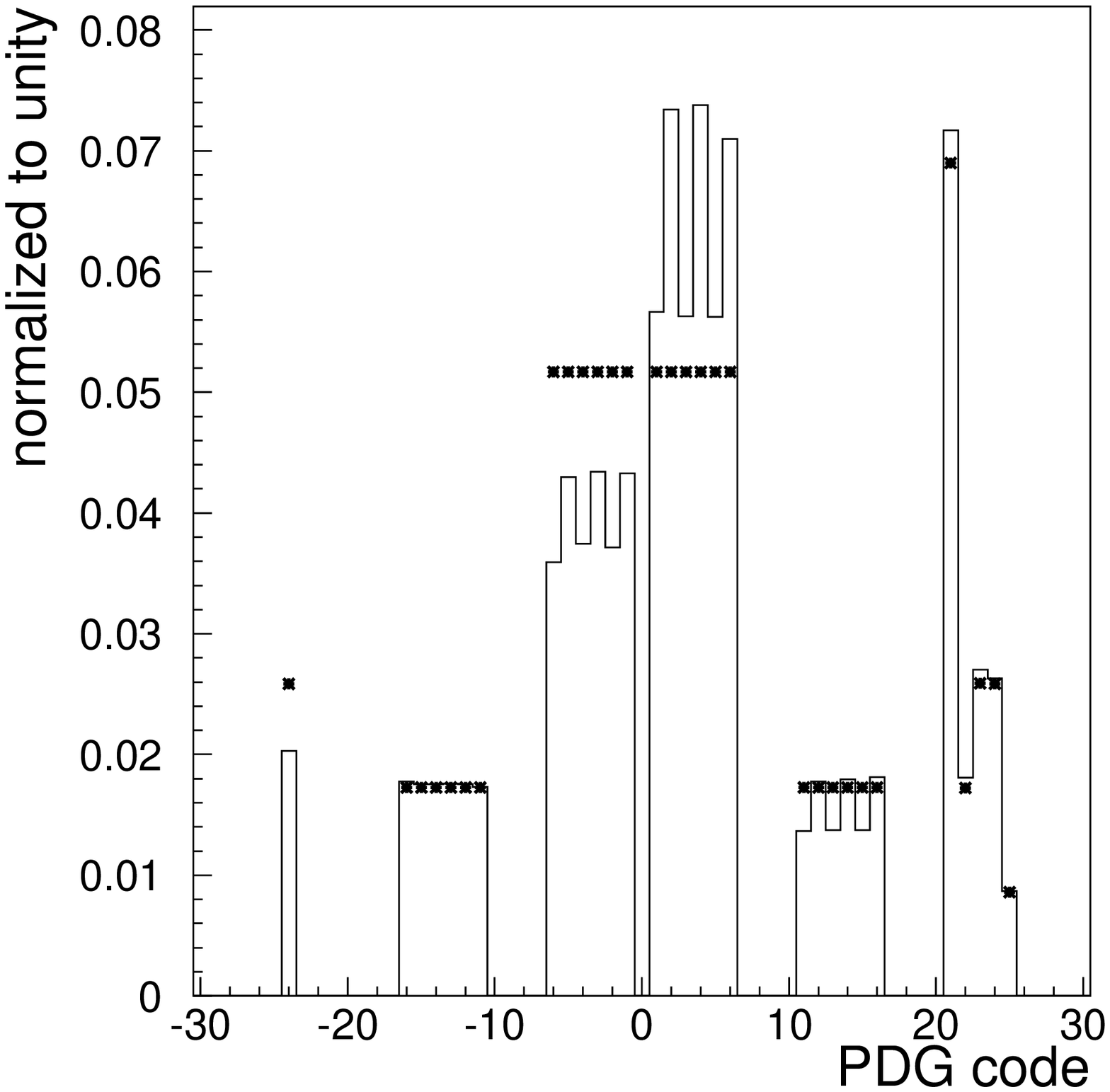}}
\resizebox{0.43\textwidth}{!}{\includegraphics{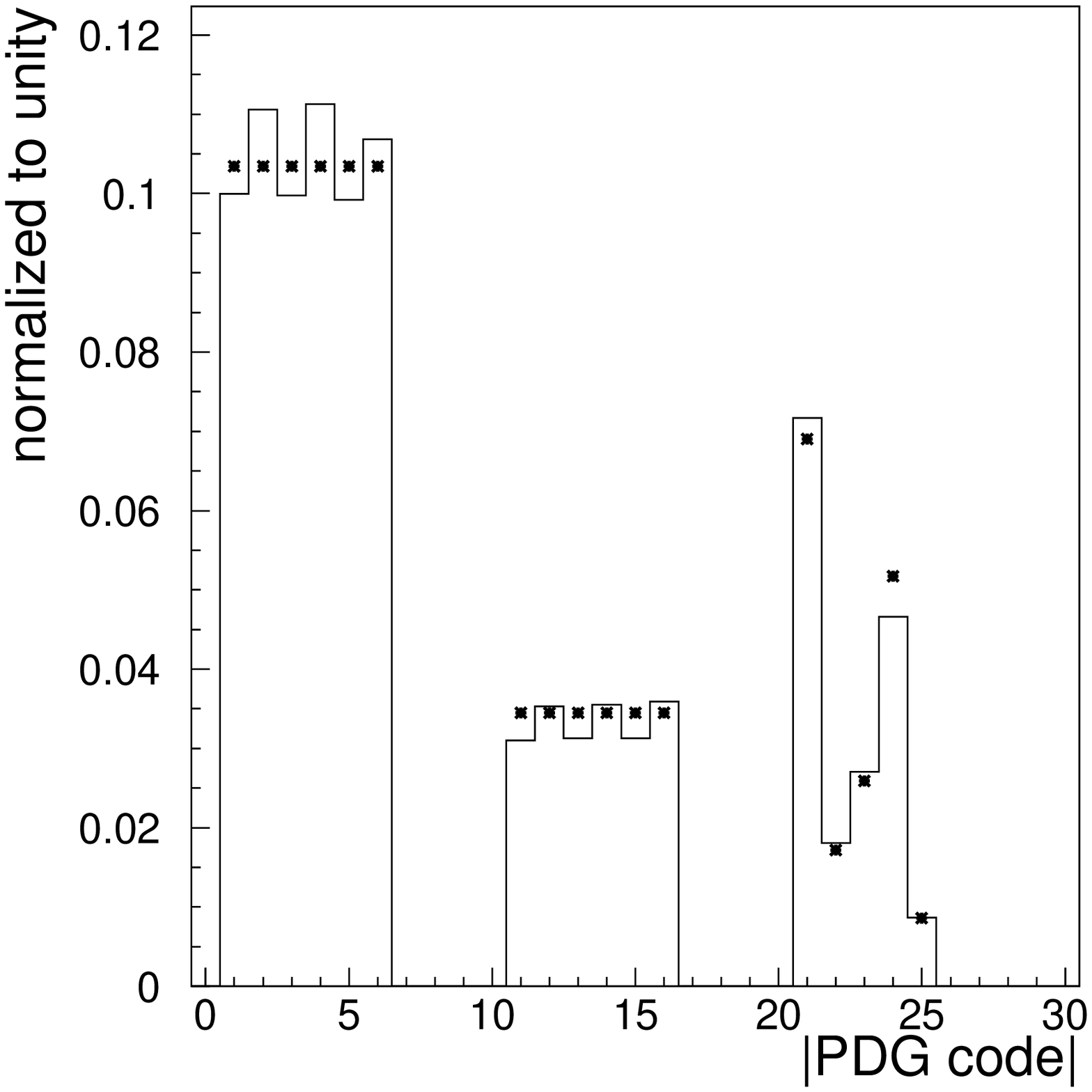}}
\end{center}
\hspace{0.7cm} (e) PDG code for (1,3,9) \hspace{2.6cm} (f) $|$PDG code$|$ for (1,3,9)

\hspace{1.4cm} in no charge and spin conservation. \hspace{0.5cm} in no charge and spin conservation.
\vspace{-0.8cm}
\begin{center}
\resizebox{0.43\textwidth}{!}{\includegraphics{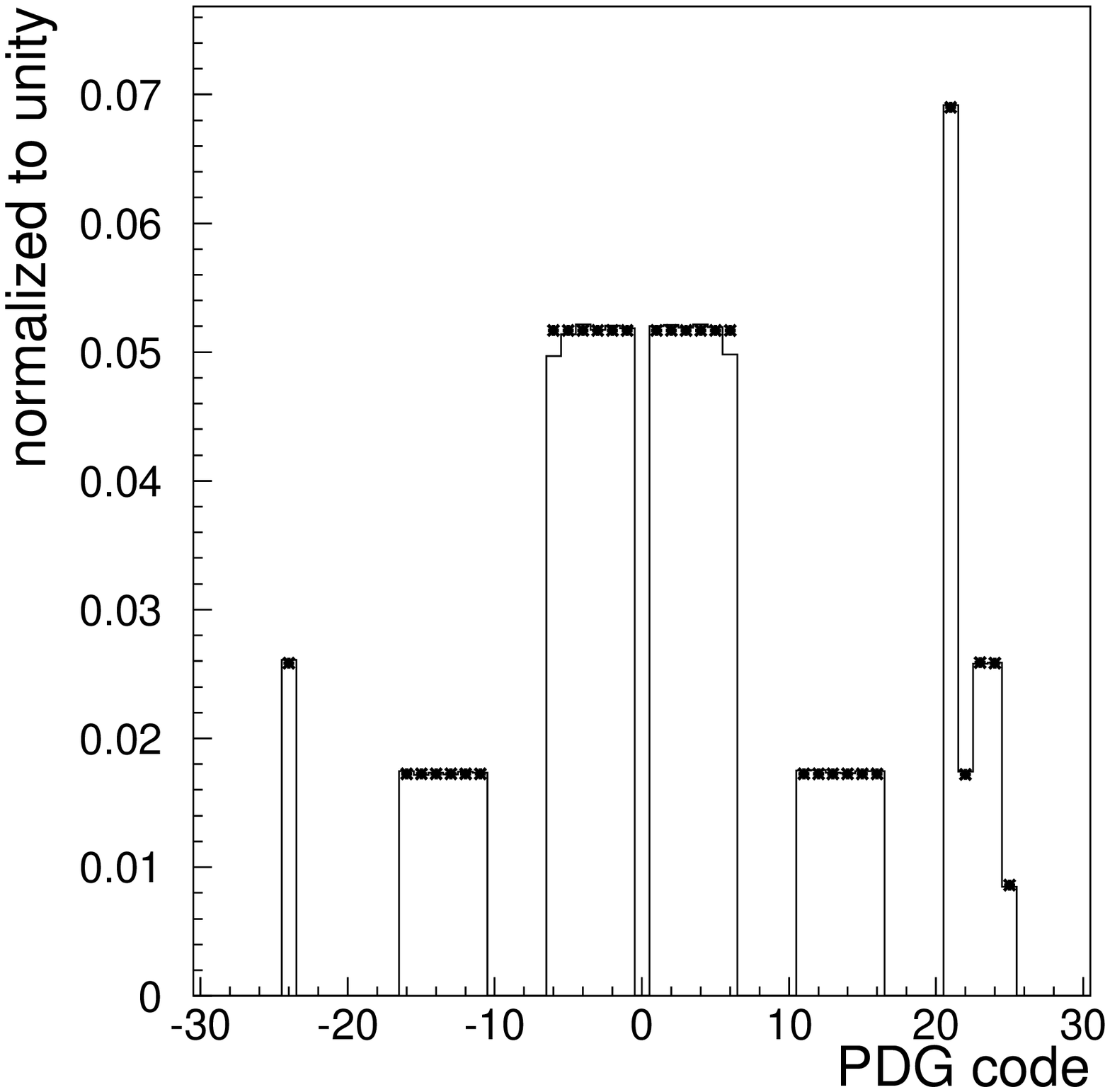}}
\resizebox{0.43\textwidth}{!}{\includegraphics{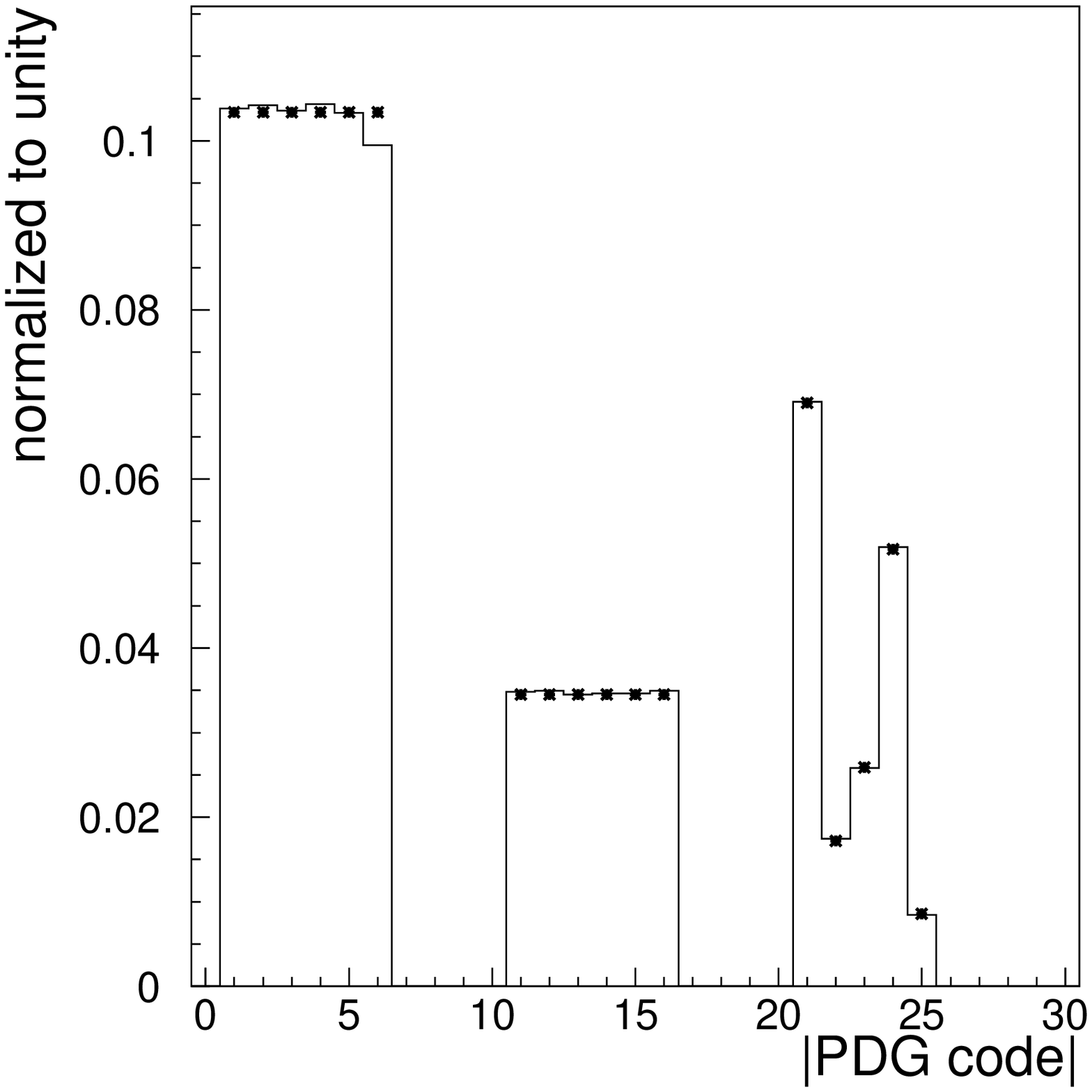}}
\end{center}
\caption{Distributions of PDG code of decay products of black holes.
The left shows PDG code and the right shows absolute value of PDG code.
The points of * show the values at Table~\ref{table:democratic-decay}.}
\label{fig:ratios-decay-products}
\end{figure}

The distribution of particle types in the decay 
of black holes are shown in Figures~\ref{fig:ratios-decay-products}.
As expected, there are more quarks than anti-quarks~(Figures~\ref{fig:ratios-decay-products}~(a) and (b))
since the LHC is a proton-proton collider.
We have also generated signal samples with ($\Mp$, $n$, $\fbhth$)=(1,3,9) in order to analyze events with high multiplicity of decay products.
As the multiplicity of decay products becomes larger,
the proportion of the different types of products becomes closer to the values of Table~\ref{table:democratic-decay},
even when many conservations are imposed, as can be seen shown in Figures~\ref{fig:ratios-decay-products}~(c) and (d).
When there is no charge and spin conservation,
the ratios are almost the same as the values of Table~\ref{table:democratic-decay},
as shown at Figures~\ref{fig:ratios-decay-products}~(e) and (f).

\begin{figure}[H]
\hspace{0.3cm} (a) $\pt$ for (1,3,1) \hspace{1.8cm} (b) $\pz$ \hspace{3.8cm} (c) Energy
\vspace{-0.6cm}
\begin{center}
\resizebox{0.31\textwidth}{!}{\includegraphics{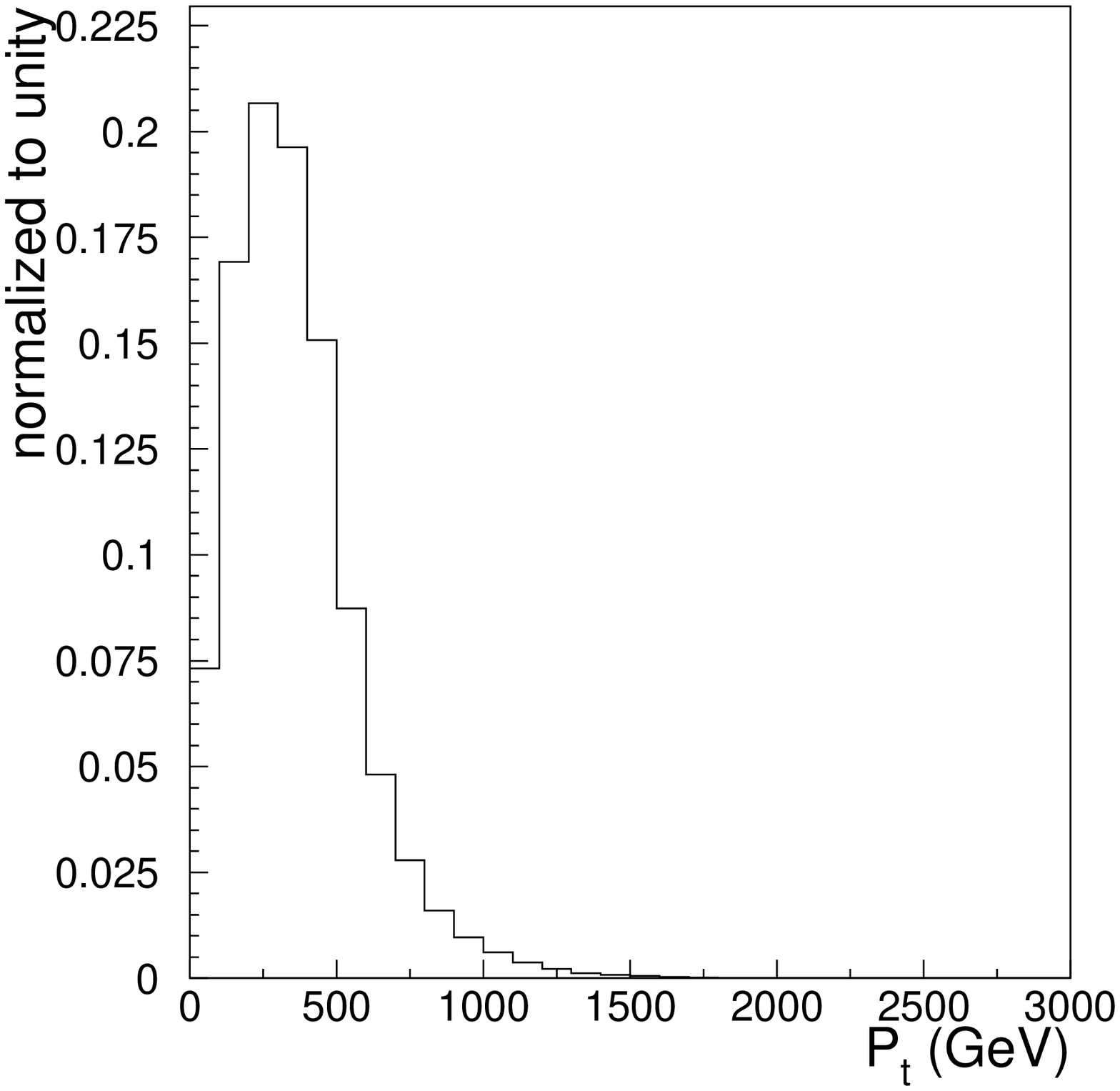}}
\resizebox{0.31\textwidth}{!}{\includegraphics{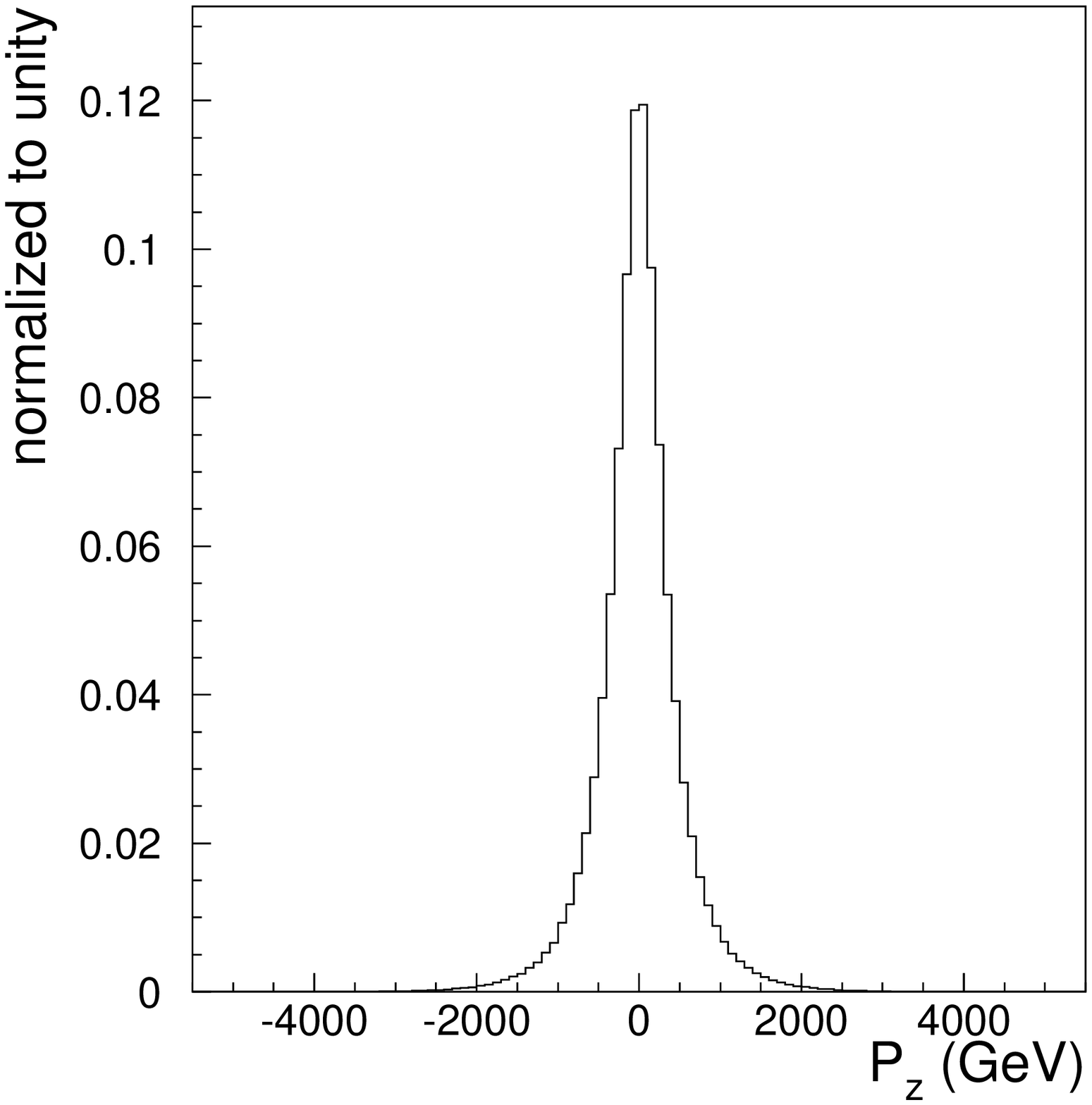}}
\resizebox{0.31\textwidth}{!}{\includegraphics{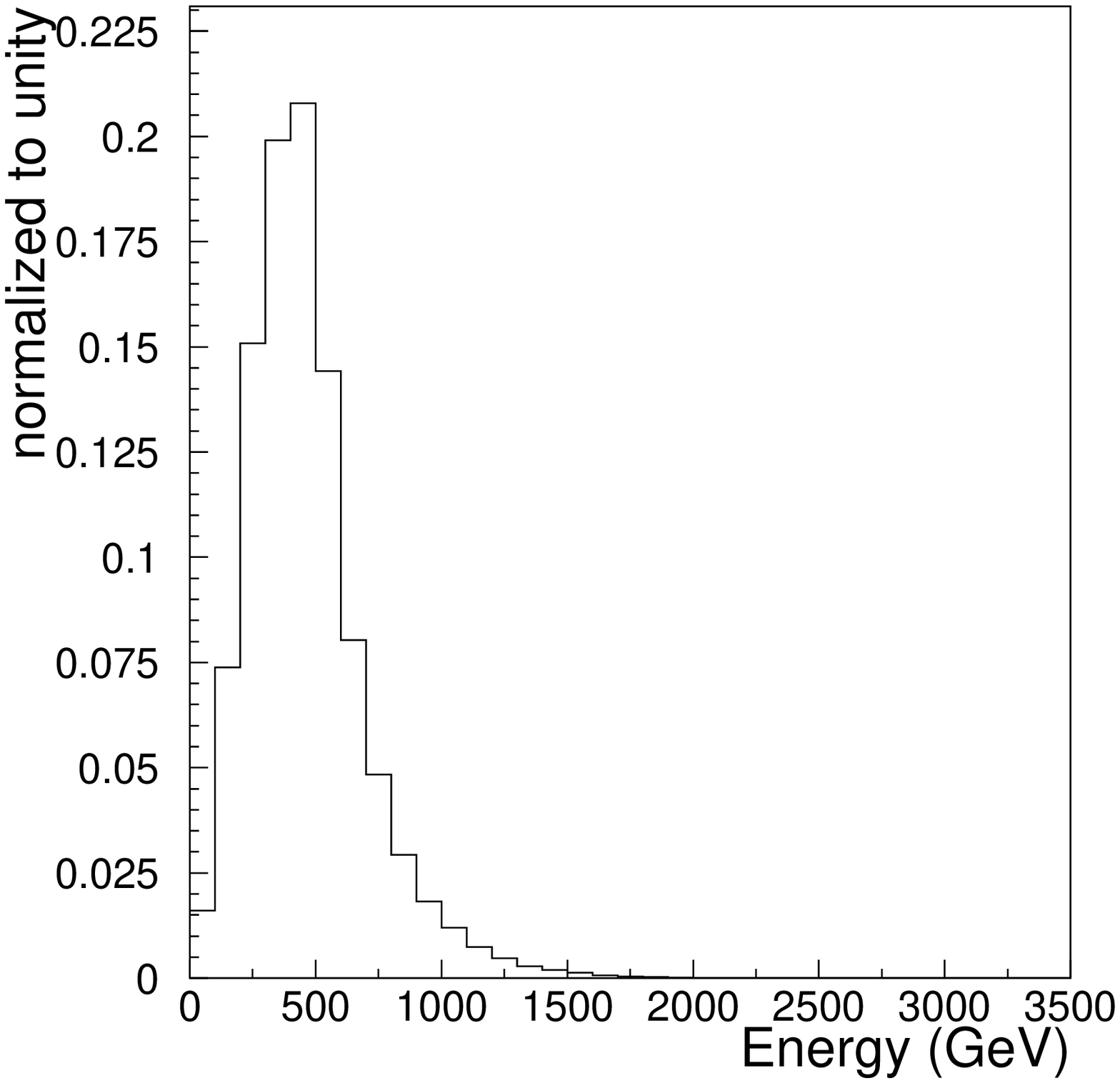}}
\end{center}
\hspace{0.3cm} (d) $\eta$ \hspace{4.1cm} (e) $\phi$ \hspace{3.8cm} (f) $\pt$ for (1,3,9)
\vspace{-0.6cm}
\begin{center}
\resizebox{0.31\textwidth}{!}{\includegraphics{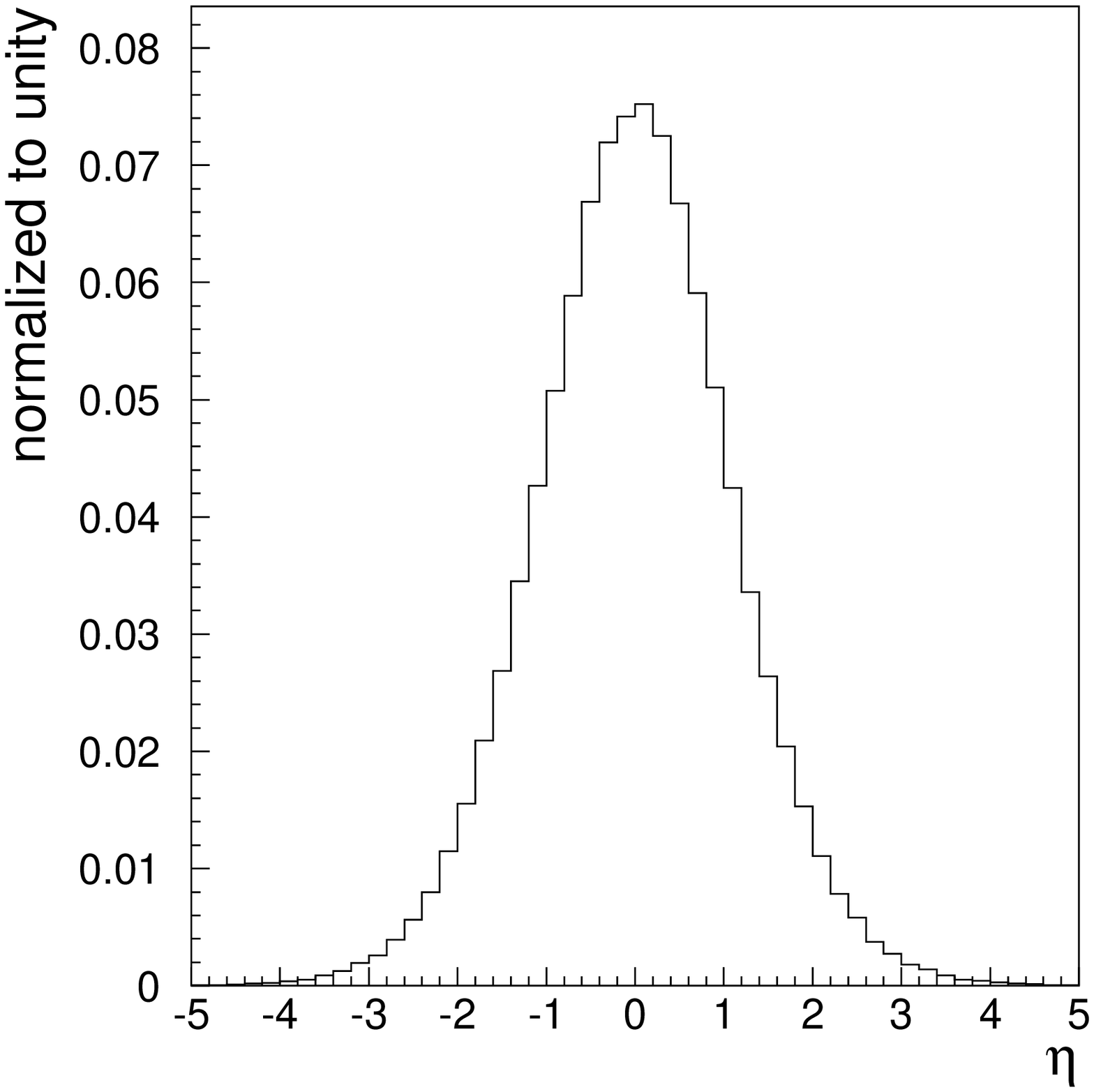}}
\resizebox{0.31\textwidth}{!}{\includegraphics{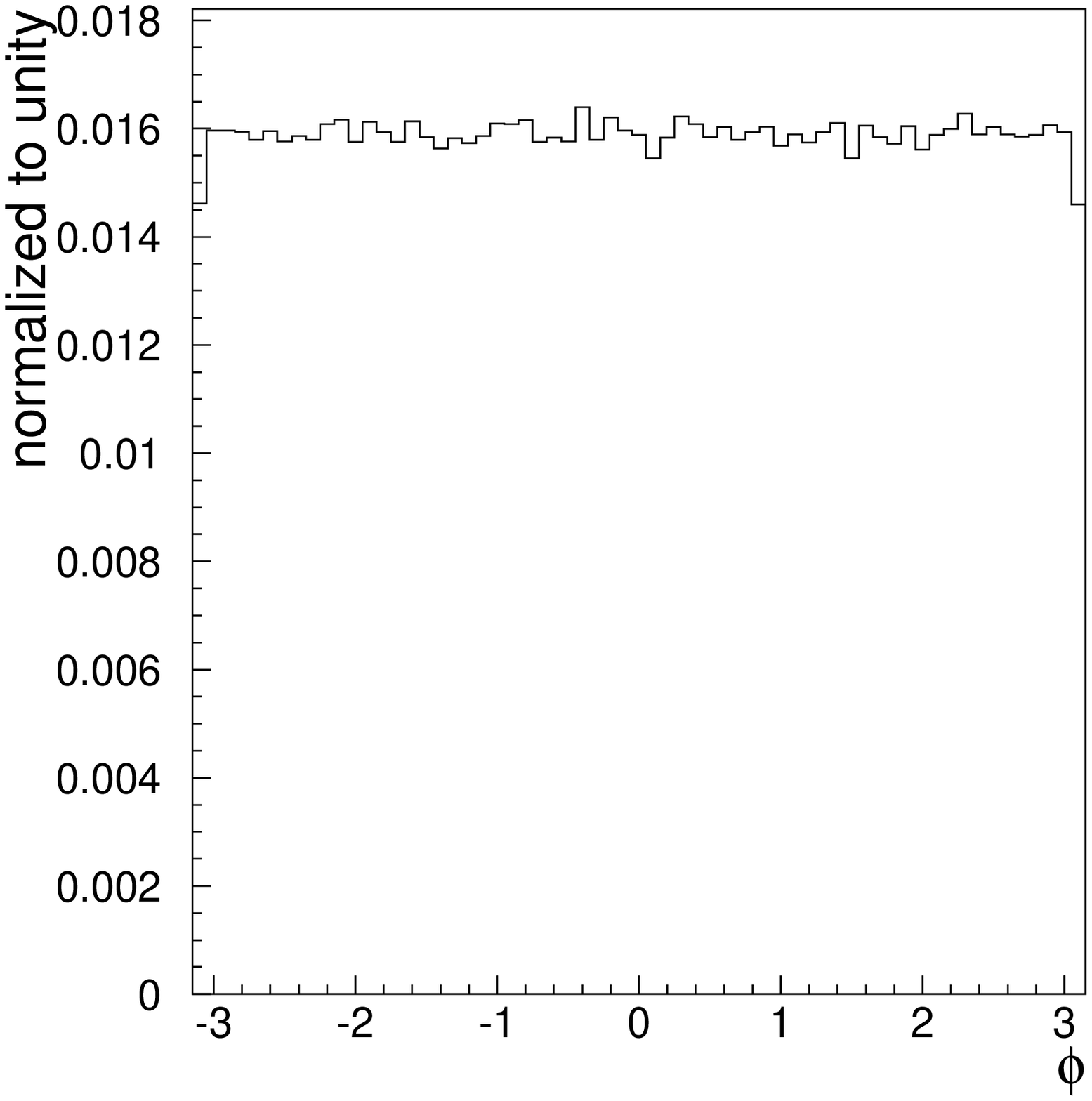}}
\resizebox{0.31\textwidth}{!}{\includegraphics{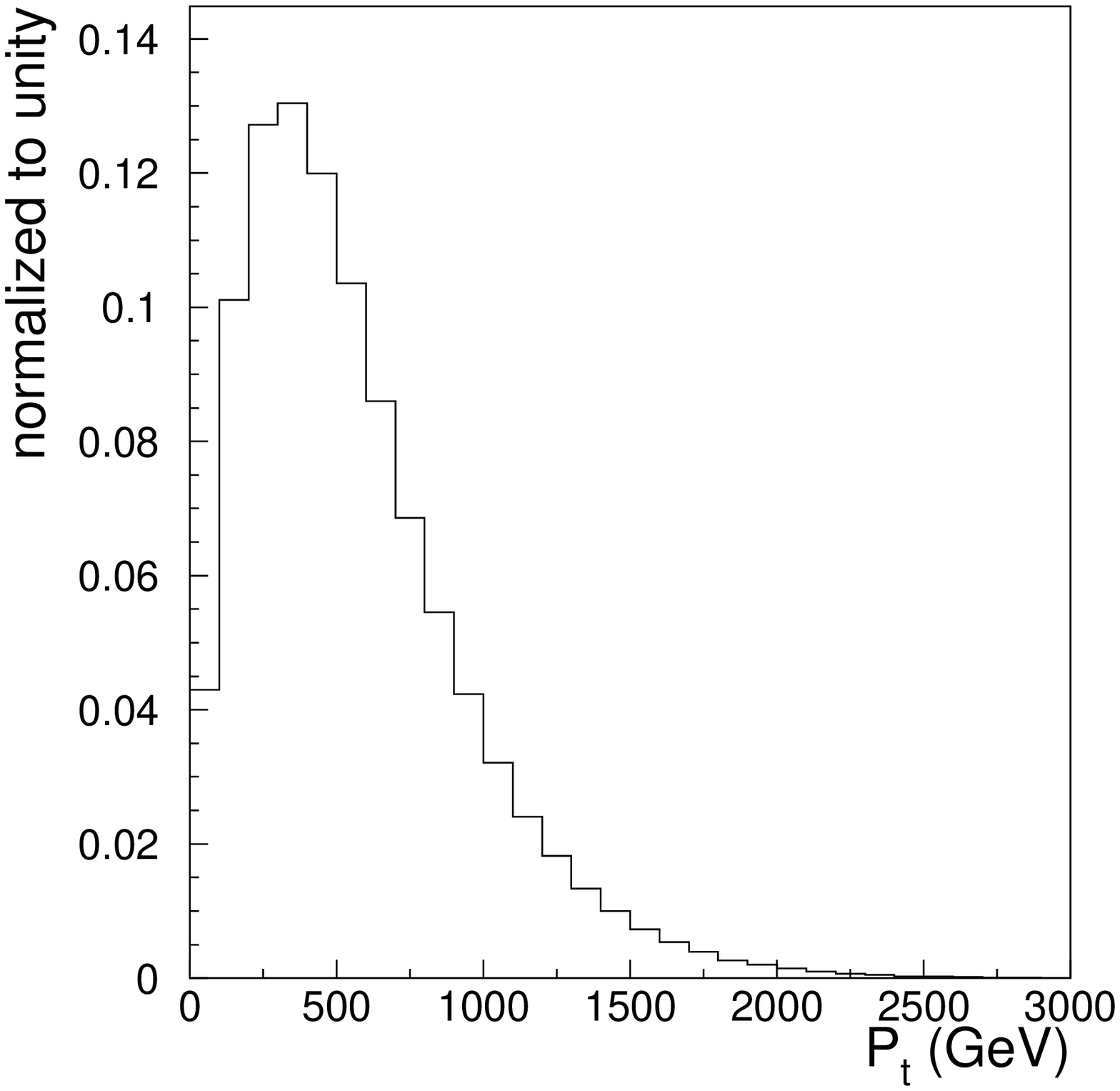}}
\end{center}
\hspace{0.3cm} (g) Energy \hspace{3.0cm} (h) $\pt$ for (7,3,1) \hspace{1.7cm} (i) Energy
\vspace{-0.6cm}
\begin{center}
\resizebox{0.31\textwidth}{!}{\includegraphics{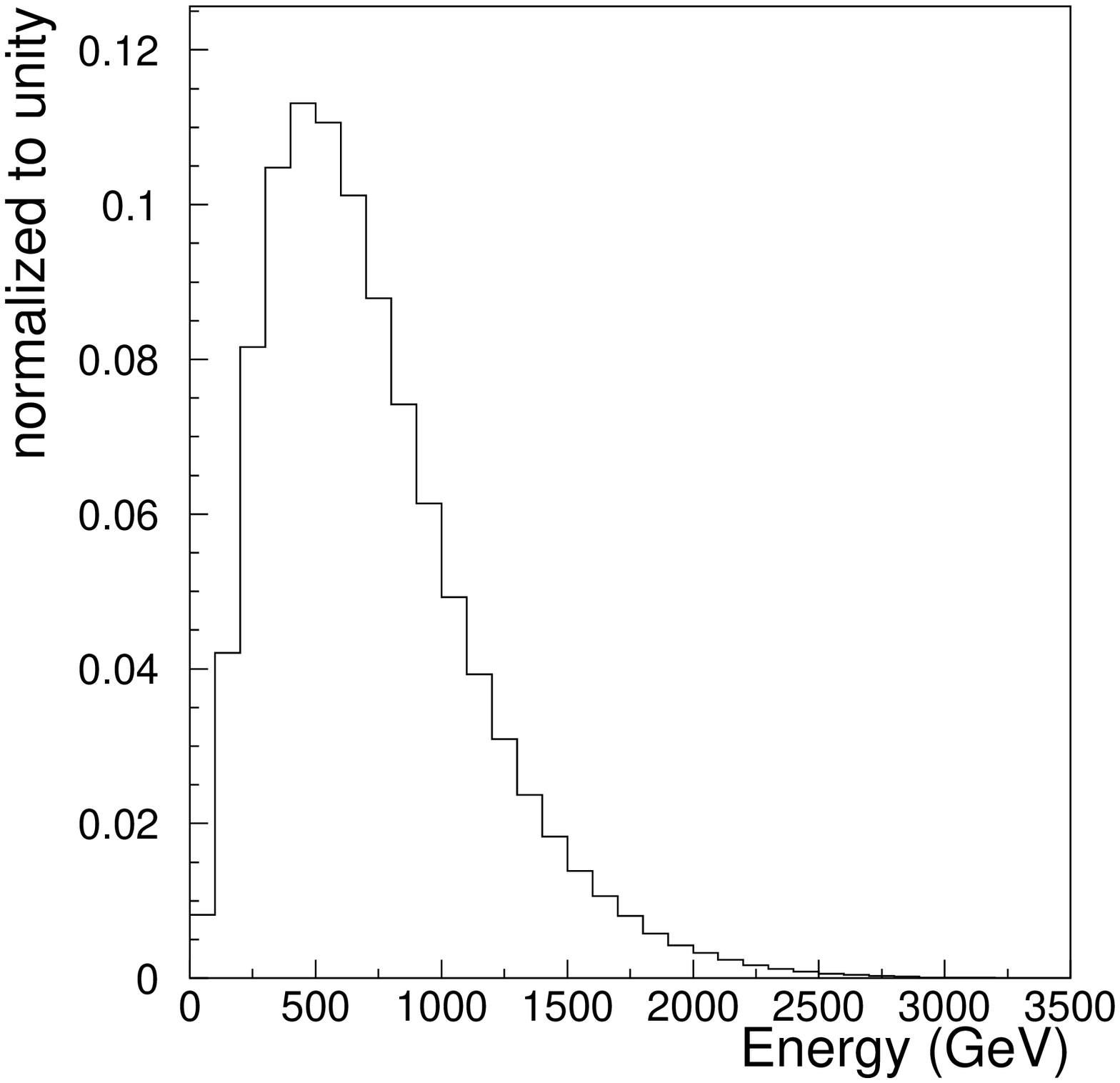}}
\resizebox{0.31\textwidth}{!}{\includegraphics{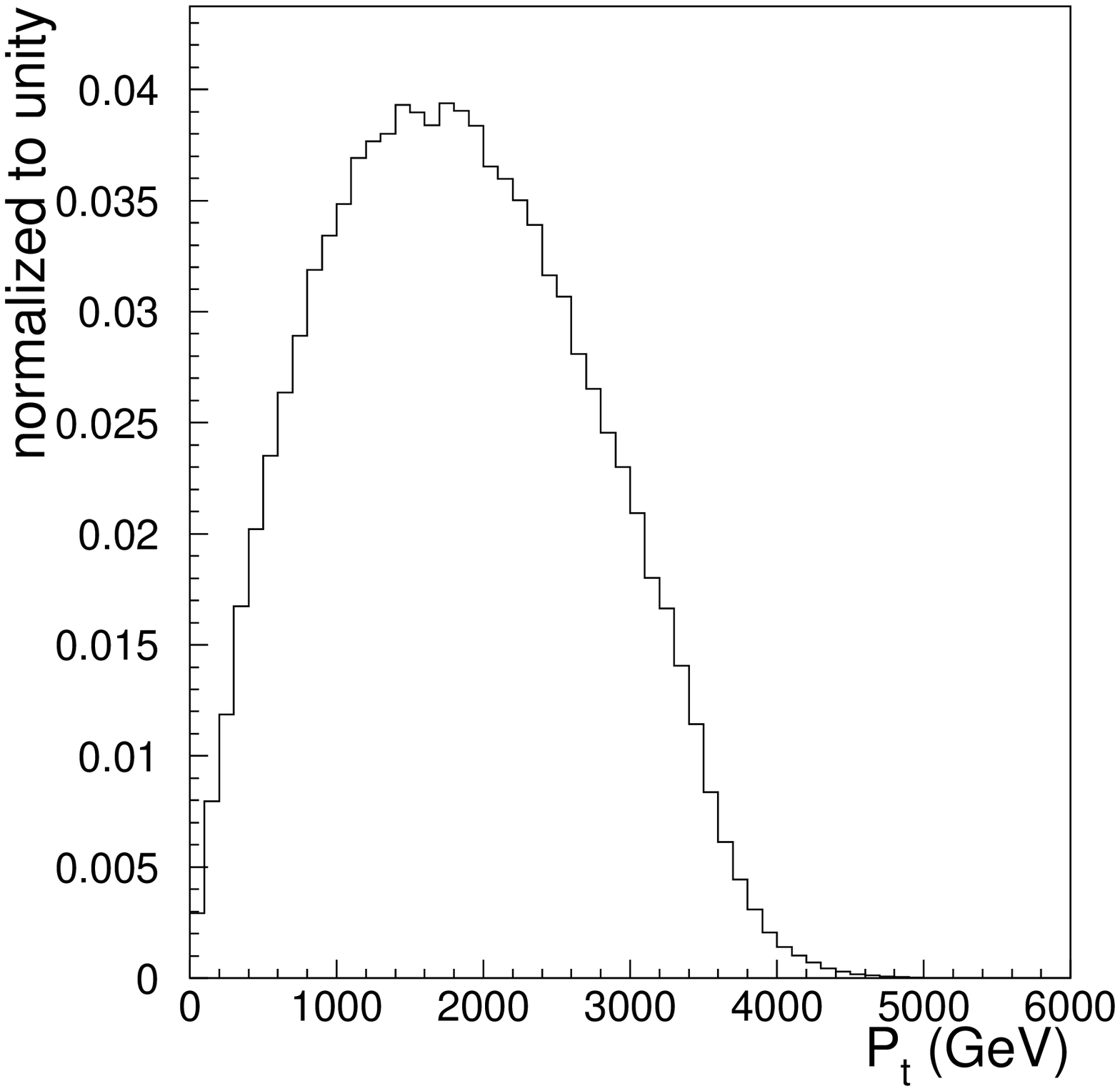}}
\resizebox{0.31\textwidth}{!}{\includegraphics{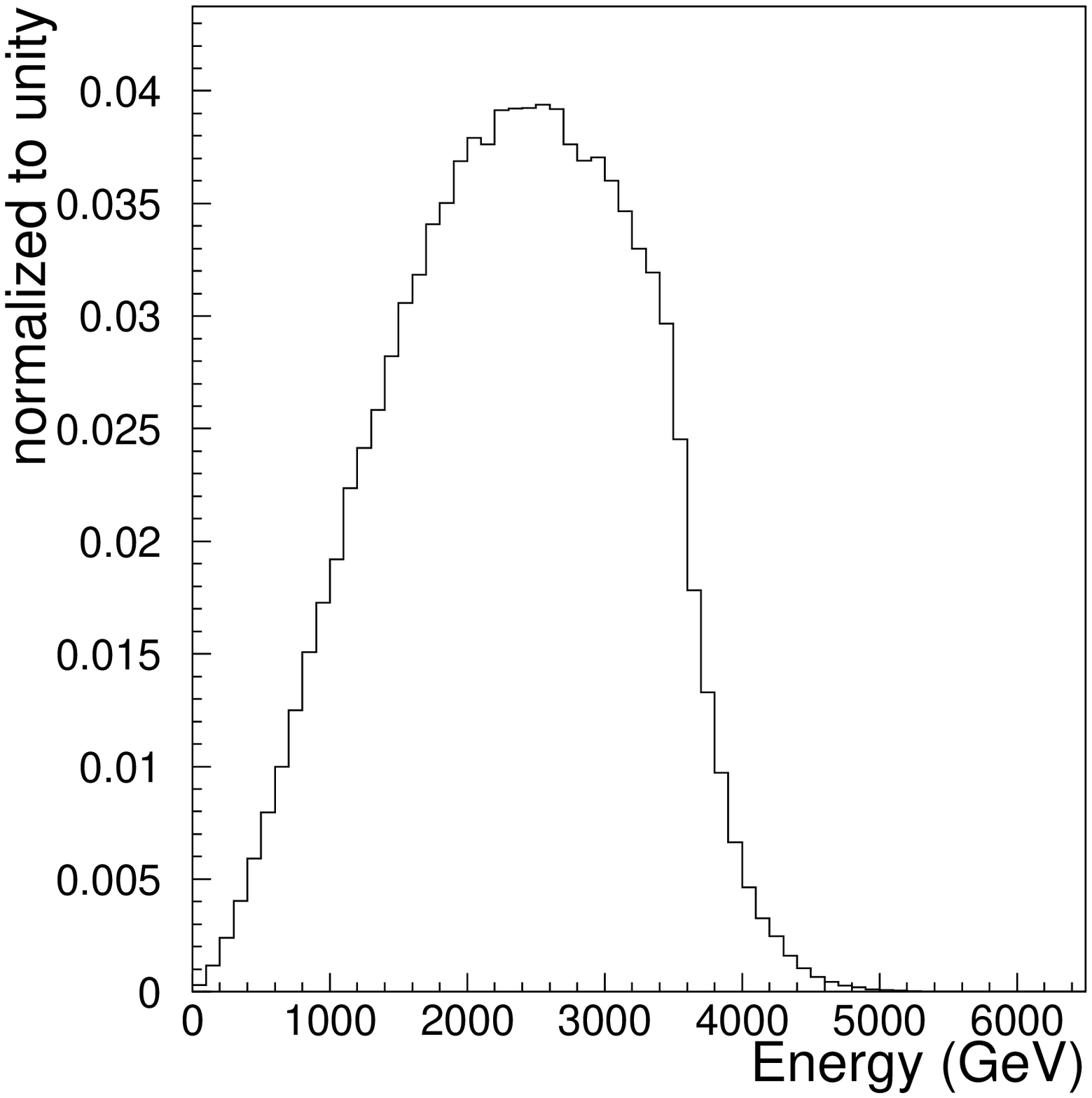}}
\end{center}
\caption{Distributions of various properties of decay products of black holes:
$\pt$, $\pz$, energy, $\eta$ and $\phi$}
\label{fig:bh-products-properties}
\end{figure}

Figures~\ref{fig:bh-products-properties} show various distributions of the decay products:
$\pt$, $\pz$, energy, $\eta$ and $\phi$.
Distributions of $\pz$, $\eta$ and $\phi$ of the products are symmetric as expected.

%% file: analysis.tex
\section{Analysis}\label{sec:analysis}
Although the expressions for black hole production given
in Section~\ref{sec:intro} are not valid at $\mbh \sim \TPs$,
our BH generator applies them in all the regions where
$\mbh > \TPs$ in this study.

\subsection{Selection Criteria and Reconstruction}\label{sec:criteria}
The following cuts are applied in order to select and reconstruct black holes~(BHs):
\begin{itemize}
\item For a precise reconstruction of the BH, it is necessary to remove particles
produced in the stage of initial state radiations~(ISRs).
Because the particles from ISRs tend to have small $\pt$ and large $|\eta|$,
as shown in Figures.~\ref{fig:pt} and \ref{fig:eta},
the following requirements are applied to particles in each event
~(ISR-cut)~:\\ \\
\hspace*{0.75cm} $\bullet  \hspace{0.25cm} \pt ~>~30~\textrm{GeV} \hspace{0.75cm} \textrm{for} \ \mu, e$ \\
\hspace*{0.75cm} $\bullet  \hspace{0.25cm} \pt ~>~ 50~\textrm{GeV} \hspace{0.75cm}\textrm{for} \ \gamma,$ jet\\ 

\begin{figure}[H]
\begin{center}
\resizebox{0.48\textwidth}{!}{\includegraphics{figs/bh-atlas-note-pt3.epsi}}
\resizebox{0.50\textwidth}{!}{\includegraphics{figs/bh-atlas-note-pt2.epsi}}
\end{center}
\caption{$\pt$ distributions~:~comparison between particles from BH and
initial state radiation~($\TPs=1$~TeV, $n$=3).}
\label{fig:pt}
\end{figure}

\begin{figure}[H]
\begin{center}
\resizebox{0.48\textwidth}{!}{\includegraphics{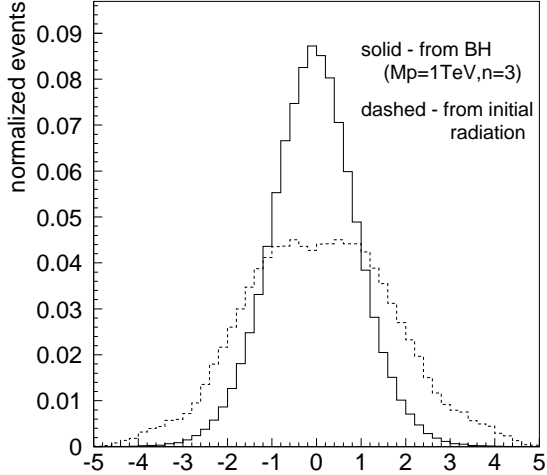}}
\end{center}
\caption{$\eta$ distributions~:~comparison between particles from BH and initial state radiation~($\TPs=1$~TeV, $n$=3).}
\label{fig:eta}
\end{figure}

\item Of all the particles passing the ISR-cut,
more than three
are required to have energy larger than 300~GeV~($E>300$~GeV),
and moreover, at least one of them has to be 
either an electron or a photon. 
This latter requirement is to suppress backgrounds.
Figure~\ref{fig:n-particles} shows the distributions of the number
of energetic particles in each event, where \bggj ~event is one of
the largest contribution in the backgrounds as shown in Table~\ref{table:bg}.
It is seen that this multiplicity cut on energetic particles is very effective.

\item We require the event shape variable $R_2$ to be less than 0.8~($R_2<0.8$).
$R_2$ represents an event topology and it is defined by 
Fox-Wolfram moments as follows:
\[R_2 \equiv H_2 /H_0\]
\[H_i \equiv \sum _{j,k} \frac{\mid p_j ^* \mid
\mid p_k ^* \mid}{E^2} P_i
(cos \phi _{jk})\]

where $H_i$ is the $i$-th Fox-Wolfram moments, $j$ and $k$ are
ID numbers of tracks, $p_j^*$ is the momentum of the
track $j$ in the rest frame of BH, $\phi _{jk}$ is the opening angle between 
the tracks $j$ and $k$ ,
and $P_i(x)$ is the Legendre polynomial.
E is obtained by summing up the energies of particles passing the ISR-cut,
calculated in the rest frame of BH.
$R_2$ ranges from 0 to 1 ($0\leq R_2 \leq 1$) as can be seen from
Figure~\ref{fig:r2} which shows the $R_2$ distribution for signal and $qq$ events with or without selection criteria.
Since lower value of $R_2$ indicates more spherical event,
we remove $qq$ events here.

\begin{figure}[H]
\begin{center}
\resizebox{0.48\textwidth}{!}{\includegraphics{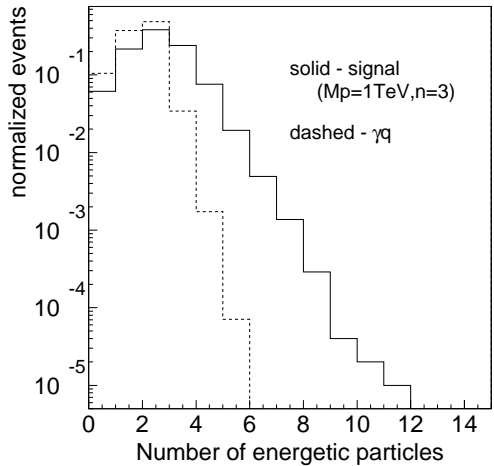}}
\end{center}
\caption{Shape of the distribution of the number of energetic particles with $E$ $>$ 300~GeV~:~comparison between signal and \bggj ~background~($\TPs=1$~TeV, $n$=3).}
\label{fig:n-particles}
\end{figure}

\begin{figure}[H]
\begin{center}
\resizebox{0.48\textwidth}{!}{\includegraphics{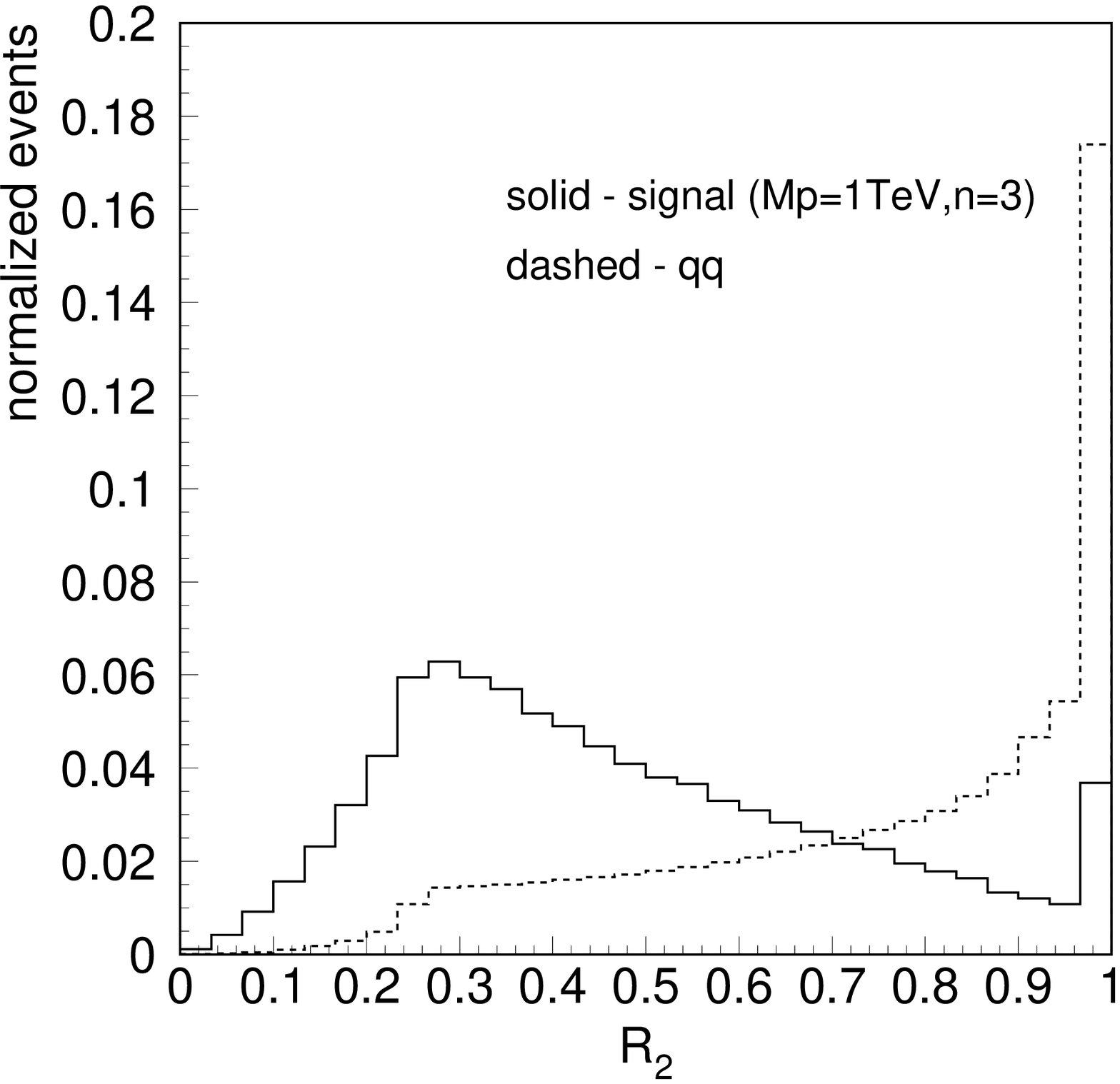}}
\resizebox{0.48\textwidth}{!}{\includegraphics{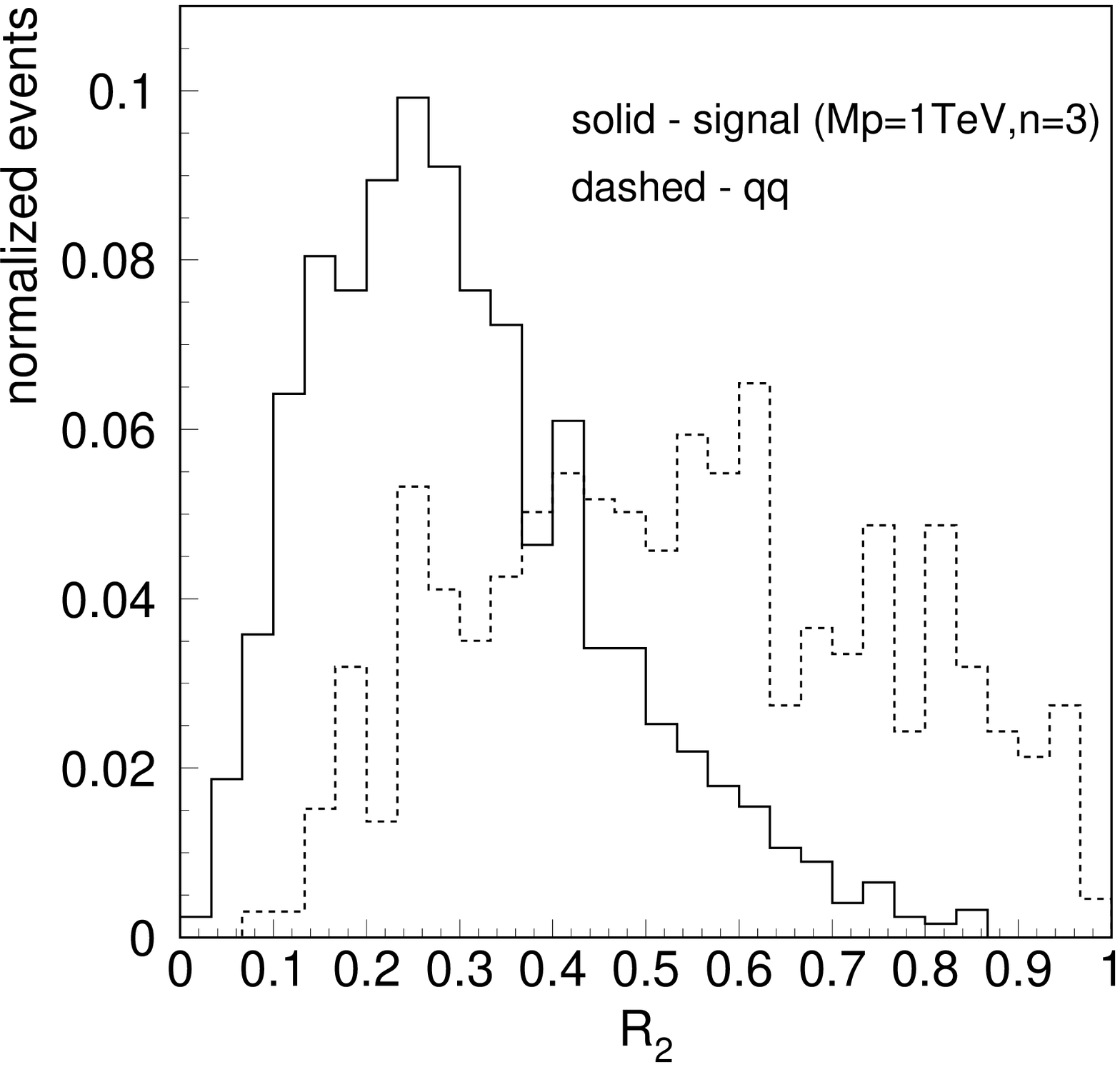}}
\end{center}
\caption{$R_2$ Distributions~:~comparison between signal event and 
$qq$~event~($\TPs=1$~TeV, $n$=3).
Left: without any selection criteria and
right: with selection criteria}
\label{fig:r2}
\end{figure}


\item $\mET<100$~GeV.
For a precise calculation of BH mass, events with high missing energy
are rejected here.
\end{itemize}

The black hole mass is then reconstructed from the 4-momenta of the remaining muons, electrons, gammas and jets as follows:
\[
p_{\mathrm{BH}} = \sum_{i=\mu,e,\gamma,jet}p_i
\]
\[
\mbh = \sqrt{p_{\mathrm{BH}}^2},
\]
where $p_i$ is a reconstructed four-momentum of each particle.

The quality of the reconstructed mass of the BH is discussed in Appendix~\ref{app:mass}.

\subsection{Discovery Potential}

We have evaluated the discovery potential for the following cases of ($\TPs$ ,$n$):
\begin{center}
$
\begin{cases}
\TPs =1,3,4,5,6,7~\mathrm{TeV} \\
n = 2,3,5,7 
\end{cases}
$
\end{center}
by using the selection criteria described in Section~\ref{sec:criteria}.
Because $\Mp$ is an unknown parameter, we cannot set the proper cut value
for the lower limit of $\mbh$.
Therefore we consider various values as follows and evaluate
$S/\sqrt{B}$ in each case:
\begin{center}
$\mbh>\mbhmin = 1,2,3,4,5$~TeV. 
\end{center}

The $\mbh$ distributions are shown in Figure~\ref{fig:bh-mass-dist}
for different values of $\TPs$ with $n=3$.
These are obtained after all the selection criteria have been applied.
The histograms represent the sum of signals and backgrounds
with the cross-hatched part showing only backgrounds.
Note that $\mbhmin$ is set to 1~TeV for this figure.

\begin{figure}[H]
\hspace{1.2cm} (a) $\Mp$=1~TeV \hspace{5.0cm} (b) $\Mp$=3~TeV
\vspace{-1.0cm}
\begin{center}
\resizebox{0.43\textwidth}{!}{\includegraphics{figs/bh-atlas-note-lboff-mbh-131.epsi}}
\resizebox{0.43\textwidth}{!}{\includegraphics{figs/bh-atlas-note-lboff-mbh-331.epsi}}
\end{center}
\hspace{1.2cm} (c) $\Mp$=4~TeV \hspace{5.0cm} (d) $\Mp$=5~TeV
\vspace{-1.0cm}
\begin{center}
\resizebox{0.43\textwidth}{!}{\includegraphics{figs/bh-atlas-note-lboff-mbh-431.epsi}}
\resizebox{0.43\textwidth}{!}{\includegraphics{figs/bh-atlas-note-lboff-mbh-531.epsi}}
\end{center}
\hspace{1.2cm} (e) $\Mp$=6~TeV \hspace{5.0cm} (f) $\Mp$=7~TeV
\vspace{-1.0cm}
\begin{center}
\resizebox{0.43\textwidth}{!}{\includegraphics{figs/bh-atlas-note-lboff-mbh-631.epsi}}
\resizebox{0.43\textwidth}{!}{\includegraphics{figs/bh-atlas-note-lboff-mbh-731.epsi}}
\end{center}
\caption{$\mbh$ distributions in the case of $n$=3 and $\mbhmin$=1~TeV.~(solid line~$:$~ signal plus background, cross hatched~$:$~background only)}
\label{fig:bh-mass-dist}
\end{figure}

Tables~\ref{table:bg}-\ref{table:signal-event-number2} show 
the number of events passing the selection criteria for background~(BG) 
processes and signal events with various ($\TPs$ ,$n$).
The number is normalized for \ILu =10~\Fb.
The values of efficiency are listed only for $\mbhmin = 1$~TeV.
As can be seen in Table~\ref{table:bg}, \bggj~process has the 
highest efficiency, but the background is dominated by the
$qq$ process owing to its large cross section.
The total number of BG events is found to be 
$\sim 10^3$~-~$10^4$ in the case of $\mbhmin = $~1~-~2~TeV 
and $\sim 10$~-~$10^2$ in the case of 
$\mbhmin =  $~4~-~5~TeV.

According to Tables~\ref{table:signal-event-number1} and
\ref{table:signal-event-number2}, the efficiency for signal events 
is $\sim$ 1\% 
in all cases of ($\TPs$ ,$n$). 
It is relatively lower
when $\TPs = 1$~TeV or $7$~TeV
because signal events are more likely to be rejected by the $\mbhmin$~cut
in the case of smaller $\TPs$, or they have a large $\mET$ for larger
$\TPs$.

\input{ana-table-number}

\input{ana-table-sb}

We calculate $S/\sqrt{B}$ from the data of Tables~\ref{table:bg}-
\ref{table:signal-event-number2} and estimate
\ILuD~---~the integrated luminosity
with which discovery is achieved.
Here the condition of discovery is set as:
\[S/\sqrt{B}~ \geq ~ 5.0 
\hspace{0.75cm}
\textrm{and} 
\hspace{0.75cm}
S~\geq ~10 \]
which is a conventional condition, as used in the analysis of Higgs events 
with ATLAS.
The results for $S/\sqrt{B}$ and \ILuD\
are shown in Table~\ref{table:soverrootb1}-\ref{table:soverrootb2} 
and \ref{table:ldiscovery} respectively.
We require that $\mbhmin$ be larger than $\TPs$ to calculate
integrated luminosities for BH discovery.
In Table~\ref{table:ldiscovery}, the shaded values 
indicate the most favorable cut in $\mbhmin$
in each case of ($\TPs$ ,$n$).
From these tables, we see that the discovery can be accomplished 
within \ILuD~$\le~1$~\Fb~ in all cases of 
$n$ if $\TPs$ is less than $\sim$ 5~TeV. 

\input{ana-table-ldis}

Figure~\ref{fig:contours} gives a contour plot
for \ILuD\ in ($\TPs$ ,$n$) plane. 
We find that the discovery potential hardly depends on $n$
but has a strong dependence on $\TPs$.
This is due to the fact that the
cross section is a strong function of $\TPs$
but not of $n$,
as shown in Tables~\ref{table:signal-event-number1} and \ref{table:signal-event-number2}.
From this relation between $\TPs$ and \ILuD,
$\TPs$ can be determined by how early the discovery is accomplished.

\begin{figure}[H]
\begin{center}
\resizebox{0.70\textwidth}{!}{\includegraphics{figs/bh-atlas-note-contour-c.epsi}}
\end{center}
\caption{Contours of \ILuD\ in ($\TPs$,~$n$) plane. }
\label{fig:contours}
\end{figure}

In a quantitative point of view,
it is found that the excess of events is detected
in $\sim$ 1~month at low luminosity~(\ILu$=$1~\Fb) if $\TPs < 5$~TeV,
and discovery within only one day~(\ILu =100~\Pb)
can be expected if $\TPs < 4$~TeV.


As was previously mentioned, the BH model we assumed here is valid
only when $\mbh \gg \TPs$.
As $\mbh$ approaches $\TPs$, the theory of BH production
becomes very complex.
If we consider events with reconstructed $\mbh>5$~TeV in the case of 
$\TPs=1$~TeV, we find that we need $\sim$5~\Pb\ instead of $\sim$0.1~\Pb\
to discover black holes.
However the events include black holes
whose generated mass was less then 5~TeV.
When black holes with mass greater than 5~TeV are generated, we find that we need $\sim$10~\Pb.

Considering that the model of BH formation and decay is valid only for
$\mbh \gg \TPs$, we show, in Figure~\ref{fig:contours2}, an evaluation of the discovery potential
of BHs when the cut $\mbhmin > \TPs +1$~TeV is applied. 
Although more integrated luminosity is required, it is clear that an excess of events will
still be easily observed within a few days of running, for $\TPs$ values up to a few TeV.

\begin{figure}[H]
\begin{center}
\resizebox{0.70\textwidth}{!}{\includegraphics{figs/bh-atlas-note-contour-b.epsi}}
\end{center}
\caption{Contours of \ILuD\ in ($\TPs$,~$n$) plane in case of $\mbhmin > \TPs +1$~TeV.}
\label{fig:contours2}
\end{figure}

%% file: ana-table-number.tex
\begin{table}
\caption{The number of remaining BG events at \ILu =10~\Fb.
$\epsmin$ is the value of efficiency for $\mbhmin = 1$~TeV.}

\begin{center}
\begin{tabular}{cc|ccc} \hline \hline
\multicolumn{2}{c|}{} & \bgqq & \bgtt & \bgww \\ \hline
\multicolumn{2}{c|}{$\sigma$~(pb)} & \enm{1.29}{4} & 493 & 0.468 \\
\multicolumn{2}{c|}{$\epsmin$} & \enm{1.67}{-5} & \enm{1.39}{-5} & \enm{8.5}{-5} \\
       & 1.0 & \enm{2.15}{3} & 68.5 & 0.40 \\
 $\mbhmin$ & 2.0 & \enm{1.18}{3} & 13 & 0.1 \\
       & 3.0 & \enm{4.3}{2} & 2 & 0.02 \\
 (TeV) & 4.0 &  77 & 0.5 & 0 \\
       & 5.0 & \enm{3}{} & 0 & 0 \\ \hline \hline
\end{tabular}
\end{center}


\begin{center}
\begin{tabular}{cc|cccc} \hline \hline
\multicolumn{2}{c|}{} & \bgwz & \bgzz & \bggg & \bggv \\ \hline
\multicolumn{2}{c|}{$\sigma$~(pb)} & 25.9 & 10.6 & 229 & 280 \\
\multicolumn{2}{c|}{$\epsmin$} & \enm{5}{-7} & \enm{1.5}{-6} &  0 & \enm{7}{-7} \\
       & 1.0 & 0.1 & 0.084 & 0 & 2 \\
 $\mbhmin$ & 2.0 & 0.05 & 0.04 & 0 & 1 \\
       & 3.0 & 0 & 0 & 0 & 0 \\
 (TeV) & 4.0 & 0 & 0 & 0 & 0 \\
       & 5.0 & 0 & 0 & 0 & 0 \\ \hline \hline
\end{tabular}
\end{center}

\begin{center}
\begin{tabular}{cc|cccc} \hline \hline
\multicolumn{2}{c|}{} & \bgwj & \bgzj & \bggj & total \\ \hline
\multicolumn{2}{c|}{$\sigma$~(pb)} & 73.4 & 31.5 & 23.5 & \\
\multicolumn{2}{c|}{$\epsmin$} & \enm{4.7}{-5} & \enm{1.40}{-4} & \enm{1.484}{-3} & \\
       & 1.0 & 35 & 44.2 & 349.3 & \enm{2.66}{3} \\
 $\mbhmin$ & 2.0 & 18 & 21 & 195 & \enm{1.43}{3} \\
       & 3.0 & 4 & 3.5 & 49.2 & \enm{4.9}{2} \\
 (TeV) & 4.0 & 0.7 & 0.3 & 7.8 & 86 \\
       & 5.0 & 0 & 0 & 2.8 & \enm{3}{} \\ \hline \hline
\end{tabular}
\end{center}
\label{table:bg}
\end{table}

\begin{table}
\caption{The number of remaining signal events at \ILu =10~\Fb.
$\epsmin$ is the value of efficiency for $\mbhmin = 1$~TeV.}
\begin{center}
\begin{tabular}{|c|c||c|c|c|} \hline
\multicolumn{2}{|c||}{($\TPs, n$)} & (1,2) & (3,2) & (4,2) \\ \hline
\multicolumn{2}{|c||}{$\sigma$~(pb)} & 9450 & 25.51 & 3.738 \\ \hline
\multicolumn{2}{|c||}{$\epsmin$} & $9.63\times 10^{-3}$ & $2.747\times 10^{-2}$ & $2.733\times 10^{-2}$ \\ \hline
 & 1.0 & $9.10\times 10^{5}$ & 7008 & 1022 \\ \cline{2-5}
 $\mbh ^{\textrm{min}}$ & 2.0 & $6.50\times 10^{5}$ & 7002 & 1021 \\ \cline{2-5}
 & 3.0 & $2.52\times 10^{5}$ & 6594 & 1016 \\ \cline{2-5}
 (TeV) & 4.0 & $7.5\times 10^{4}$& 2987 & 943.5 \\ \cline{2-5}
 & 5.0 & $2.3\times 10^{4}$ & $1.06\times 10^{3}$ & 404.5 \\ \hline
\end{tabular}
\end{center}

\begin{center}
\begin{tabular}{|c|c||c|c|c|} \hline
\multicolumn{2}{|c||}{($\TPs, n$)} & (5,2) & (6,2) & (7,2) \\ \hline
\multicolumn{2}{|c||}{$\sigma$~(pb) } & 0.6622 & 0.1249 & $2.292\times 10^{-2}$ \\ \hline
\multicolumn{2}{|c||}{$\epsmin$} & $2.601\times 10^{-2}$ & $2.287\times 10^{-2}$ & $2.022\times 10^{-2}$ \\ \hline
 & 1.0 & 172.2 & 28.56 & 4.634 \\ \cline{2-5}
 $\mbh ^{\textrm{min}}$ & 2.0 & 172.2 & 28.54 & 4.634 \\ \cline{2-5}
 & 3.0 & 171.7 & 28.53 & 4.632 \\ \cline{2-5}
 (TeV) & 4.0 & 170.1 & 28.46 & 4.616 \\ \cline{2-5}
 & 5.0 & 154.2 & 28.15 & 4.593 \\ \hline
\end{tabular}
\end{center}

\begin{center}
\begin{tabular}{|c|c||c|c|c|} \hline
\multicolumn{2}{|c||}{($\TPs, n$)} & (1,3) & (3,3) & (4,3) \\ \hline
\multicolumn{2}{|c||}{$\sigma$~(pb)} & 8256 & 22.89 & 3.381 \\ \hline
\multicolumn{2}{|c||}{$\epsmin$} & $8.46\times 10^{-3}$ & $2.725\times 10^{-2}$ & $2.605\times 10^{-2}$ \\ \hline
 & 1.0 & $6.98\times 10^{5}$ &  6238 & 880.8 \\ \cline{2-5}
 $\mbh ^{\textrm{min}}$ & 2.0 & $4.82\times 10^{5}$ & 6224 & 879.4 \\ \cline{2-5}
 & 3.0 & $2.01\times 10^{5}$ & 5878 & 874.0 \\ \cline{2-5}
 (TeV) & 4.0 &$5.8\times 10^{4}$& 2745 & 810.8 \\ \cline{2-5}
 & 5.0 &$1.2\times 10^{4}$& 929 & 354.3 \\ \hline
\end{tabular}
\end{center}

\begin{center}
\begin{tabular}{|c|c||c|c|c|} \hline
\multicolumn{2}{|c||}{($\TPs, n$)} & (5,3) & (6,3) & (7,3) \\ \hline
\multicolumn{2}{|c||}{$\sigma$~(pb) } & 0.6027 & 0.1142 & $2.105\times 10^{-2}$ \\ \hline
\multicolumn{2}{|c||}{$\epsmin$} & $2.574\times 10^{-2}$ & $2.347\times 10^{-2}$ & $2.071\times 10^{-2}$ \\ \hline
 & 1.0 & 155.1 & 26.80 & 4.359 \\ \cline{2-5}
 $\mbh ^{\textrm{min}}$ & 2.0 & 155.1 & 26.80 & 4.357 \\ \cline{2-5}
 & 3.0 & 154.4 & 26.76 & 4.351 \\ \cline{2-5}
 (TeV) & 4.0 & 152.7& 26.59 & 4.341 \\ \cline{2-5}
 & 5.0 & 140.4 & 26.23 & 4.317 \\ \hline
\end{tabular}
\end{center}
\label{table:signal-event-number1}
\end{table}

\begin{table}
\caption{The number of remaining signal events at \ILu =10~\Fb.
$\epsmin$ is the value of efficiency for $\mbhmin = 1$~TeV.}
\begin{center}
\begin{tabular}{|c|c||c|c|c|} \hline
\multicolumn{2}{|c||}{($\TPs, n$)} & (1,5)& (3,5) & (4,5) \\ \hline
\multicolumn{2}{|c||}{$\sigma$~(pb)} & 8244 & 23.42 & 3.487 \\ \hline
\multicolumn{2}{|c||}{$\epsmin$} & $7.24\times 10^{-3}$ & $2.563\times 10^{-2}$ & $2.779\times 10^{-2}$ \\ \hline
 & 1.0 & $5.97\times 10^{5}$ & 6003 & 969.0 \\ \cline{2-5}
 $\mbh ^{\textrm{min}}$ & 2.0 & $4.24\times 10^{5}$ & 5993 & 969.0 \\ \cline{2-5}
 & 3.0 & $1.72\times 10^{5}$ & 5600 & 964.5 \\ \cline{2-5}
 (TeV) & 4.0 &$5.0\times 10^{4}$& 2497 & 892.0 \\ \cline{2-5}
 & 5.0 &$1.4\times 10^{4}$& 775 & 377.3 \\ \hline
\end{tabular}
\end{center}

\begin{center}
\begin{tabular}{|c|c||c|c|c|} \hline
\multicolumn{2}{|c||}{($\TPs, n$)} & (5,5)& (6,5) & (7,5) \\ \hline
\multicolumn{2}{|c||}{$\sigma$~(pb)} & 0.6252 & 0.1191 & $2.203\times 10^{-2}$ \\ \hline
\multicolumn{2}{|c||}{$\epsmin$} & $2.501\times 10^{-2}$ & $2.244\times 10^{-2}$ & $2.139\times 10^{-2}$ \\ \hline
 & 1.0 & 156.4 & 26.73 & 4.712 \\ \cline{2-5}
 $\mbh ^{\textrm{min}}$ & 2.0 & 156.3 & 26.73 & 4.712 \\ \cline{2-5}
 & 3.0 & 155.7 & 26.69 & 4.706 \\ \cline{2-5}
 (TeV) & 4.0 & 154.9 & 26.61 & 4.688 \\ \cline{2-5}
 & 5.0 & 139.2 & 26.33 & 4.662 \\ \hline
\end{tabular}
\end{center}

\begin{center}
\begin{tabular}{|c|c||c|c|c|} \hline
\multicolumn{2}{|c||}{($\TPs, n$)} & (1,7)& (3,7) & (4,7) \\ \hline
\multicolumn{2}{|c||}{$\sigma$~(pb)} & 9053 & 26.02 & 3.887 \\ \hline
\multicolumn{2}{|c||}{$\epsmin$} & $6.54\times 10^{-3}$ & $2.577\times 10^{-2}$ & $2.630\times 10^{-2}$ \\ \hline
 & 1.0 & $5.92\times 10^{5}$ & 6705 & 1022 \\ \cline{2-5}
 $\mbh ^{\textrm{min}}$ & 2.0 & $4.22\times 10^{5}$ & 6698 & 1020 \\ \cline{2-5}
 & 3.0 & $1.57\times 10^{5}$ & 6310 & 1017 \\ \cline{2-5}
 (TeV) & 4.0 &$4.5\times 10^{4}$& 2789 & 943.0 \\ \cline{2-5}
 & 5.0 &$5\times 10^{3}$& 892 & 382 \\ \hline
\end{tabular}
\end{center}

\begin{center}
\begin{tabular}{|c|c||c|c|c|} \hline
\multicolumn{2}{|c||}{($\TPs, n$)} & (5,7)& (6,7) & (7,7) \\ \hline
\multicolumn{2}{|c||}{$\sigma$~(pb) } & 0.6991 & 0.1335 & $2.474\times 10^{-2}$ \\ \hline
\multicolumn{2}{|c||}{$\epsmin$} & $2.577\times 10^{-2}$ & $2.319\times 10^{-2}$ & $2.090\times 10^{-2}$ \\ \hline
 & 1.0 & 180.2 & 30.96 & 5.171 \\ \cline{2-5}
 $\mbh ^{\textrm{min}}$ & 2.0 & 180.2 & 30.93 & 5.171 \\ \cline{2-5}
 & 3.0 & 179.7 & 30.89 & 5.168 \\ \cline{2-5}
 (TeV) & 4.0 & 178.3 & 30.72 & 5.151 \\ \cline{2-5}
 & 5.0 & 164.7 & 30.48 & 5.124 \\ \hline
\end{tabular}
\end{center}
\label{table:signal-event-number2}
\end{table}

%% file: ana-table-sb.tex
\begin{table}
\caption{The value of $S/\sqrt{B}$ at \ILu =10~\Fb.}
\begin{center}
\begin{tabular}{|c|c||c|c|c|} \hline
\multicolumn{2}{|c||}{($\TPs, n$)} & (1,2)& (3,2) & (4,2) \\ \hline
\multicolumn{2}{|c||}{$\sigma$~(pb)} & 9450 & 25.51 & 3.738 \\ \hline
 & 1.0 & $1.76\times 10^{4}$ & 136 & 19.8 \\ \cline{2-5}
 $\mbh ^{\textrm{min}}$ & 2.0 & $1.72\times 10^{4}$ & 185 & 27.0 \\ \cline{2-5}
 & 3.0 & $1.1\times 10^{4}$ & $3.0\times 10^{2}$ & 46 \\ \cline{2-5}
 (TeV) & 4.0 &$8.1\times 10^{3}$& $3.2\times 10^{2}$ & $1.0\times 10^{2}$ \\ \cline{2-5}
 & 5.0 &$4\times 10^{3}$& $2\times 10^{2}$ & $7\times 10$ \\ \hline
\end{tabular}
\end{center}

\begin{center}
\begin{tabular}{|c|c||c|c|c|} \hline
\multicolumn{2}{|c||}{($\TPs, n$)} & (5,2)& (6,2) & (7,2) \\ \hline
\multicolumn{2}{|c||}{$\sigma$~(pb) } & 0.6622 & 0.1249 & $2.292\times 10^{-2}$ \\ \hline
 & 1.0 & 3.34 & 0.554 & 0.0899 \\ \cline{2-5}
 $\mbh ^{\textrm{min}}$ & 2.0 & 4.55 & 0.755 & 0.123 \\ \cline{2-5}
 & 3.0 & 7.8 & 1.3 & 0.21 \\ \cline{2-5}
 (TeV) & 4.0 & 18 & 3.1 & 0.50 \\ \cline{2-5}
 & 5.0 & $3\times 10$ & 5 & 0.8 \\ \hline
\end{tabular}
\end{center}

\begin{center}
\begin{tabular}{|c|c||c|c|c|} \hline
\multicolumn{2}{|c||}{($\TPs, n$)} & (1,3)& (3,3) & (4,3) \\ \hline
\multicolumn{2}{|c||}{$\sigma$~(pb)} & 8256 & 22.89 & 3.381 \\ \hline
 & 1.0 & $1.35\times 10^{4}$ & 121 & 17.1 \\ \cline{2-5}
 $\mbh ^{\textrm{min}}$ & 2.0 & $1.27\times 10^{4}$ & 165 & 23.3 \\ \cline{2-5}
 & 3.0 & $9.1\times 10^{3}$ & $2.7\times 10^{2}$ & 39 \\ \cline{2-5}
 (TeV) & 4.0 &$6.3\times 10^{3}$& $3.0\times 10^{2}$ & 87 \\ \cline{2-5}
 & 5.0 &$2\times 10^{3}$& $2\times 10^{2}$ & $6\times 10$ \\ \hline
\end{tabular}
\end{center}

\begin{center}
\begin{tabular}{|c|c||c|c|c|} \hline
\multicolumn{2}{|c||}{($\TPs, n$)} & (5,3)& (6,3) & (7,3) \\ \hline 
\multicolumn{2}{|c||}{$\sigma$~(pb) } & 0.6027 & 0.1142 & $2.105\times 10^{-2}$ \\ \hline
 & 1.0 & 3.01 & 0.520 & 0.0845 \\ \cline{2-5}
 $\mbh ^{\textrm{min}}$ & 2.0 & 4.10 & 0.709 & 0.115 \\ \cline{2-5}
 & 3.0 & 7.0 & 1.2 & 0.20 \\ \cline{2-5}
 (TeV) & 4.0 & 16 & 2.9 & 0.47 \\ \cline{2-5}
 & 5.0 & $3\times 10$ & 5 & 0.8 \\ \hline
\end{tabular}
\end{center}
\label{table:soverrootb1}
\end{table}

\begin{table}
\caption{The value of $S/\sqrt{B}$ at \ILu =10~\Fb.}
\begin{center}
\begin{tabular}{|c|c||c|c|c|} \hline
\multicolumn{2}{|c||}{($\TPs, n$)} & (1,5)& (3,5) & (4,5) \\ \hline 
\multicolumn{2}{|c||}{$\sigma$~(pb)} & 8244 & 23.42 & 3.487 \\ \hline
 & 1.0 & $1.16\times 10^{4}$ & 116 & 18.8 \\ \cline{2-5}
 $\mbh ^{\textrm{min}}$ & 2.0 & $1.12\times 10^{4}$ & 158 & 25.6 \\ \cline{2-5}
 & 3.0 & $7.8\times 10^{3}$ & $2.5\times 10^{2}$ & 44 \\ \cline{2-5}
 (TeV) & 4.0 &$5.4\times 10^{3}$& $2.7\times 10^{2}$ & 96 \\ \cline{2-5}
 & 5.0 &$3\times 10^{3}$ & $1\times 10^{2}$ & $7\times 10$ \\ \hline
\end{tabular}
\end{center}

\begin{center}
\begin{tabular}{|c|c||c|c|c|} \hline
\multicolumn{2}{|c||}{($\TPs, n$)} & (5,5)& (6,5) & (7,5) \\ \hline 
\multicolumn{2}{|c||}{$\sigma$~(pb) } & 0.6252 & 0.1191 & $2.203\times 10^{-2}$ \\ \hline
 & 1.0 & 3.03 & 0.518 & 0.0914 \\ \cline{2-5}
 $\mbh ^{\textrm{min}}$ & 2.0 &4.13 & 0.707 & 0.125 \\ \cline{2-5}
 & 3.0 & 7.0 & 1.2 & 0.21 \\ \cline{2-5}
 (TeV) & 4.0 & 17 & 2.9 & 0.51 \\ \cline{2-5}
 & 5.0 & $3\times 10$ & 5 & 0.9 \\ \hline
\end{tabular}
\end{center}

\begin{center}
\begin{tabular}{|c|c||c|c|c|} \hline
\multicolumn{2}{|c||}{($\TPs, n$)} & (1,7)& (3,7) & (4,7) \\ \hline 
\multicolumn{2}{|c||}{$\sigma$~(pb)} & 9053 & 26.02 & 3.887 \\ \hline
 & 1.0 & $1.15\times 10^{4}$ & 130 & 19.8 \\ \cline{2-5}
 $\mbh ^{\textrm{min}}$ & 2.0 & $1.11\times 10^{4}$ & 177 & 27.0 \\ \cline{2-5}
 & 3.0 & $7.1\times 10^{3}$ & $2.9\times 10^{2}$ & 46 \\ \cline{2-5}
 (TeV) & 4.0 &$4.9\times 10^{3}$& $3.0\times 10^{2}$ & $1.0\times 10^{2}$ \\ \cline{2-5}
 & 5.0 &$9\times 10^{2}$& $2\times 10^{2}$ & $7\times 10$ \\ \hline
\end{tabular}
\end{center}

\begin{center}
\begin{tabular}{|c|c||c|c|c|} \hline
\multicolumn{2}{|c||}{($\TPs, n$)} & (5,7)& (6,7) & (7,7) \\ \hline 
\multicolumn{2}{|c||}{$\sigma$~(pb) } & 0.6991 & 0.1335 & $2.474\times 10^{-2}$ \\ \hline
 & 1.0 & 3.49 & 0.600 & 0.100 \\ \cline{2-5}
 $\mbh ^{\textrm{min}}$ & 2.0 & 4.77 & 0.818 & 0.137 \\ \cline{2-5}
 & 3.0 & 8.1 & 1.4 & 0.23 \\ \cline{2-5}
 (TeV) & 4.0 & $19 $ & 3.3 & 0.56 \\ \cline{2-5}
 & 5.0 & $3\times 10$ & 6 & 0.9 \\ \hline
\end{tabular}
\end{center}
\label{table:soverrootb2}
\end{table}

%% file: ana-table-ldis.tex
\begin{table}
\caption{Integrated luminosity~(\Fb) required for discovery,
for various values of ($\TPs, n$), as a function of $\mbhmin$~($\mbhmin\ge\TPs$).
The shaded entry gives the most favorable value in each case of ($\TPs, n$).}
\begin{center}
\begin{tabular}{|c|c||c|c|c|c|} \hline
\multicolumn{2}{|c||}{($\TPs, n$)} & (1,2)& (3,2) & (4,2) & (5,2) \\ \hline 
\multicolumn{2}{|c||}{$\sigma$~(pb)} & 9450 & 25.51 & 3.738 & 0.6622 \\ \hline
 & 1.0 & \colorbox{mycolor}{$1.10\times 10^{-4}$} & - & - & - \\ \cline{2-6}
 $\mbhmin$ & 2.0 & $1.54\times 10^{-4}$ & - & - & - \\ \cline{2-6}
 & 3.0 & $3.97\times 10^{-4}$ & \colorbox{mycolor}{$1.517\times 10^{-2}$} & - & - \\ \cline{2-6}
 (TeV) & 4.0 &$1.3\times 10^{-3}$& $3.348\times 10^{-2}$ & \colorbox{mycolor}{0.1060} & - \\ \cline{2-6}
 & 5.0 & $4.3\times 10^{-3}$& $9.43\times 10^{-2}$ & 0.2472 & \colorbox{mycolor}{0.6485} \\ \hline
\end{tabular}
\end{center}

\begin{center}
\begin{tabular}{|c|c||c|c|c|c|} \hline
\multicolumn{2}{|c||}{($\TPs, n$)} & (1,3)& (3,3) & (4,3) & (5,3) \\ \hline 
\multicolumn{2}{|c||}{$\sigma$~(pb)} & 8256 & 22.89 & 3.381 & 0.6027 \\ \hline
 & 1.0 & \colorbox{mycolor}{$1.43\times 10^{-4}$} & - & - & - \\ \cline{2-6}
 $\mbhmin$ & 2.0 & $2.07\times 10^{-4}$ & - & - & - \\ \cline{2-6}
 & 3.0 & $4.98\times 10^{-4}$ & \colorbox{mycolor}{$1.701\times 10^{-2}$} & - & - \\ \cline{2-6}
 (TeV) & 4.0 & $1.7\times 10^{-3}$& $3.643\times 10^{-2}$ 
                             & \colorbox{mycolor}{0.1233} & - \\ \cline{2-6}
 & 5.0 & $8.3\times 10^{-3}$ & 0.108 & 0.2822 & \colorbox{mycolor}{0.7123} \\ \hline
\end{tabular}
\end{center}

\begin{center}
\begin{tabular}{|c|c||c|c|c|c|} \hline
\multicolumn{2}{|c||}{($\TPs, n$)} & (1,5)& (3,5) & (4,5) & (5,5) \\ \hline 
\multicolumn{2}{|c||}{$\sigma$~(pb)} & 8244 & 23.42 & 3.487 & 0.6252 \\ \hline
 & 1.0 & \colorbox{mycolor}{$1.68\times 10^{-4}$} & - & - & - \\ \cline{2-6}
 $\mbhmin$ & 2.0 & $2.36\times 10^{-4}$ & - & - & - \\ \cline{2-6}
 & 3.0 & $5.81\times 10^{-4}$ & \colorbox{mycolor}{$1.786\times 10^{-2}$} & - & - \\ \cline{2-6}
 (TeV) & 4.0 & $2.0\times 10^{-3}$& $4.005\times 10^{-2}$ & \colorbox{mycolor}{0.1121} & - \\ \cline{2-6}
 & 5.0 & $7.1\times 10^{-3}$& 0.129 & $0.2650$ & \colorbox{mycolor}{0.7184} \\ \hline
\end{tabular}
\end{center}

\begin{center}
\begin{tabular}{|c|c||c|c|c|c|} \hline
\multicolumn{2}{|c||}{($\TPs, n$)} & (1,7)& (3,7) & (4,7) & (5,7) \\ \hline 
\multicolumn{2}{|c||}{$\sigma$~(pb)} & 9053 & 26.02 & 3.887 & 0.6991 \\ \hline
 & 1.0 & \colorbox{mycolor}{$1.69\times 10^{-4}$} & - & - & - \\ \cline{2-6}
 $\mbhmin$ & 2.0 & $2.37\times 10^{-4}$ & - & - & - \\ \cline{2-6}
 & 3.0 & $6.37\times 10^{-4}$ & \colorbox{mycolor}{$1.585\times 10^{-2}$} & - & - \\ \cline{2-6}
 (TeV) & 4.0 &$2.2\times 10^{-3}$& $3.586\times 10^{-2}$ & \colorbox{mycolor}{0.1060} & - \\ \cline{2-6}
 & 5.0 & $2\times 10^{-2}$& 0.112 & 0.262 & \colorbox{mycolor}{0.6072} \\ \hline
\end{tabular}
\end{center}

\label{table:ldiscovery}
\end{table}

%% file: conclusion.tex
\section{Conclusion}\label{sec:conclusion}
We have studied the potential to observe black holes with the ATLAS detector at the LHC.
We developed a generator of black holes with simple assumptions
for the production and decay process, taking
the expressions in Section~\ref{sec:intro}
as valid in the full region of the black hole mass~$\mbh>\TPs$.
We find an excess of events from the Standard Model backgrounds
for the integrated luminosity of 1~\Fb~($\sim$ 1~month at low luminosity) if $\TPs < 5$~TeV and
with the integrated luminosity of 100~\Pb~($\sim$ 1~day) if $\TPs < 4$~TeV.
If we assume validity when $\mbh > \TPs + 1$~TeV,
the required luminosities are slightly larger.

We also proposed a simple method to estimate the Planck scale~$\TPs$.
It will be necessary to find a way to estimate
the number of large extra dimensions~$n$ and to identify the excess as the black holes.
The development of more realistic black hole generators is also important
for more detail studies.

%% file: acknowledgments.tex
\\
\\
{\Large \bf Acknowledgments}
\\
We thank to G.~Azuelos and L.~Poggioli for thier suggestions for our analysis.

%% file: mass.tex
\section{Reconstructed Mass of Black Hole}\label{app:mass}
The resolution in the reconstructed mass of the BH depends on various factors.
We investigate reasons of the overestimation and the underestimation
on the reconstructed mass~$\mbh$ as follows.

\begin{figure}[H]
\hspace{0.0cm} (a) ($\Mp$, $n$, $\fbhth$)=(1,3,1) \hspace{0.05cm} (b) ($\Mp$, $n$, $\fbhth$)=(1,3,6) \hspace{0.05cm} (c) ($\Mp$, $n$, $\fbhth$)=(1,3,9)
\vspace{-0.6cm}
\begin{center}
\resizebox{0.31\textwidth}{!}{\includegraphics{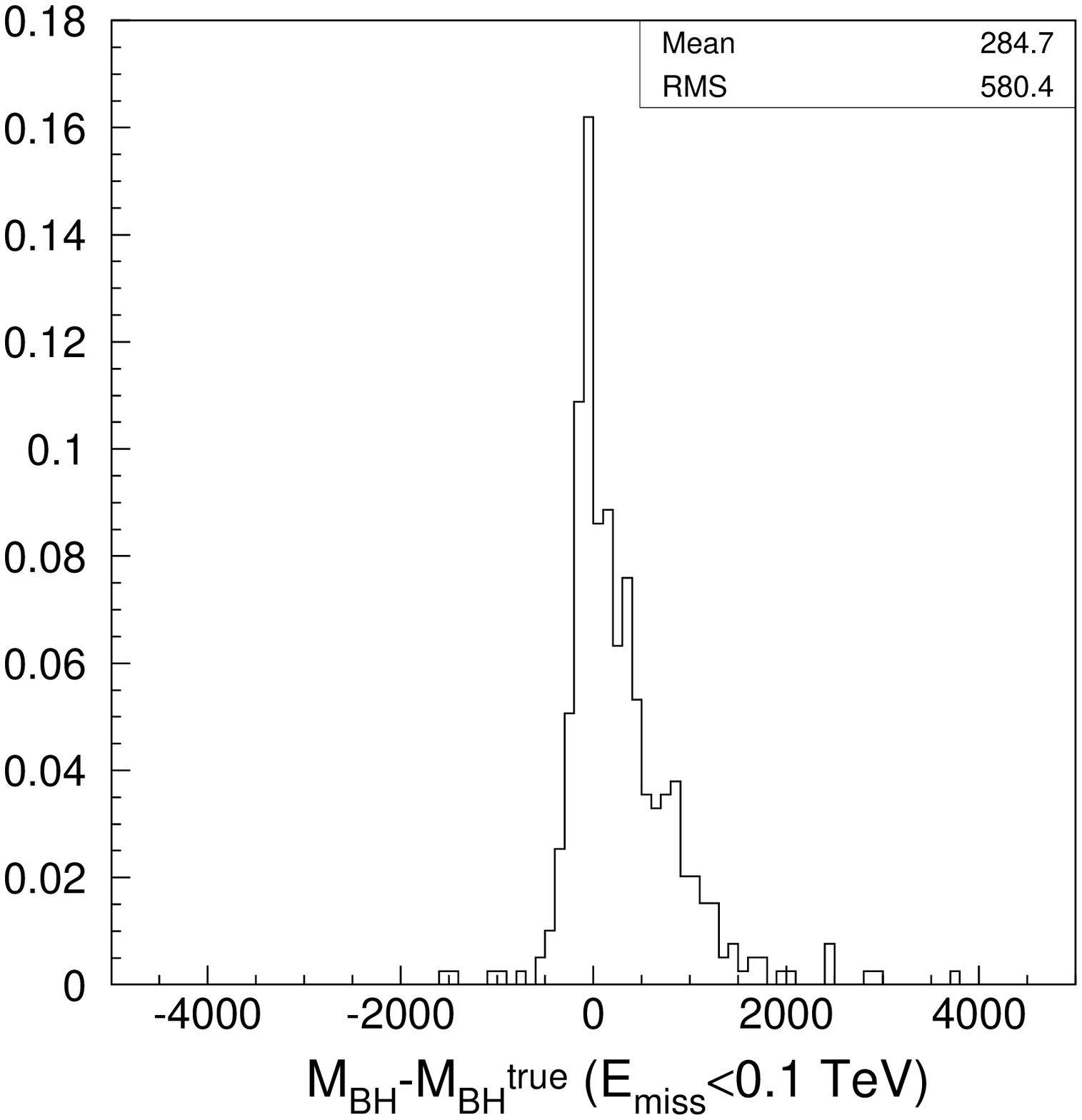}}
\resizebox{0.31\textwidth}{!}{\includegraphics{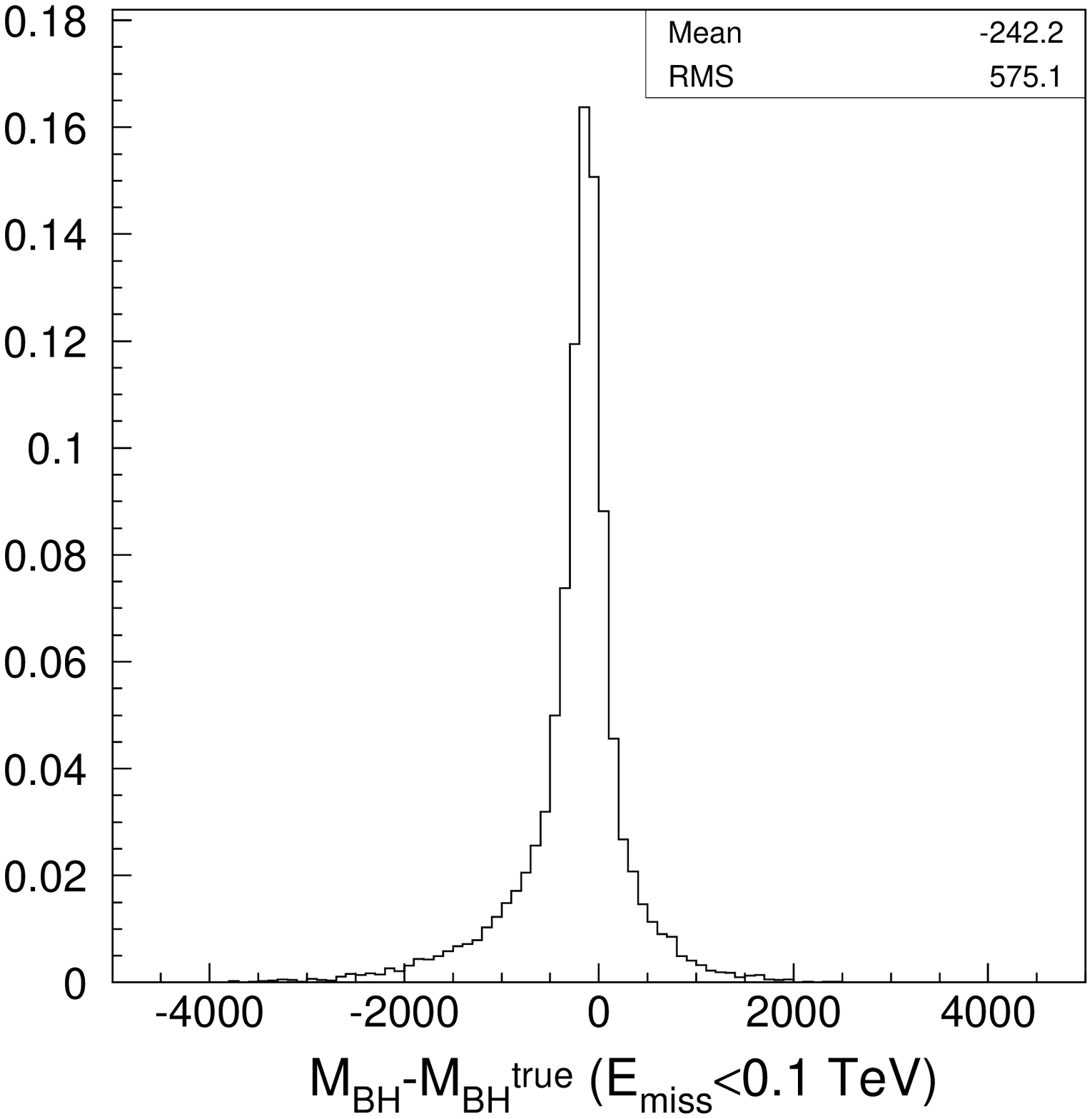}}
\resizebox{0.31\textwidth}{!}{\includegraphics{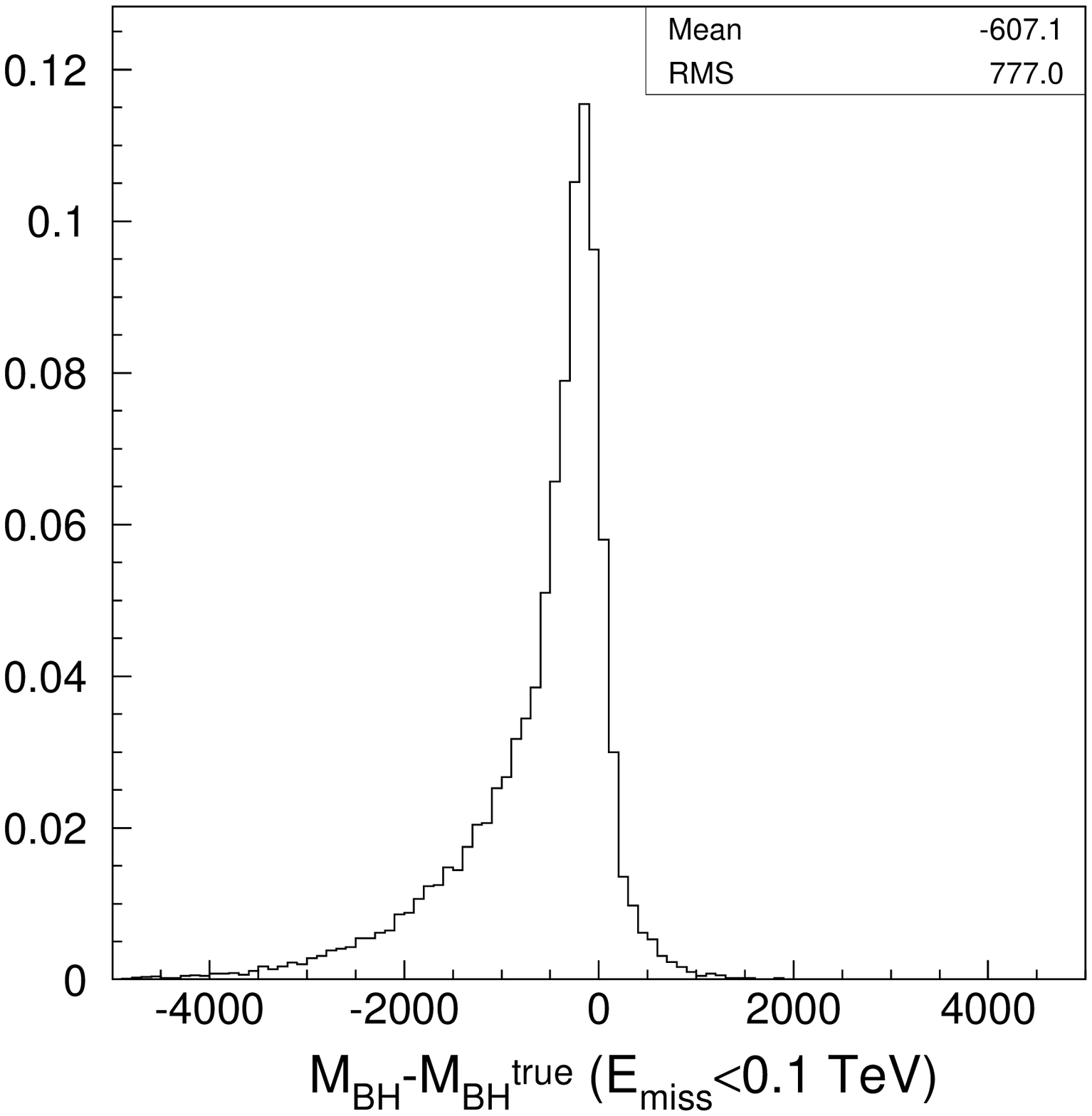}}
\end{center}
\caption{Distributions of the difference between a reconstructed and a generated mass of a black hole~$\mbh-\MBHTru$.
Our selection criteria is applied to reconstruct a black hole.
$E_{miss}$ in the figure means\; $\mET$.}
\label{fig:dif-mbh-dist}
\end{figure}

Figures~\ref{fig:dif-mbh-dist} show distributions of $\mbh-\MBHTru$
for BH events with $\MBHTru \geq 1.0, 6.0,$ and $9.0$~TeV
in case of ($\TPs,n$) = (1,3), respectively.

\begin{figure}[H]
\hspace{0.0cm} (a) ($\Mp$, $n$, $\fbhth$)=(1,3,1) \hspace{0.05cm} (b) ($\Mp$, $n$, $\fbhth$)=(1,3,6) \hspace{0.05cm} (c) ($\Mp$, $n$, $\fbhth$)=(1,3,9)
\vspace{-0.6cm}
\begin{center}
\resizebox{0.31\textwidth}{!}{\includegraphics{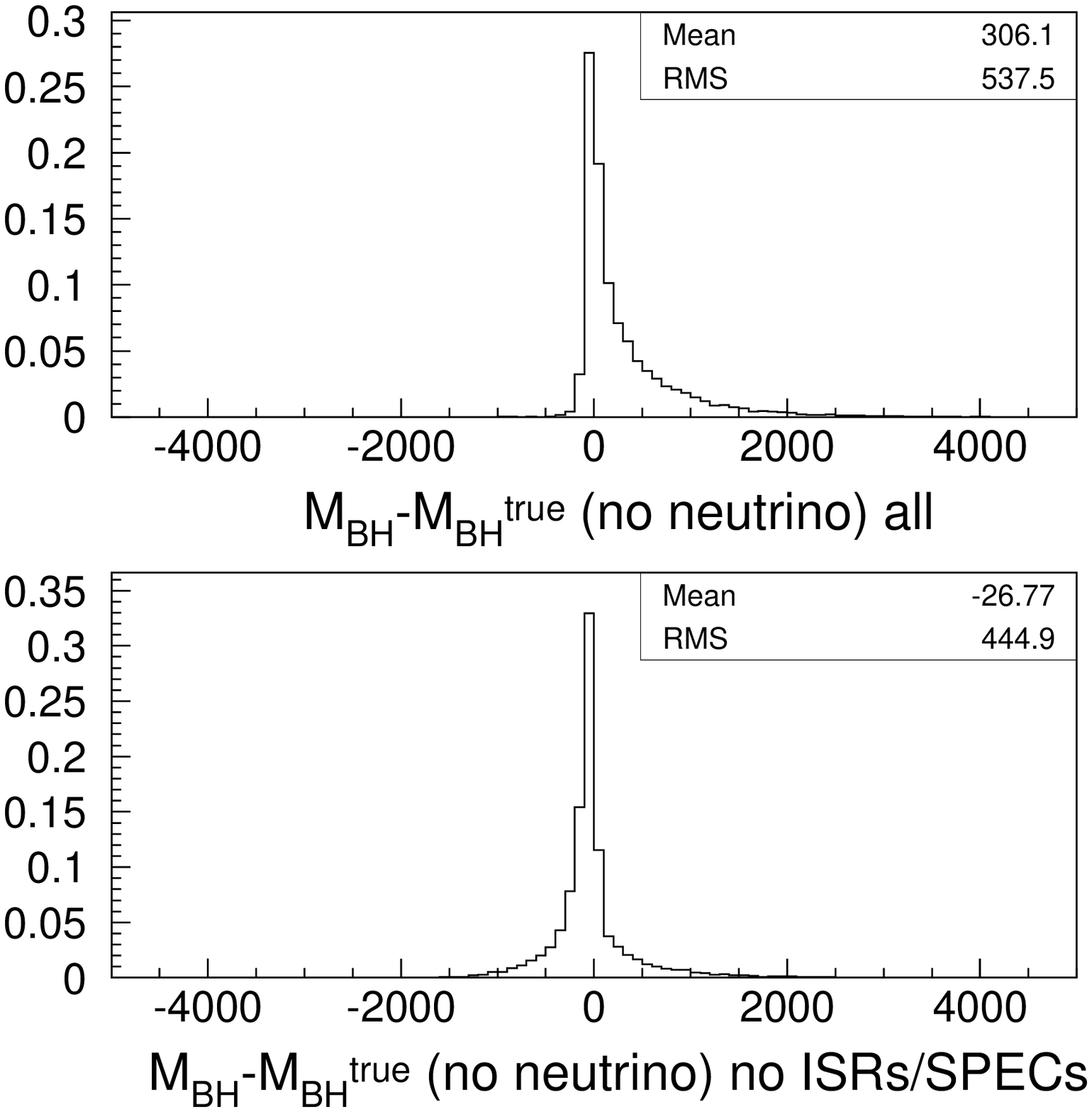}}
\resizebox{0.31\textwidth}{!}{\includegraphics{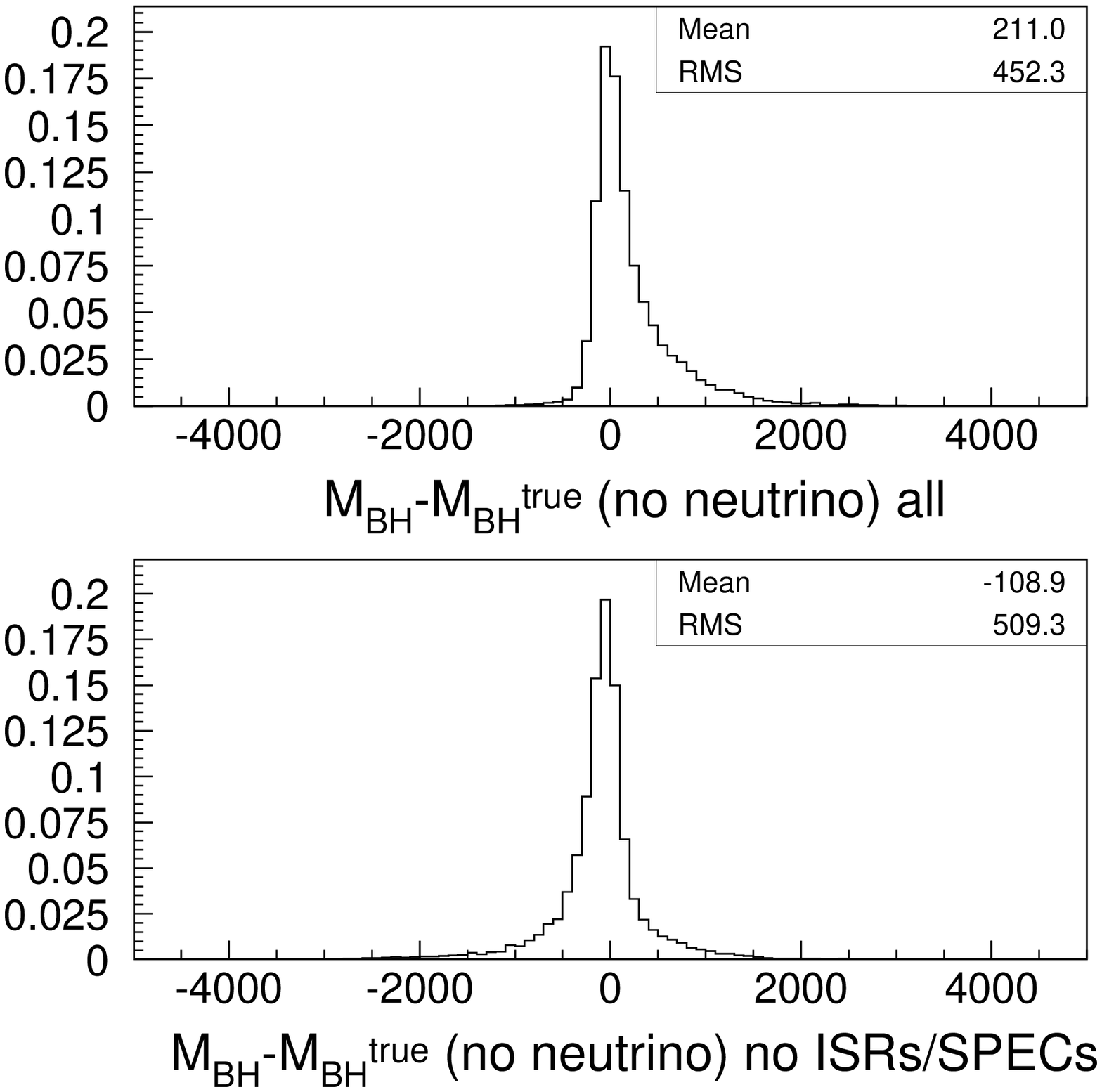}}
\resizebox{0.31\textwidth}{!}{\includegraphics{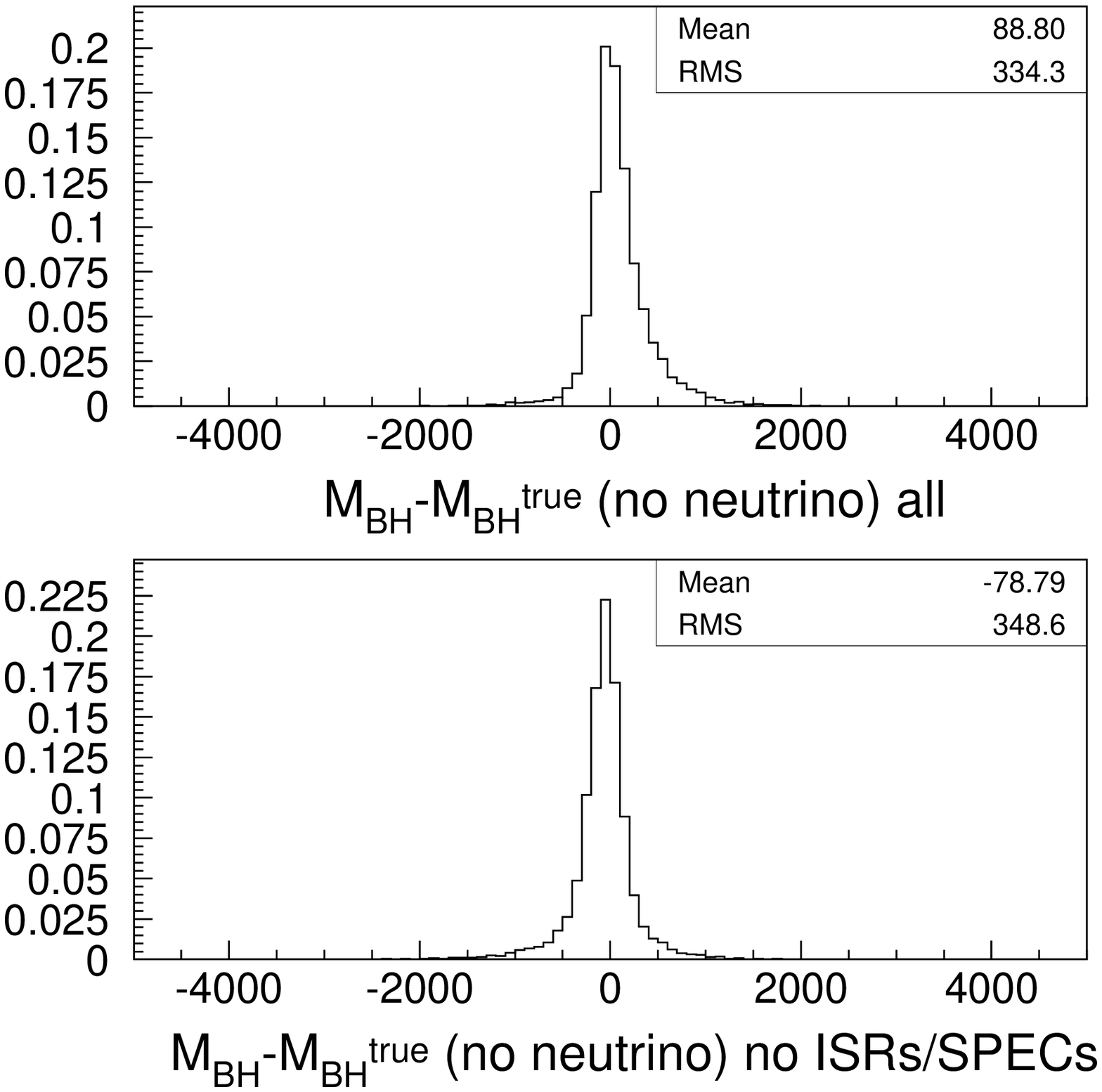}}
\end{center}
\caption{$\mbh-\MBHTru$ Distributions. We use events with no neutrino.
The upper figure is obtained from all reconstructed particles and jets and
the lower is from them except for ISRs/SPECs. 
Details are given in the text.}
\label{fig:mbh-no-neutrino}
\end{figure}

There are contaminations from particles arising from the initial state radiation
and from the spectator proton fragments of the hard scattering.
We call these particles ``ISRs/SPECs''.
Figures~\ref{fig:mbh-no-neutrino} show $\mbh-\MBHTru$ distributions of 
the events with no neutrino.
We compare the difference in mass when all particles are used and
when ISRs/SPECs are excluded.
The generator information is used to determine whether a particle comes from ISRs/SPECs.
We can see that the overestimation of a black hole mass is due to
the contaminations of ISRs/SPECs.

\begin{figure}[H]
\hspace{0.0cm} (a) ($\Mp$, $n$, $\fbhth$)=(1,3,1) \hspace{0.05cm} (b) ($\Mp$, $n$, $\fbhth$)=(1,3,6) \hspace{0.05cm} (c) ($\Mp$, $n$, $\fbhth$)=(1,3,9)
\vspace{-0.6cm}
\begin{center}
\resizebox{0.31\textwidth}{!}{\includegraphics{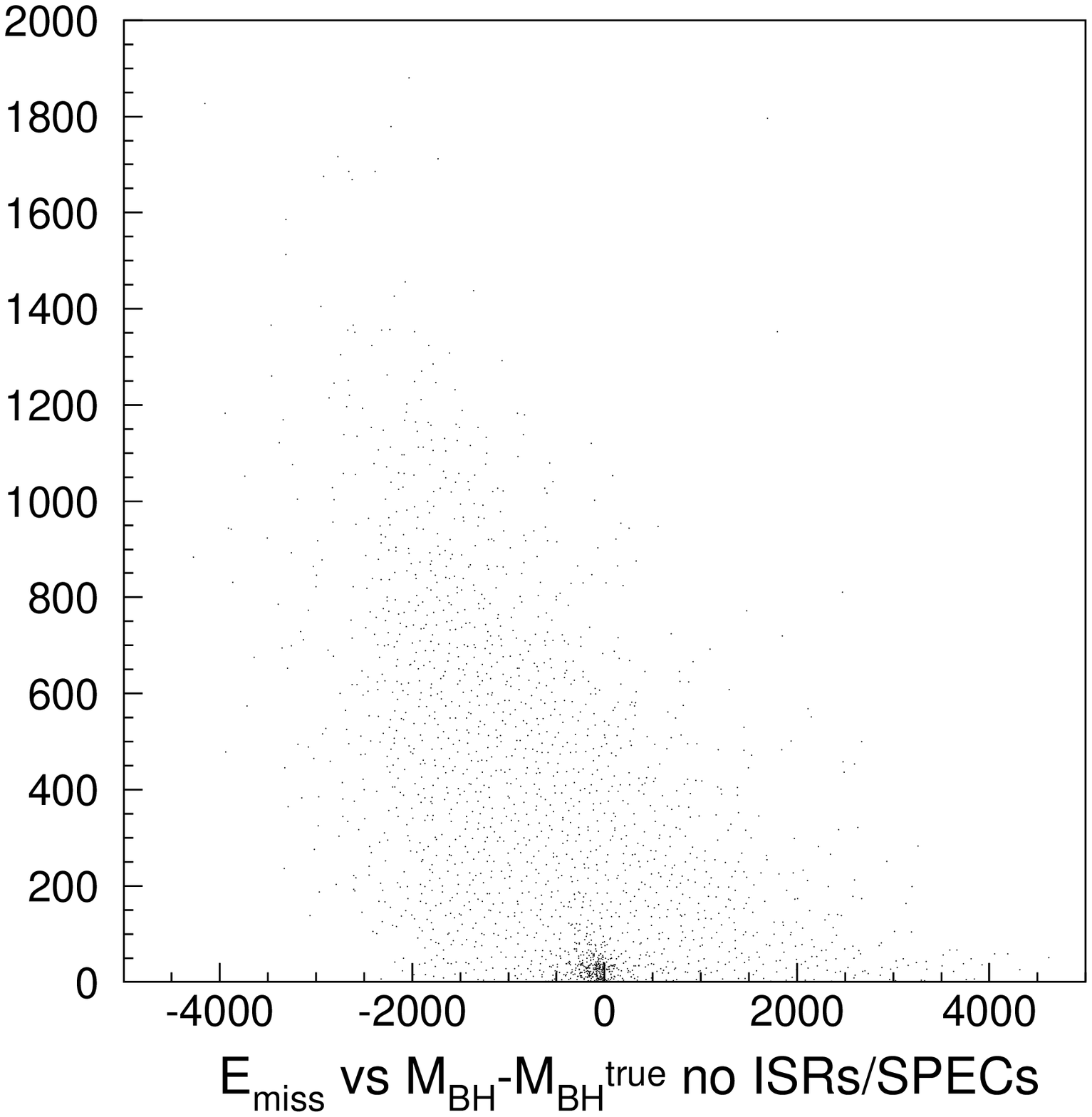}}
\resizebox{0.31\textwidth}{!}{\includegraphics{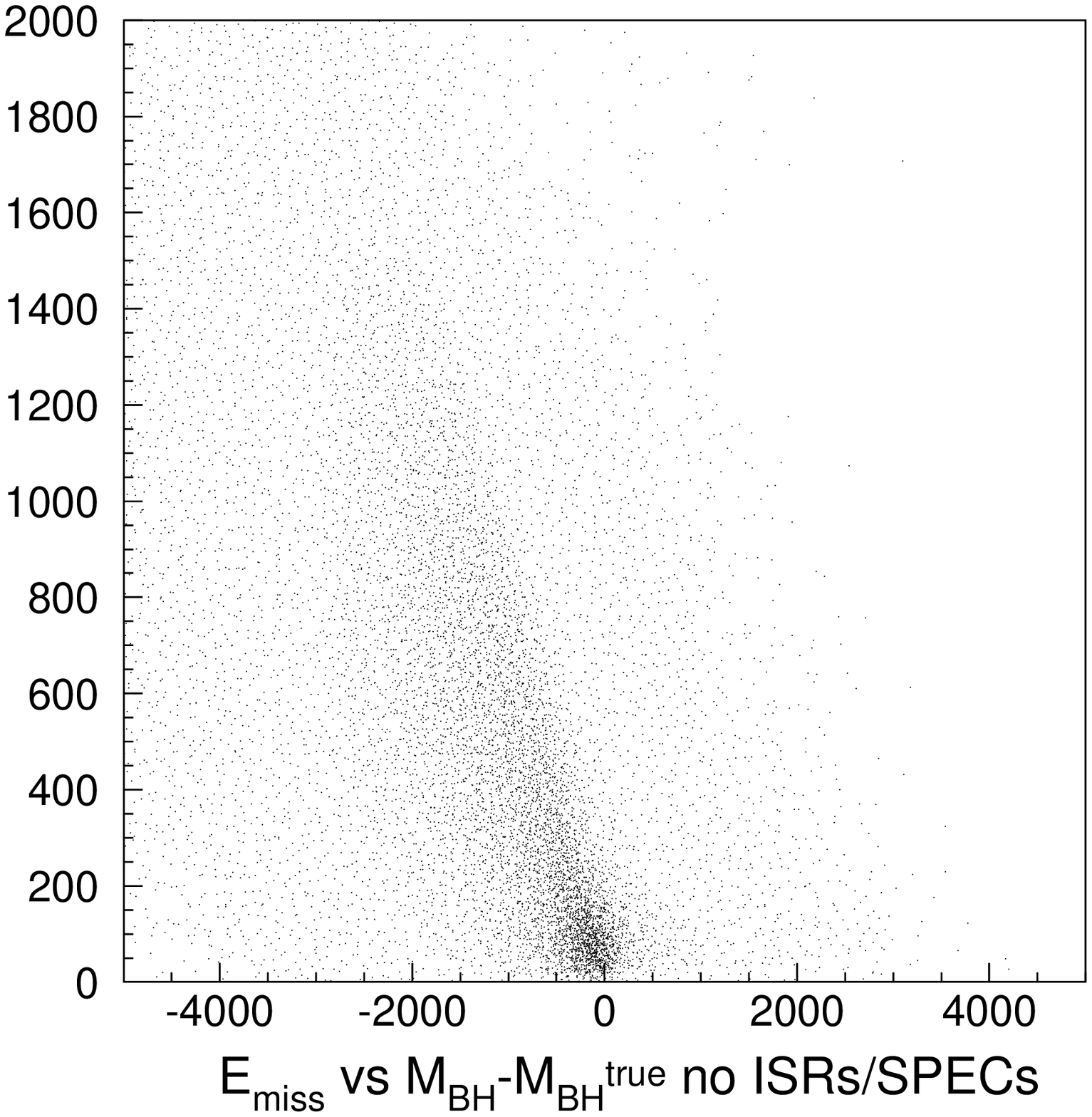}}
\resizebox{0.31\textwidth}{!}{\includegraphics{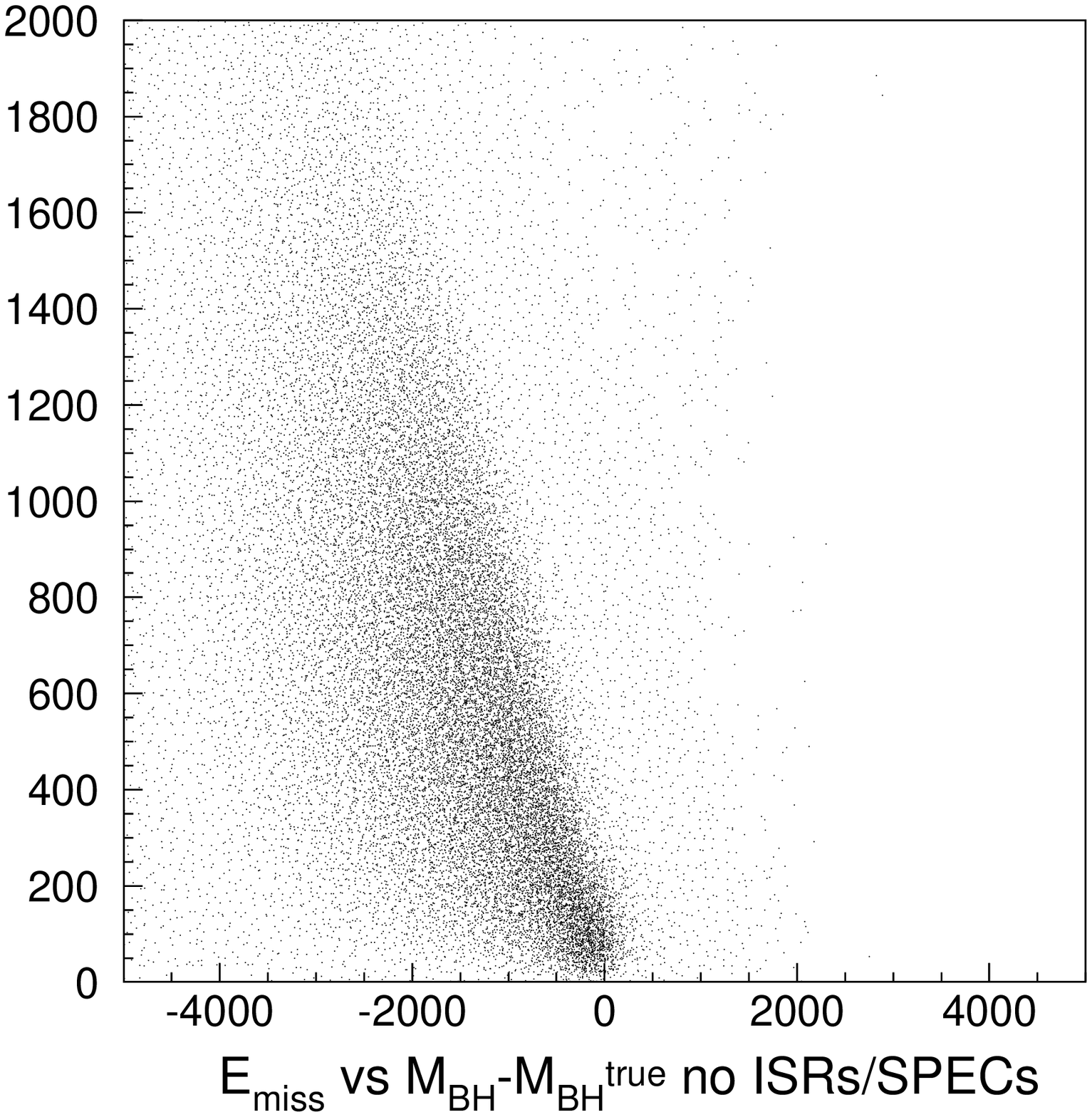}}
\end{center}
\caption{Measured missing energy $\mET$ versus $\mbh-\MBHTru$.
The mass of a black hole is obtained from reconstructed particles and jets except for ISRs/SPECs.}
\label{fig:emiss-vs-mbh}
\end{figure}

Figure~\ref{fig:emiss-vs-mbh} shows the correlation between
a $\mbh-\MBHTru$ and a measured missing energy~$\mET$.
We can see clearly that
we underestimate a black hole mass if the missing energy is important.

\begin{figure}[H]
\hspace{0.0cm} (a) ($\Mp$, $n$, $\fbhth$)=(1,3,1) \hspace{0.05cm} (b) ($\Mp$, $n$, $\fbhth$)=(1,3,6) \hspace{0.05cm} (c) ($\Mp$, $n$, $\fbhth$)=(1,3,9)
\vspace{-0.6cm}
\begin{center}
\resizebox{0.31\textwidth}{0.51\textwidth}{\includegraphics{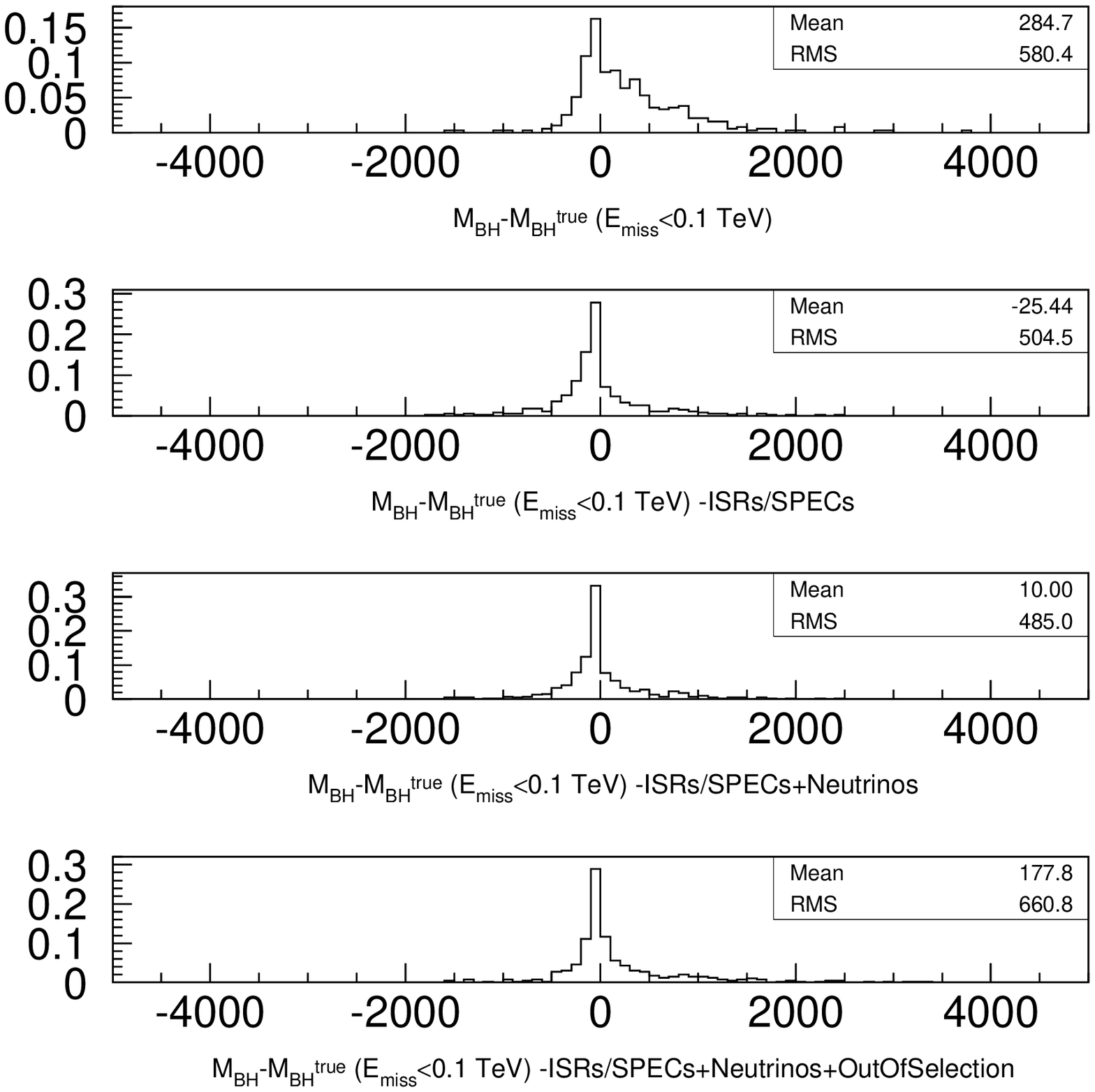}}
\resizebox{0.31\textwidth}{0.51\textwidth}{\includegraphics{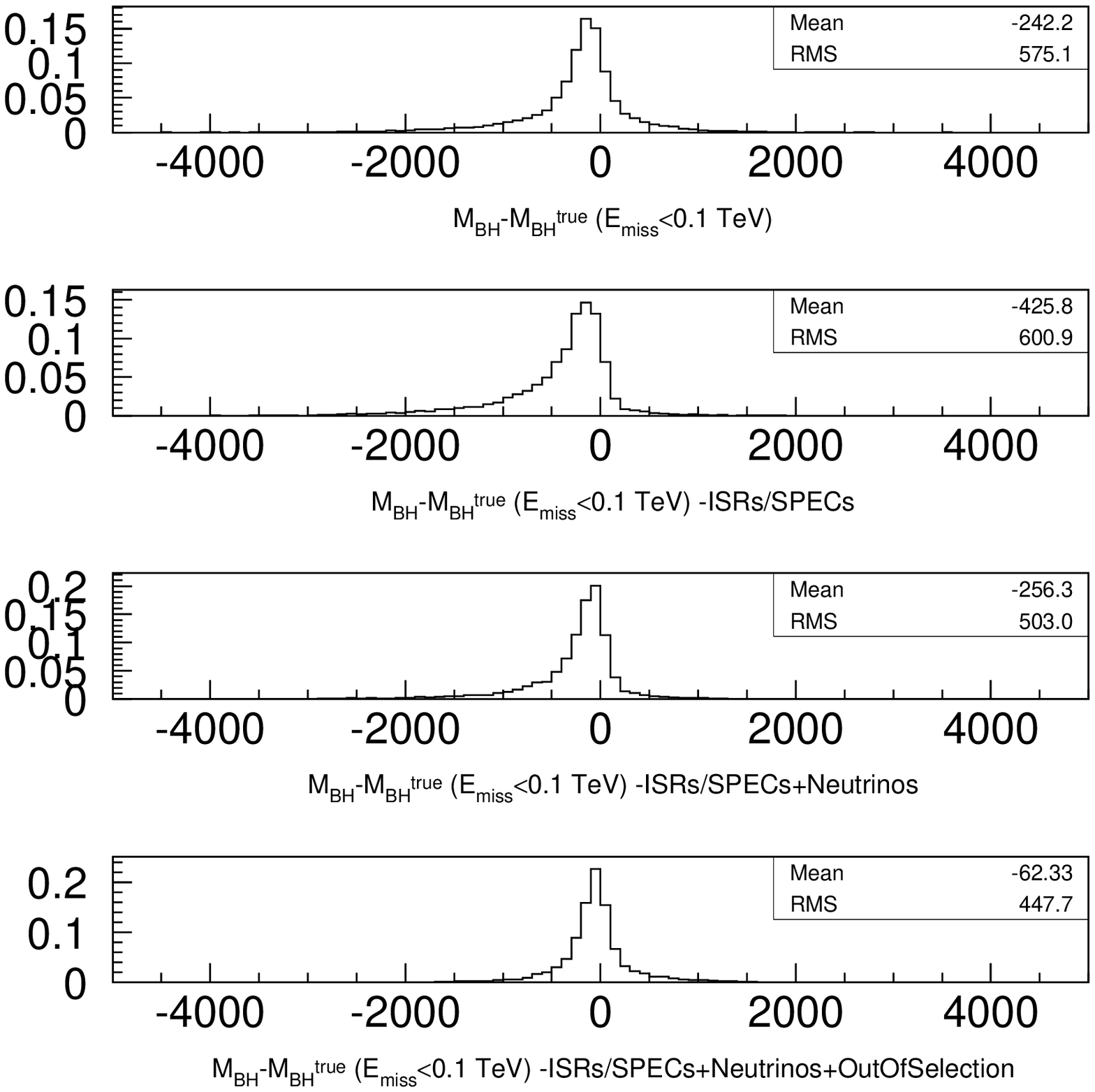}}
\resizebox{0.31\textwidth}{0.51\textwidth}{\includegraphics{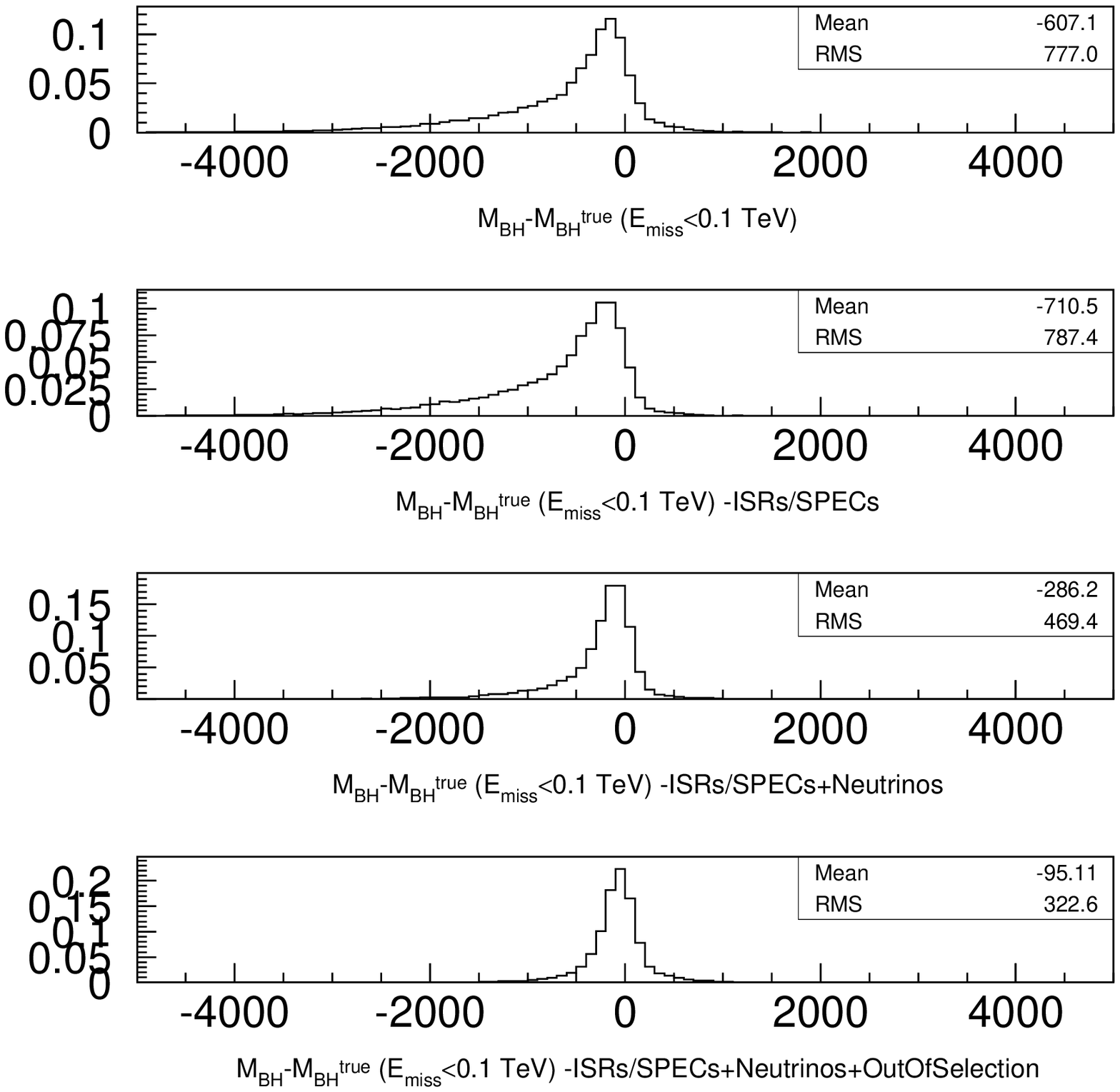}}
\end{center}
\caption{$\mbh-\MBHTru$ Distributions.
The top figures are the same with Figures~\ref{fig:dif-mbh-dist}.
The 2nd, 3rd and bottom figures show $\mbh-\MBHTru$ distributions corrected step-by-step.
Details are given in the text.}
\label{fig:mbhs-3types}
\end{figure}

We perform various corrections:(1) we subtract the momentum of particles
which are selected by our criteria but come from ISRs/SPECs,
from the reconstructed mass~$\mbh$,
(2) add the momentum of neutrinos,
(3) add the momentum of particles which do not come from ISRs/SPECs
but are removed by our selection criteria,
as shown at Figures~\ref{fig:mbhs-3types}.
We can see that the reconstructed mass is improved by these corrections
as we expected.
From these results we understand that
the overestimation of $\mbh$ is caused by the contamination of
particles from the initial state radiation and the spectators
while the underestimation is due to the missing energy of neutrinos.

